\documentclass[preprintnumbers,amsmath,amssymb,nofootinbib,superscriptaddress,floatfix,onecolumn,reprint,groupedaddress,eqsecnum,amsfonts,aps]{revtex4}
\pdfoutput=1
\usepackage[normalem]{ulem}
\usepackage{graphicx}
\usepackage[colorlinks,citecolor=blue,urlcolor=blue,linkcolor=blue]{hyperref}
\usepackage{latexsym}
\usepackage{amssymb}
\usepackage{mathrsfs}
\usepackage{amsmath}
\usepackage{amsthm}
\usepackage{color}
\usepackage{dcolumn}
\usepackage{bm}

\usepackage{graphicx}
\usepackage{caption}
\usepackage{subcaption}
\usepackage{upgreek}

\usepackage[normalem]{ulem}

\usepackage{enumitem}
\usepackage[normalem]{ulem}
\usepackage{subeqnarray}
\usepackage{cases}
\usepackage{verbatim}
\usepackage{latexsym}
\usepackage{mathrsfs}
\usepackage{amsmath}
\usepackage{amssymb}
\usepackage{amsfonts}
\usepackage{amsthm}
\usepackage{color}
\usepackage{dcolumn}
\usepackage{bm}

\newcommand{\arcsinh}{\mathrm{arcsinh}\,}

\newcommand{\arctanh}{\mathrm{arctanh}\,}
\newcommand{\arccoth}{\mathrm{arccoth}\,}
\newcommand{\F}{\mathrm{F}\,}
\newcommand{\K}{\mathrm{K}\,}
\newcommand{\am}{\mathrm{am}\,}
\newcommand{\scn}{\mathrm{cn}\,}
\newcommand{\ssn}{\mathrm{sn}\,}
\newcommand{\ssc}{\mathrm{sc}\,}

\definecolor{dyellow}{rgb}{1.,0.8,.0}
\definecolor{myblue}{rgb}{.1,.1,.7}
\definecolor{dcyan}{rgb}{.0,.6,.6}
\definecolor{dmagenta}{rgb}{0.6,0.0,0.6}
\definecolor{brown}{rgb}{0.6,0.2,0.}
\definecolor{darkblue}{rgb}{.0,.0,0.5}
\definecolor{darkred}{rgb}{0.75,0.0,0.0}
\definecolor{orange}{rgb}{1.,.6,.0}
\definecolor{dorange}{rgb}{0.8,.4,.0}
\definecolor{darkgreen}{rgb}{0.0,0.6,0.0}
\definecolor{purple}{rgb}{.4,.0,.4}
\definecolor{grey}{rgb}{0.5,0.5,0.5}

\begin{document}
\hyphenpenalty=1000
\title{Black hole images under spherical-shell and circular-annulus accretion models in Schwarzschild spacetime: a semianalytical approach}

\author{Bofeng Wu}
\affiliation{Guangxi Key Laboratory for Relativistic Astrophysics, School of Physical Science and Technology, Guangxi University, Nanning 530004, China}

\author{En-Wei Liang}
\affiliation{Guangxi Key Laboratory for Relativistic Astrophysics, School of Physical Science and Technology, Guangxi University, Nanning 530004, China}

\author{Xiao Zhang}
\affiliation{Guangxi Center for Mathematical Research, Guangxi University, Nanning 530004,  China}
\affiliation{Academy of Mathematics and Systems Science, Chinese Academy of Sciences, Beijing 100190, China}
\affiliation{School of Mathematical Sciences, University of Chinese Academy of Sciences, Beijing 100049, China}

\begin{abstract}
In the static and infalling spherical-shell models of optically thin accretion on Schwarzschild black hole, the formulas for the integrated intensities observed by a distant observer are derived, and by taking the monochromatic emission pattern with a $1/r^{2}$ radial profile as example, the black hole images for the spherical shell with different boundaries are plotted. For these BH images, the geometric and luminosity features are summarized, and the qualitative explanations of the luminosity variations between the static and infalling spherical-shell models are provided. A notable feature of the black hole image in the infalling spherical-shell model is that when the inner boundary of the spherical shell  is far from the bound photon orbit, the observed luminosity near the exterior of the shadow is enhanced. The circular-annulus models of optically and geometrically thin accretion on Schwarzschild black hole are further explored. For a lightlike geodesic, the analytical forms of the transfer functions working for all impact parameter values are first given, and the redshift factors in the static, infalling, and rotating circular-annulus models are then deduced. With these results, in the three situations, the formulas for the integrated intensities observed by a distant observer viewing the circular annulus at an inclination angle are derived, and the corresponding black hole images for each emission pattern provided in Phys. Rev. D \textbf{100} (2019) 024018 are plotted. Finally, for the BH images of arbitrary order,  the geometric and luminosity features are also summarized, and the qualitative explanations of the luminosity variations between different CA models are also given.
\end{abstract}
\maketitle
\section{Introduction}
The images~\cite{EventHorizonTelescope:2019dse,EventHorizonTelescope:2019uob,EventHorizonTelescope:2019jan,EventHorizonTelescope:2019ths,EventHorizonTelescope:2019pgp,EventHorizonTelescope:2019ggy,EventHorizonTelescope:2021srq,EventHorizonTelescope:2021bee,EventHorizonTelescope:2022wkp,EventHorizonTelescope:2022apq,
EventHorizonTelescope:2022wok,EventHorizonTelescope:2022exc,EventHorizonTelescope:2022urf,EventHorizonTelescope:2022xqj} of the supermassive black holes (BHs) at the centers of  M87 galaxy and the Milky Way obtained by the Event Horizon Telescope (EHT)  have increased the interest for the study of the visualization of BH. Normally, in a BH image, the dark central region, namely the BH shadow, is surrounded by a bright ring, and the optical appearance of the BH image is determined by the nature of the BH and the accretion model. In order to visualize BHs, multiple numerical algorithms are employed in a large amount of Refs.~\cite{Zeng:2020dco,Gan:2021pwu,Guo:2021bwr,Hu:2022lek,Wen:2022hkv,Hu:2023pyd}, which implies that numerical method is the universal approach to the presentation of a BH image. For some gravitational theories like General Relativity (GR), the equations of lightlike geodesics in spherically symmetric spacetimes have been available~\cite{chandrasekhar1985mathematical,Cadez:2004cg,Hackmann:2008zz,Gibbons:2011rh,munoz2014orbits}, and when the results are applied to the ray-tracing problems, the calculations are found to be more accurate and considerably faster than commonly used numerical integrations~\cite{Cadez:2004cg}, which indicates that these results could serve as useful tools for the visualization of BH.

In this paper, we intend to make use of the equations of lightlike geodesics to generate the Schwarzschild BH images under spherical-shell and circular-annulus accretion models and analyze their features. In general, for various cases of emissions from accreting matters, we are actually interested in the region near the considered BH to a distant observer, so in the actual applications, the lightlike geodesics reaching infinity are what we need to take into account. The equations of such lightlike geodesics can directly be found in Refs.~\cite{chandrasekhar1985mathematical,Cadez:2004cg,munoz2014orbits}.
Before a BH image is generated, the formula for the integrated intensity observed by a distant observer needs to be first presented. Therefore, how to derive the formula for the observed integrated intensity by means of the equations of lightlike geodesics is a crucial problem of this paper.

\subsection{The BH images in the static and infalling spherical-shell models of optically thin accretion on Schwarzschild BH}

In the study of BH physics, the spherical accretion models play important roles~\cite{Zeng:2020dco,Gan:2021pwu,Guo:2021bwr,Hu:2022lek,Wen:2022hkv,Saurabh:2020zqg,Qin:2020xzu,Jusufi:2020zln,He:2021aeo}. The static and infalling spherical models of optically thin accretion on Schwarzschild BH are explored in Ref.~\cite{Narayan:2019imo}, and the properties of the BH images are studied. It is indicated that for accreting matters that fill in the entire space outside the horizon of the BH, the geometric feature of the BH images is that the edge curve equation of the shadow is always $b=b_{\text{cri}}$, and the shadow in the infalling model is so much deeper than that in the static model. Here, $b_{\text{cri}}:=3\sqrt{3}m=3\sqrt{3}GM/c^{2}$ is the critical impact parameter, where $G$ is the gravitational constant, $M$ is the mass of the BH, and $c$ is the speed of light in vacuum.
However, in view that accreting matters should be distributed within a limited area, and in the infalling model, they can not freely fall from the infinity, it is necessary to extend the spherical accretion models to more realistic accretion models.

The first purpose of this paper is to generate the Schwarzschild BH images in the static and infalling spherical-shell (SS) models of optically thin accretion and analyze their features, where accreting matters located within a SS around the BH are considered. The radiation is assumed to be emitted isotropically in the rest frame of the accreting matter, and every lightlike geodesic can travel without being absorbed or scattered because accreting matters are optically thin. In the SS accretion models, the backward ray tracing method based on equation of radiative transfer~\cite{Jaroszynski:1997bw,Vincent:2011wz,younsi2012general,Bambi:2013nla,Pu:2016eml} can be used to calculate the integrated intensity observed by a distant observer, but since the calculations always involve the integral along a lightlike geodesic within the boundaries of the SS, the integrated intensity will vary with the change of the boundaries of the SS. In this paper, for the SS with different boundaries in the static and infalling models, the formulas for calculating the observed integrated intensities are presented, and by use of them, the Schwarzschild BH images for any emission pattern in the static and infalling SS models of optically thin accretion can be generated.

As an application example of the above formulas, the BH images in the SS accretion models for the monochromatic emission pattern with a $1/r^{2}$ radial profile are plotted. These images unveil that although the size and shape of the shadow still have nothing to do with the boundaries of the SS,  their features become richer and more diverse compared to those in the corresponding spherical accretion models. In this paper, based on a comprehensive analysis, the geometric and luminosity features of the BH images in the SS accretion models are summarized in detail. It is shown that these features strongly depend on the boundary positions of the SS. Firstly, the boundary positions of the SS play crucial roles in defining the geometric features of the BH images. For example, when the outer boundary of the SS lies outside the bound photon orbit, it determines the outer edge of the bright region, and when the inner boundary of the SS lies outside the bound photon orbit, it determines the radial position of the luminosity peak. Secondly, the boundary positions of the SS significantly influence the luminosity variations of the BH images between the static and infalling SS models. For the shadow, our images not only confirm the conclusion that the Doppler beam reduces the observed luminosity in the infalling SS model~\cite{Narayan:2019imo}, but also point out that the luminosity will be further reduced as the initial radial position of accreting matters increases. For the small region near the peak, it is indicated that when the SS is thin enough, the observed luminosity in the infalling SS model is also reduced, and it will be also further reduced as the initial radial position of accreting matters increases. As a contrast, when the SS is thick enough, the observed luminosity of the small region near the peak in the infalling SS model only has slight variations compared to that in the static model, and the luminosity is not sensitive to the initial radial position of accreting matters. The above conclusions suggest that in the infalling SS model, the luminosity of a BH image is reduced under most situations, whereas our images display that when the inner boundary of the SS is far from the bound photon orbit, the observed luminosity near the exterior of the shadow is enhanced, and it will be further enhanced as the initial radial position of accreting matters increases. Given that the unusual luminosity enhancement phenomenon is rare, it could be viewed as a notable luminosity feature of the BH image in the infalling SS model. The qualitative explanations of the above luminosity variations of the BH images between the static and infalling SS models are provided from the perspective of the gravitational and Doppler shift effects.

\subsection{The BH images in the static, infalling, and rotating circular-annulus models of optically and geometrically thin accretion on Schwarzschild BH}

In addition to the spherical accretion models, the disk accretion models are also important in the study of BH physics~\cite{Hu:2023pyd,Luminet:1979nyg,Gyulchev:2019tvk,Gyulchev:2021dvt,Paul:2019trt,Rahaman:2021kge,Liu:2021lvk,Guo:2023grt,Hu:2023bzy}. In Ref.~\cite{Gralla:2019xty}, a basic method to evaluate the integrated intensity observed by a distant observer at the face-on orientation is provided for the static disk model of optically and geometrically thin accretion on Schwarzschild BH, and the BH images for three emission patterns are presented. The results illustrate that the size of the shadow is very much dependent on the emission pattern, and the luminosities of BH images are mainly contributed by the first and second order emissions. These conclusions reveal some significant features of the BH images for thin disk accretion models, but because the observer can not always view the accretion disk at the face-on orientation, and accreting matters can not always be static, the results in Ref.~\cite{Gralla:2019xty} need to be further generalized. For the rotating disk accretion model, the BH images are plotted by means of analytical methods in Ref.~\cite{Luminet:1979nyg}, and motivated by the techniques in this reference, we intend to extend the static disk model in Ref.~\cite{Gralla:2019xty} to more realistic accretion models.

The second purpose of this paper is to generate the Schwarzschild BH images in the static, infalling, and rotating circular-annulus (CA) models of optically and geometrically thin accretion and analyze their features, where accreting matters located within a CA centered at the center of the BH are considered. Suppose that the CA lies in the equatorial plane and the radiation is emitted isotropically in the frame comoving with the accreting matter. When a lightlike geodesic is traced from a distant observer backwards towards the region near the BH, it could intersect the equatorial plane many times. If the emitting point of the geodesic is the $k$th $(k=1,2,3,\cdots)$ intersection point and the geodesic only picks up luminosity from the emitting point, the geodesic is referred to as the $k$th order lightlike geodesic and the corresponding emission is called the $k$th order emission. In the spirit of Ref.~\cite{Gralla:2019xty}, for a $k$th order lightlike geodesic emitted from the CA,  the radial coordinate of the emitting point should be determined by the polar coordinates of the receiving point on the screen of the observer with the help of the $k$th order transfer function. In this work, by means of the equations of the lightlike geodesics, the analytical forms of the transfer functions working for all impact parameter values are presented, where it should be noted that the similar results for the rotating disk accretion model in Ref.~\cite{Luminet:1979nyg} only hold for the case of $b>b_{\text{cri}}$.

In order to derive the integrated intensities observed by a distant observer viewing the CA at an inclination angle in the static, infalling, and rotating CA models, the redshift factors in the three situations need to be taken into account. The redshift factors for the static and rotating models can be easily handled~\cite{Gralla:2019xty,Luminet:1979nyg,Liu:2021lvk,Tian:2019yhn}, but the evaluation of the redshift factor in the infalling model is a little subtle. When accreting matters radially move in towards the BH, the redshift factor is dependent on whether the emitting point is on the inward or outward segment of the considered lightlike geodesic, which implies that for a lightlike geodesic with two segments, while the periastron coincides with the emitting point, how to determine the value of the impact parameter is crucial. In the present work, with the aid of the equations of the lightlike geodesics this difficulty is overcome, and the redshift factor for the infalling model is derived.
With the transfer functions and the redshift factors, after considering the contributions from the emissions at all orders, the formulas for deriving the observed integrated intensities in the static, infalling, and rotating CA models are achieved by means of the preliminary method in Ref.~\cite{Gralla:2019xty}. On the basis of these formulas, once emitted specific intensity in a CA accretion model is given, the observed integrated intensity can be evaluated.

If the BH image plotted only based on the $k$th order emission is called the $k$th order BH image, the above formulas indicate that the luminosity of a complete BH image in the CA accretion models should contain the contributions from the BH images at all orders. In fact, as stated in Ref.~\cite{Gralla:2019xty}, for the fourth and higher order BH images, the bright regions are extremely demagnified so that their contributions are totally negligible. Thus, by just considering the first three order emissions, for each emission pattern provided  in Ref.~\cite{Gralla:2019xty}, the BH images in the static, infalling, and rotating CA models of optically and geometrically thin accretion are presented. These images display that the boundary positions of the CA also play crucial roles in defining the geometric features of the BH images. For example, for a given order BH image, they determine the size and shape of the bright region, and in particular, for a complete BH image, the inner boundary position of the CA determines the internal structure of the shadow. In general, the shadow of a BH image in the CA models usually consists of several distinct dark parts with zero luminosities, which is different from the case in the SS accretion models.

The positions of the boundaries of the CA can also influence the luminosity variations of the BH images between different CA accretion models. Based on an analysis on the BH images in the CA models, we present the following conclusions. For BH images of arbitrary order, if the outer boundary of the CA is sufficiently far from the bound photon orbit, the luminosity near the outer edge of the bright region in the infalling model is higher than that in the static model, and it will be further enhanced as the initial radial position of accreting matters increases. On the contrary,  if the inner boundary of the CA is extended to a region near the bound photon orbit, the luminosity near the inner edge of the bright region in the infalling model is lower than that in the static model, and it will be further reduced as the initial radial position of accreting matters increases. It should be emphasized that the above two points do not apply to the entire region contributed by the lightlike geodesics that travel around the BH less than $\pi/2$ in the first order BH image. Within this entire region in the first order BH image, the luminosity in the infalling model is always lower than that in the static model, and it will be further reduced
with the increase of the initial radial position of accreting matters. As to the luminosity variations of the BH images between the static and rotating CA models, it is shown that for BH images of arbitrary order, in a region near the position where the polar angle on the observational screen is equal to $\pi/2$ or $3\pi/2$,  the observed luminosity in the rotating model is always lower or higher than that in the static model, and as the polar angle tends toward this position, the luminosity will gradually decrease or increase. While presenting the above conclusions, the qualitative explanations of the above luminosity variations of the BH images between different CA models are also provided from the perspective of the gravitational and Doppler shift effects.

This paper is organized as follows. In Sec.~\ref{Sec:second}, the Schwarzschild BH images in the static and infalling SS models of optically thin accretion are generated and their features are analyzed. In Sec.~\ref{Sec:third}, the Schwarzschild BH images in the static, infalling, and rotating CA models of optically and geometrically thin accretion are generated and their features are analyzed. In Sec.~\ref{Sec:fourth}, the summary and the related discussions are presented.  In the Appendix, the equations of the lightlike geodesics in Schwarzschild spacetime are reviewed, and they are presented in a form suitable for practical application.

Throughout this paper, the international system of units is used, and the signature of the metric $g_{\mu\nu}$ is $(-,+,+,+)$. The spacetime indices are marked by Greek indices $\alpha,\beta,\gamma,\cdots$ running from 0 to 3 and the space indices are denoted by Latin indices $i,j,k,\cdots$ running from 1 to 3. The sum should be taken over when the repeated indices appear within a term.
\section{The BH images in the static and infalling SS models of optically thin accretion~\label{Sec:second}}
In the study of BH physics, the spherical accretion models play important roles~\cite{Zeng:2020dco,Gan:2021pwu,Guo:2021bwr,Hu:2022lek,Wen:2022hkv,Saurabh:2020zqg,Qin:2020xzu,Jusufi:2020zln,He:2021aeo}. As shown in Ref.~\cite{Narayan:2019imo}, the static and infalling spherical models of optically thin accretion on Schwarzschild BH are explored, and some properties of the BH images are presented. The results indicate that for accreting matters that fill in the entire space outside the horizon of the BH, the geometric feature of the BH images is that the shadow edge is located at $b=b_{\text{cri}}$, and the luminosity of the shadow in the infalling model is so much deeper than that in the static model. In more realistic situations, accreting matters  should be distributed within a limited area, and for the infalling model, they can not freely fall from the infinity. Therefore, it is necessary to extend the spherical accretion models to more realistic accretion models. In this section, we will make use of the equations of the lightlike geodesics to generate the Schwarzschild BH images in the static and infalling spherical-shell (SS) models of optically thin accretion and analyze their features, where accreting matters located within a SS around the BH are considered.

\subsection{The formulas for the integrated intensity observed by a distant observer}

The radiation emitted from the SS is assumed to be isotropically in the rest frame of the accreting matter. By following the method presented in Refs.~\cite{Jaroszynski:1997bw,Bambi:2013nla}, the specific intensity on the screen of a distant observer could be evaluated by adopting the backward ray tracing method based on equation of radiative transfer~\cite{Jaroszynski:1997bw,Vincent:2011wz,younsi2012general,Bambi:2013nla,Pu:2016eml}. This method states that the specific intensity of radiation $I(b,\nu_{\text{o}})$ observed by the observer at the frequency $\nu_{\text{o}}$ along a lightlike geodesic with the impact parameter $b$ is given by
\begin{eqnarray}
\label{equ2.1}I(b,\nu_{\text{o}})=\int_{\text{geodesic}}g^{3}j(\nu_{\text{e}})\text{d}l,
\end{eqnarray}
where $g:=\nu_{\text{o}}/\nu_{\text{e}}$ is the redshift factor, and in the rest frame of the emitter, $\nu_{\text{e}}$ is the photon frequency, $j(\nu_{\text{e}})$ is the emissivity per unit volume, and $\text{d}l$ is the infinitesimal proper length of the lightlike geodesic measured in the frame comoving with the emitter. As mentioned above, formula~(\ref{equ2.1}) originates from equation of radiative transfer. It is because accreting matters are assumed to be optically thin that any lightlike geodesic can travel without being absorbed or scattered, which means that in the equation of radiative transfer, only the term related to the emissivity per unit volume needs to be kept. As a consequence, by performing the integral along the lightlike geodesic, formula~(\ref{equ2.1}) is found. With $I(b,\nu_{\text{o}})$, the integrated intensity of radiation on the screen of the observer can be achieved by~\cite{Wen:2022hkv,Gralla:2019xty}
\begin{eqnarray}
\label{equ2.2}F(b)=\int_{0}^{+\infty}I(b,\nu_{\text{o}})\text{d}\nu_{\text{o}}=\int_{0}^{+\infty}gI(b,\nu_{\text{o}})\text{d}\nu_{\text{e}}.
\end{eqnarray}

\begin{figure}[tbp]
	\centering
	\begin{subfigure}{0.32\linewidth}
		\centering
		\includegraphics[width=1\linewidth]{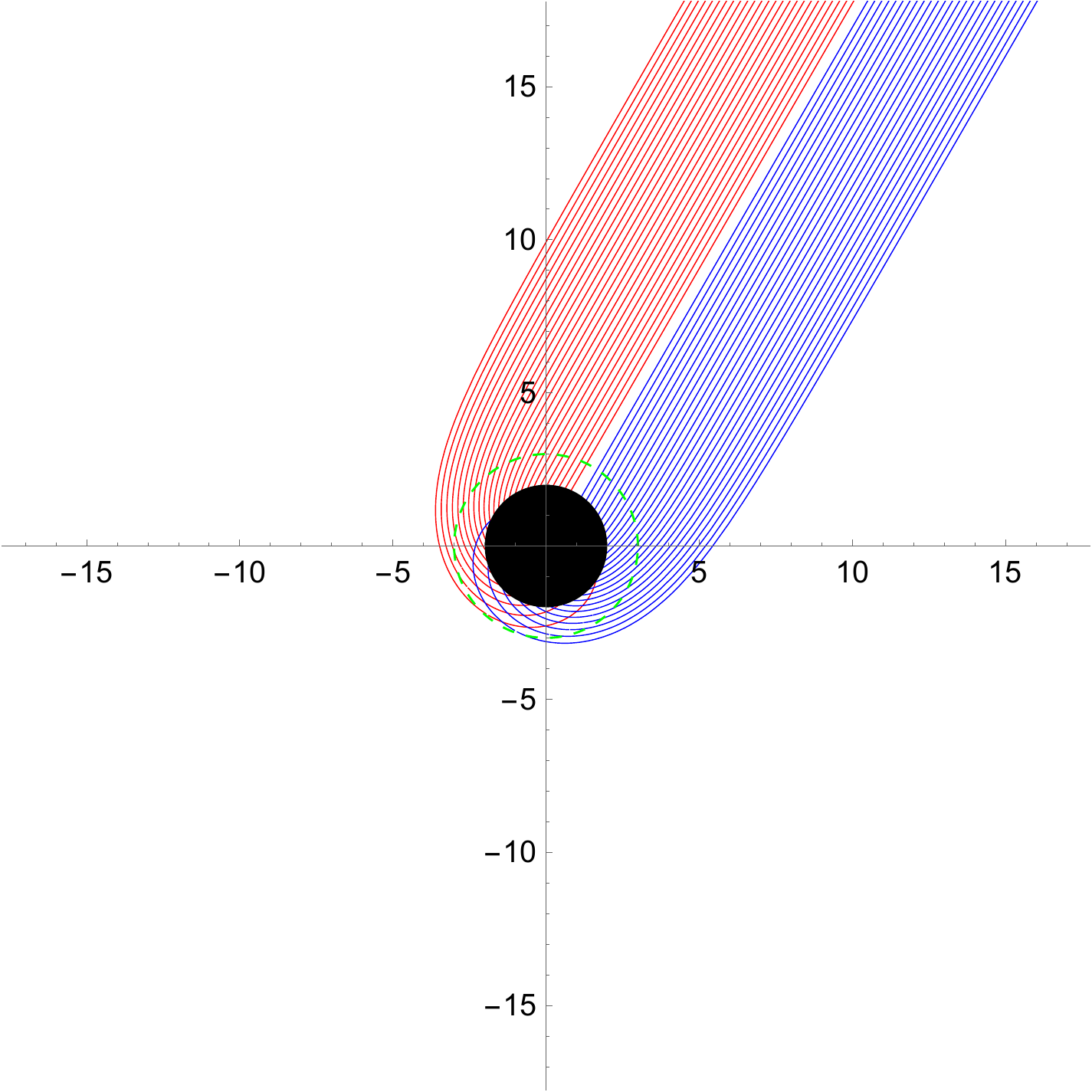}
	\end{subfigure}
	\centering
	\begin{subfigure}{0.32\linewidth}
		\centering
		\includegraphics[width=1\linewidth]{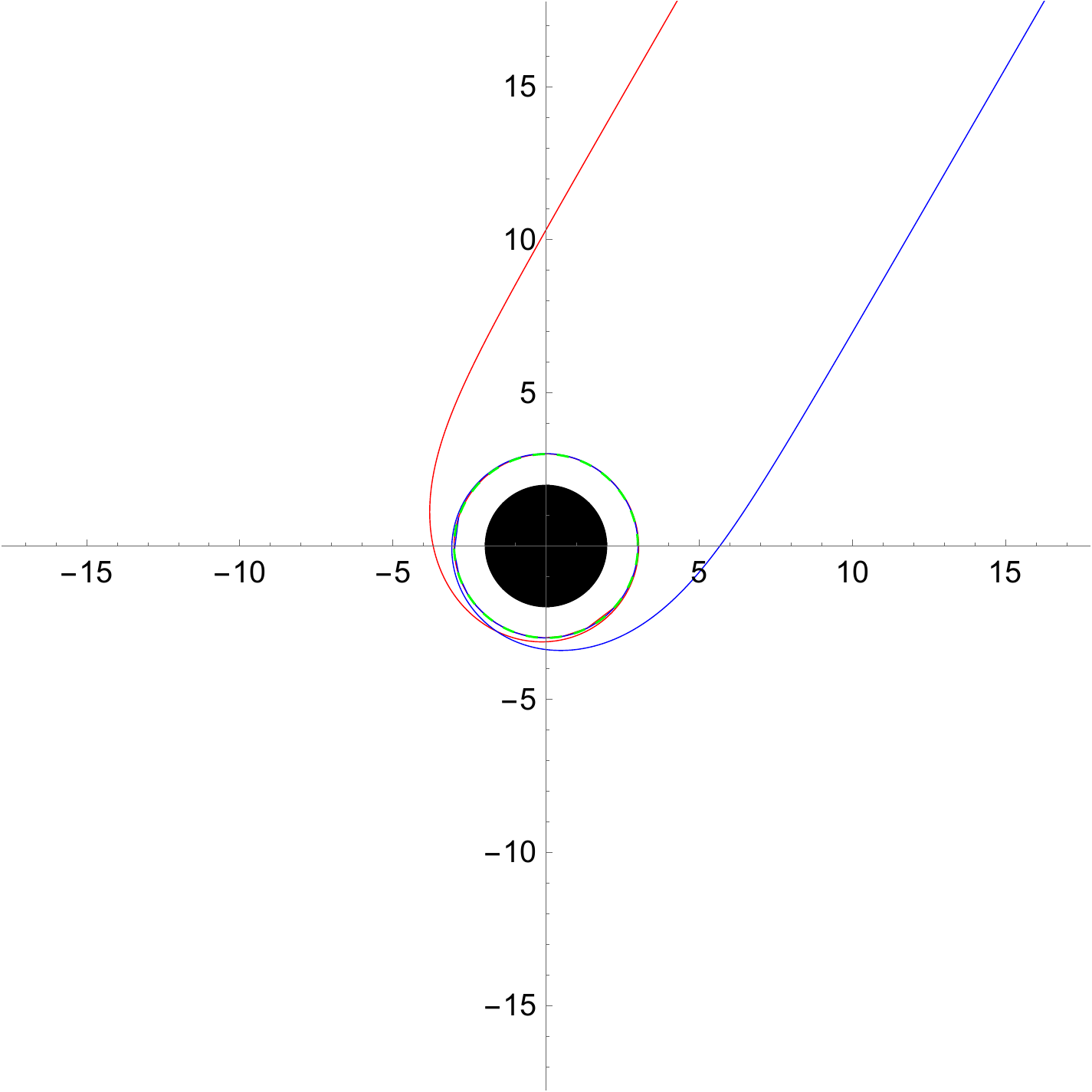}
	\end{subfigure}
    \centering
	\begin{subfigure}{0.32\linewidth}
		\centering
		\includegraphics[width=1\linewidth]{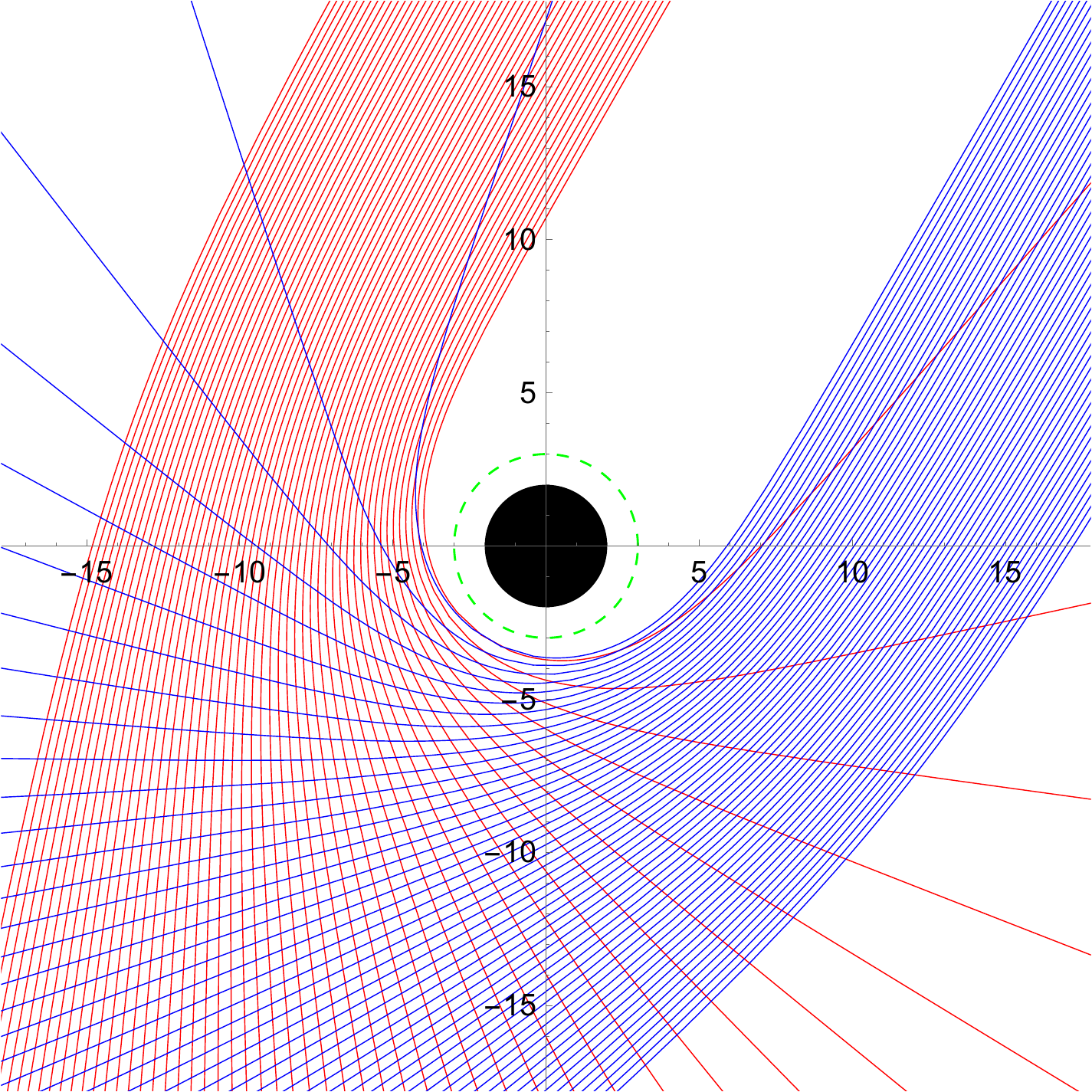}
	\end{subfigure}
    \caption{The trajectories of the lightlike geodesics through the spatial point $(+\infty,0,\pi/3)$ under the cases of
    $0\leqslant b<b_{\text{cri}}$ (left), $b=b_{\text{cri}}$ (middle), and $b>b_{\text{cri}}$ (right) in the Cartesian coordinate system on the screen of the observer. The spacing in impact parameter is $0.2$ for all lightlike geodesics. The red and blue lines, respectively, represent the clockwise and counterclockwise lightlike geodesics, while photons move towards the point $(+\infty,0,\pi/3)$. The BH and the bound photon orbit are shown as a black disk and a dashed green circle. $\phantom{11111111111111111111111111111111111}$}
    \label{fig1}
\end{figure}

Let $r^{\text{SS}}_{\text{inn}}$ and $r^{\text{SS}}_{\text{out}}$ be the radial coordinates of the inner and outer boundaries of the SS.  In general, since $r^{\text{SS}}_{\text{inn}}$ does not extend to the horizon of the BH, and $r^{\text{SS}}_{\text{out}}$ does not extend to infinity, compared with the cases in the spherical models of optically thin accretion in Ref.~\cite{Narayan:2019imo}, the upper and lower limits in the integral of formula~(\ref{equ2.1}) will change. The first task of this section is to determine the integral upper and lower limits. Before the formal discussions, it should be emphasized that only the lightlike geodesics reaching infinity are what we need to take into account, because the observer is far away from the BH. As examples, by means of Eqs.~(\ref{equA54}), (\ref{equA56}), and (\ref{equA57}) in the appendix, the images of the trajectories for the lightlike geodesics through the spatial point $(+\infty,0,\pi/3)$ under the cases of $0\leqslant b<b_{\text{cri}}$, $b=b_{\text{cri}}$, and $b>b_{\text{cri}}$ are plotted in Fig.~\ref{fig1}.
Explicitly, apart from the radial coordinates of the inner and outer boundaries of the SS, the nature of the lightlike geodesics will also influence the values of the integral upper and lower limits in formula~(\ref{equ2.1}).
When $r^{\text{SS}}_{\text{inn}}<r^{\text{SS}}_{\text{out}}<r_{\text{pho}}$, only the lightlike geodesics with $0\leqslant b<b_{\text{cri}}$ pass through the SS, and all the photons are always moving away from the BH, where $r_{\text{pho}}:=3m$ is the equation of the bound photon orbit. So, if denote $\mathcal{P}(r)$ as a point along the considered lightlike geodesic, formula~(\ref{equ2.1}) can be cast as
\begin{eqnarray}
\label{equ2.3}I(b,\nu_{\text{o}})=\left\{
\begin{array}{ll}
\displaystyle\int_{\mathcal{P}\left(r^{\text{SS}}_{\text{inn}}\right)}^{\mathcal{P}\left(r^{\text{SS}}_{\text{out}}\right)}\left(g^{(\text{outw})}\right)^{3}j(\nu_{\text{e}})\text{d}l^{(\text{outw})},&\quad \text{for}\quad    0\leqslant b< b_{\text{cri}},\medskip\\
0,&\quad \text{for}\quad  b_{\text{cri}}\leqslant b.
\end{array}\right.
\end{eqnarray}
In this equation, we adopt the rule that for a physical quantity $A$ defined on the lightlike geodesic, $(A)^{(\text{outw})}$  or $(A)^{(\text{inw})}$ represents that it takes values at points on the outward or inward segment of the geodesic, respectively. This rule applies to the whole paper. When $r^{\text{SS}}_{\text{inn}}\leqslant r_{\text{pho}}\leqslant r^{\text{SS}}_{\text{out}}$, the behaviors of lightlike geodesics are stated as follows.
1) The lightlike geodesics with $0\leqslant b<b_{\text{cri}}$ and $b=b_{\text{cri}}$ pass through the SS, and all the photons are always moving away from the BH. Note that in the situation of $b=b_{\text{cri}}$, photons may travel around the BH an infinite number of times within the SS; 2) For a lightlike geodesic with $b>b_{\text{cri}}$, if the radial coordinate of its periastron happens to be $r^{\text{SS}}_{\text{out}}$, according to Eq.~(\ref{equA30}), its impact parameter should be
\begin{eqnarray}
\label{equ2.4}b^{\text{SS}}_{\text{max}}:=\frac{r^{\text{SS}}_{\text{out}}}{\displaystyle\sqrt{1-\frac{2m}{r^{\text{SS}}_{\text{out}}}}}.
\end{eqnarray}
It can be inferred that the lightlike geodesics with $b_{\text{cri}}< b\leqslant b^{\text{SS}}_{\text{max}}$ pass through the SS, whereas the lightlike geodesics with $b>b^{\text{SS}}_{\text{max}}$ do not pass through the SS, where in the former situation,
since the periastrons of the geodesics are located inside the SS, photons will first move in toward the BH, and then move away from it. Based on these conclusions,
formula~(\ref{equ2.1}) can be written as
\begin{eqnarray}
\label{equ2.5}I(b,\nu_{\text{o}})=\left\{
\begin{array}{ll}
\displaystyle\int_{\mathcal{P}\left(r^{\text{SS}}_{\text{inn}}\right)}^{\mathcal{P}\left(r^{\text{SS}}_{\text{out}}\right)}\left(g^{(\text{outw})}\right)^{3}j(\nu_{\text{e}})\text{d}l^{(\text{outw})},&\quad \text{for}\quad   0\leqslant b< b_{\text{cri}},\medskip\\
\displaystyle+\infty,&\quad \text{for}\quad   b=b_{\text{cri}},\medskip\\
\displaystyle\int^{\mathcal{P}\left(r_{2}\right)}_{\mathcal{P}\left(r^{\text{SS}}_{\text{out}}\right)}\left(g^{(\text{inw})}\right)^{3}j(\nu_{\text{e}})\text{d}l^{(\text{inw})}+\int_{\mathcal{P}\left(r_{2}\right)}^{\mathcal{P}\left(r^{\text{SS}}_{\text{out}}\right)}\left(g^{(\text{outw})}\right)^{3}j(\nu_{\text{e}})\text{d}l^{(\text{outw})},&\quad \text{for}\quad  b_{\text{cri}}< b\leqslant b^{\text{SS}}_{\text{max}},\medskip\\
0,&\quad \text{for}\quad b^{\text{SS}}_{\text{max}}<b,
\end{array}\right.
\end{eqnarray}
and here, according to Eq.~(\ref{equA26}), $r_{2}=m/X_{2}$ is the radial coordinate of the periastron for the considered geodesic. When $r_{\text{pho}}<r^{\text{SS}}_{\text{inn}}<r^{\text{SS}}_{\text{out}}$, the expressions of $I(b,\nu_{0})$ are not exactly the same as
those in Eq.~(\ref{equ2.5}), because the behaviors of the lightlike geodesics with $b_{\text{cri}}\leqslant b\leqslant b^{\text{SS}}_{\text{max}}$ are different from those in the above case. If the radial coordinate of the periastron for a lightlike geodesic with $b>b_{\text{cri}}$ is $r^{\text{SS}}_{\text{inn}}$,
\begin{eqnarray}
\label{equ2.6}b^{\text{SS}}_{\text{min}}:=\frac{r^{\text{SS}}_{\text{inn}}}{\displaystyle\sqrt{1-\frac{2m}{r^{\text{SS}}_{\text{inn}}}}}
\end{eqnarray}
should be its impact parameter. Then, one could conclude that the periastrons of the lightlike geodesics with $b_{\text{cri}}<b< b^{\text{SS}}_{\text{min}}$  are located inside the inner boundary of the SS, and
the periastrons of the lightlike geodesics with $b^{\text{SS}}_{\text{min}}\leqslant b\leqslant b^{\text{SS}}_{\text{max}}$ are located inside the SS. As a result, in the present case, formula~(\ref{equ2.5}) needs to be modified to be
\begin{eqnarray}
\label{equ2.7}I(b,\nu_{\text{o}})=\left\{
\begin{array}{ll}
\displaystyle\int_{\mathcal{P}\left(r^{\text{SS}}_{\text{inn}}\right)}^{\mathcal{P}\left(r^{\text{SS}}_{\text{out}}\right)}\left(g^{(\text{outw})}\right)^{3}j(\nu_{\text{e}})\text{d}l^{(\text{outw})},&\quad\text{for}\quad 0\leqslant b\leqslant  b_{\text{cri}},\medskip\\
\displaystyle\int^{\mathcal{P}\left(r^{\text{SS}}_{\text{inn}}\right)}_{\mathcal{P}\left(r^{\text{SS}}_{\text{out}}\right)}\left(g^{(\text{inw})}\right)^{3}j(\nu_{\text{e}})\text{d}l^{(\text{inw})}+\int_{\mathcal{P}\left(r^{\text{SS}}_{\text{inn}}\right)}^{\mathcal{P}\left(r^{\text{SS}}_{\text{out}}\right)}\left(g^{(\text{outw})}\right)^{3}j(\nu_{\text{e}})\text{d}l^{(\text{outw})},&\quad\text{for}\quad b_{\text{cri}}<b< b^{\text{SS}}_{\text{min}},\medskip\\
\displaystyle\int^{\mathcal{P}\left(r_{2}\right)}_{\mathcal{P}\left(r^{\text{SS}}_{\text{out}}\right)}\left(g^{(\text{inw})}\right)^{3}j(\nu_{\text{e}})\text{d}l^{(\text{inw})}+\int_{\mathcal{P}\left(r_{2}\right)}^{\mathcal{P}\left(r^{\text{SS}}_{\text{out}}\right)}\left(g^{(\text{outw})}\right)^{3}j(\nu_{\text{e}})\text{d}l^{(\text{outw})},&\quad\text{for}\quad b^{\text{SS}}_{\text{min}}\leqslant b\leqslant b^{\text{SS}}_{\text{max}},\medskip\\
0,&\quad \text{for}\quad   b^{\text{SS}}_{\text{max}}<b.
\end{array}\right.
\end{eqnarray}

Next, we will employ formulas~(\ref{equ2.2}), (\ref{equ2.3}), (\ref{equ2.5}), and (\ref{equ2.7}) to derive the integrated intensities observed by the distant observer in the static and infalling SS models of optically thin accretion. Simple calculations indicate that the four-velocities of the static and infalling emitters are, respectively,
\begin{eqnarray}
\label{equ2.8}u^{\mu}_{\text{es}}&=&(u^{0}_{\text{es}},0,0,0),\\
\label{equ2.9}u^{\mu}_{\text{ef}}&=&(u^{0}_{\text{ef}},u^{r}_{\text{ef}},0,0)
\end{eqnarray}
with
\begin{eqnarray}
\label{equ2.10}u^{0}_{\text{es}}=\frac{c}{\displaystyle\sqrt{1-\frac{2m}{r}}},\quad u^{0}_{\text{ef}}=\frac{\mathcal{E}^{\text{SS}}}{\displaystyle c\left(1-\frac{2m}{r}\right)},\quad u^{r}_{\text{ef}}= -c\sqrt{\frac{(\mathcal{E}^{\text{SS}})^{2}}{c^4}-\left(1-\frac{2m}{r}\right)},
\end{eqnarray}
where if the coordinate of the initial radial position of accreting matters in the infalling model is $r^{\text{SS}}_{\text{ini}}$, there is
\begin{eqnarray}
\label{equ2.11}\mathcal{E}^{\text{SS}}:=c^{2}\sqrt{1-\frac{2m}{r^{\text{SS}}_{\text{ini}}}}.
\end{eqnarray}
In the above equations, the subscripts ``s'' and ``f'' imply that the corresponding quantities are defined in the static and infalling models, respectively, and the rule applies to this section. Since the observer is far away from the BH, his spatial coordinates could be set to be $(+\infty,0,\varphi_{0})$, and then, his four-velocity is
\begin{eqnarray}
\label{equ2.12}u^{\mu}_{\text{o}}=(c,0,0,0).
\end{eqnarray}
By virtue of Eqs.~(\ref{equ2.8})--(\ref{equ2.12}), the redshift factors in the two models are given by
\begin{eqnarray}
\label{equ2.13}g^{(\text{outw})}_{\text{s}}&=&g^{(\text{inw})}_{\text{s}}=\frac{-p_{\mu}u^{\mu}_{\text{o}}}{-p_{\nu}u^{\nu}_{\text{es}}}=\frac{1}{u^{0}_{\text{es}}/c}=\sqrt{1-\frac{2m}{r}},\\
\label{equ2.14}g^{(\text{outw})}_{\text{f}}&=&\frac{-p_{\mu}u^{\mu}_{\text{o}}}{-p_{\nu}u^{\nu}_{\text{ef}}}=\frac{1}{\displaystyle\frac{u^{0}_{\text{ef}}}{c}-\left(-\frac{p_{r}}{p_{0}}\right)^{(\text{outw})}\frac{u^{r}_{\text{ef}}}{c}}\nonumber\\
&=&\left(1-\frac{2m}{r}\right)\left(\sqrt{1-\frac{2m}{r^{\text{SS}}_{\text{ini}}}}+\sqrt{1-\frac{b^2}{r^2}\left(1-\frac{2m}{r}\right)}\sqrt{\frac{2m}{r}-\frac{2m}{r^{\text{SS}}_{\text{ini}}}}\right)^{-1},\\
\label{equ2.15}g^{(\text{inw})}_{\text{f}}&=&\frac{-p_{\mu}u^{\mu}_{\text{o}}}{-p_{\nu}u^{\nu}_{\text{ef}}}=\frac{1}{\displaystyle\frac{u^{0}_{\text{ef}}}{c}-\left(-\frac{p_{r}}{p_{0}}\right)^{(\text{inw})}\frac{u^{r}_{\text{ef}}}{c}}\nonumber\\
&=&\left(1-\frac{2m}{r}\right)\left(\sqrt{1-\frac{2m}{r^{\text{SS}}_{\text{ini}}}}-\sqrt{1-\frac{b^2}{r^2}\left(1-\frac{2m}{r}\right)}\sqrt{\frac{2m}{r}-\frac{2m}{r^{\text{SS}}_{\text{ini}}}}\right)^{-1},
\end{eqnarray}
where in the derivations, the following identities for the photon four-momentum, namely,
\begin{eqnarray}
\label{equ2.16}p_{0}&=&-\frac{E}{c}=-\frac{L}{b},\\
\label{equ2.17}\left(-\frac{p_{0}}{p^{r}}\right)^{(\text{outw})}&=&-\left(-\frac{p_{0}}{p^{r}}\right)^{(\text{inw})}=\frac{1}{\displaystyle\sqrt{1-\frac{b^2}{r^2}\left(1-\frac{2m}{r}\right)}},\\
\label{equ2.18}\left(-\frac{p_{r}}{p_{0}}\right)^{(\text{outw})}&=&-\left(-\frac{p_{r}}{p_{0}}\right)^{(\text{inw})}=\frac{\displaystyle\sqrt{1-\frac{b^2}{r^2}\left(1-\frac{2m}{r}\right)}}{\displaystyle1-\frac{2m}{r}}
\end{eqnarray}
have been utilized, and they can directly be obtained from Eqs.~(\ref{equA9}) and the definition of the impact parameter $b$. As to the infinitesimal proper length of a lightlike geodesic measured in the frame comoving with the emitter, by use of $p^{r}=\text{d}r/\text{d}\lambda$, it is provided by
\begin{eqnarray}
\label{equ2.19}\text{d}l&:=&-p_{\alpha}\left(\frac{u^{\alpha}_{\text{e}}}{c}\right)\text{d}\lambda=-\frac{p_{\alpha}}{p^{r}}\frac{u^{\alpha}_{\text{e}}}{c}\text{d}r,
\end{eqnarray}
where $u^{\alpha}_{\text{e}}$ is the four-velocity of the emitter and $\lambda$ refers to an affine parameter of the geodesic, and thus, in the static and infalling models, there are
\begin{eqnarray}
\label{equ2.20}\text{d}l_{\text{s}}^{(\text{outw})}&=&\left(-\frac{p_{0}}{p^{r}}\right)^{(\text{outw})}\frac{u^{0}_{\text{es}}}{c}\text{d}r=\frac{1}{g^{(\text{outw})}_{\text{s}}}\left(-\frac{p_{0}}{p^{r}}\right)^{(\text{outw})}\text{d}r,\\
\label{equ2.21}\text{d}l_{\text{s}}^{(\text{inw})}&=&\left(-\frac{p_{0}}{p^{r}}\right)^{(\text{inw})}\frac{u^{0}_{\text{es}}}{c}\text{d}r=\frac{1}{g^{(\text{inw})}_{\text{s}}}\left(-\frac{p_{0}}{p^{r}}\right)^{(\text{inw})}\text{d}r,\\
\label{equ2.22}\text{d}l_{\text{f}}^{(\text{outw})}&=&\left(-\frac{p_{0}}{p^{r}}\right)^{(\text{outw})}\left[\frac{u^{0}_{\text{ef}}}{c}-\displaystyle\left(-\frac{p_{r}}{p_{0}}\right)^{(\text{outw})}\frac{u^{r}_{\text{ef}}}{c}\right]\text{d}r=\frac{1}{g^{(\text{outw})}_{\text{f}}}\left(-\frac{p_{0}}{p^{r}}\right)^{(\text{outw})}\text{d}r,\\
\label{equ2.23}\text{d}l_{\text{f}}^{(\text{inw})}&=&\left(-\frac{p_{0}}{p^{r}}\right)^{(\text{inw})}\left[\frac{u^{0}_{\text{ef}}}{c}-\displaystyle\left(-\frac{p_{r}}{p_{0}}\right)^{(\text{inw})}\frac{u^{r}_{\text{ef}}}{c}\right]\text{d}r=\frac{1}{g^{(\text{inw})}_{\text{f}}}\left(-\frac{p_{0}}{p^{r}}\right)^{(\text{inw})}\text{d}r.
\end{eqnarray}
In Ref.~\cite{Bambi:2013nla}, the result corresponding to Eqs.~(\ref{equ2.22}) and (\ref{equ2.23}) is incorrect because $p^{r}$ is replaced by $p_{r}$, which results in that relevant conclusions based on this result in a large amount of references are all faulty.
After plugging Eqs.~(\ref{equ2.13})--(\ref{equ2.23}) into formulas~(\ref{equ2.3}), (\ref{equ2.5}), and (\ref{equ2.7}) and then employing formula~(\ref{equ2.2}), the integrated intensities observed by the distant observer in the two models are presented as follows.
\begin{itemize}
\item When $r^{\text{SS}}_{\text{inn}}<r^{\text{SS}}_{\text{out}}<r_{\text{pho}}$, there are
\begin{eqnarray}
\label{equ2.24}F_{\text{s}}(b)&=&\left\{
\begin{array}{ll}
\displaystyle\int_{0}^{+\infty}\int_{r^{\text{SS}}_{\text{inn}}}^{r^{\text{SS}}_{\text{out}}}\left(g_{\text{s}}^{(\text{outw})}\right)^{3}\left(-\frac{p_{0}}{p^{r}}\right)^{(\text{outw})}j(\nu_{\text{e}})\text{d}r\text{d}\nu_{\text{e}},&\quad \text{for}\quad  0\leqslant b< b_{\text{cri}},\medskip\\
0,&\quad \text{for}\quad  b_{\text{cri}}\leqslant b,
\end{array}\right.\\
\label{equ2.25}F_{\text{f}}(b)&=&\left\{
\begin{array}{ll}
\displaystyle\int_{0}^{+\infty}\int_{r^{\text{SS}}_{\text{inn}}}^{r^{\text{SS}}_{\text{out}}}\left(g_{\text{f}}^{(\text{outw})}\right)^{3}\left(-\frac{p_{0}}{p^{r}}\right)^{(\text{outw})}j(\nu_{\text{e}})\text{d}r\text{d}\nu_{\text{e}},&\quad \text{for}\quad   0\leqslant b< b_{\text{cri}},\medskip\\
0,&\quad \text{for}\quad b_{\text{cri}}\leqslant b.
\end{array}\right.
\end{eqnarray}

\item When $r^{\text{SS}}_{\text{inn}}\leqslant r_{\text{pho}}\leqslant r^{\text{SS}}_{\text{out}}$, there are
\begin{eqnarray}
\label{equ2.26}F_{\text{s}}(b)&=&\left\{
\begin{array}{ll}
\displaystyle\int_{0}^{+\infty}\int_{r^{\text{SS}}_{\text{inn}}}^{r^{\text{SS}}_{\text{out}}}\left(g_{\text{s}}^{(\text{outw})}\right)^{3}\left(-\frac{p_{0}}{p^{r}}\right)^{(\text{outw})}j(\nu_{\text{e}})\text{d}r\text{d}\nu_{\text{e}},& \quad \text{for}\quad 0\leqslant b< b_{\text{cri}},\medskip\\
\displaystyle+\infty,&\quad \text{for}\quad  b=b_{\text{cri}},\medskip\\
\displaystyle2\int_{0}^{+\infty}\int_{r_{2}}^{r^{\text{SS}}_{\text{out}}}\left(g_{\text{s}}^{(\text{outw})}\right)^{3}\left(-\frac{p_{0}}{p^{r}}\right)^{(\text{outw})}j(\nu_{\text{e}})\text{d}r\text{d}\nu_{\text{e}},& \quad \text{for}\quad b_{\text{cri}}< b\leqslant b^{\text{SS}}_{\text{max}},\medskip\\
0,&\quad \text{for}\quad b^{\text{SS}}_{\text{max}}<b,
\end{array}\right.\\
\label{equ2.27}F_{\text{f}}(b)&=&\left\{
\begin{array}{ll}
\displaystyle\int_{0}^{+\infty}\int_{r^{\text{SS}}_{\text{inn}}}^{r^{\text{SS}}_{\text{out}}}\left(g_{\text{f}}^{(\text{outw})}\right)^{3}\left(-\frac{p_{0}}{p^{r}}\right)^{(\text{outw})}j(\nu_{\text{e}})\text{d}r\text{d}\nu_{\text{e}},& \quad \text{for}\quad 0\leqslant b< b_{\text{cri}},\medskip\\
\displaystyle+\infty,&\quad \text{for}\quad b=b_{\text{cri}},\medskip\\
\displaystyle\int_{0}^{+\infty}\int_{r_{2}}^{r^{\text{SS}}_{\text{out}}}\left[\left(g_{\text{f}}^{(\text{inw})}\right)^{3}+\left(g_{\text{f}}^{(\text{outw})}\right)^{3}\right]\left(-\frac{p_{0}}{p^{r}}\right)^{(\text{outw})}j(\nu_{\text{e}})\text{d}r\text{d}\nu_{\text{e}},& \quad \text{for}\quad b_{\text{cri}}< b\leqslant b^{\text{SS}}_{\text{max}},\medskip\\
0,&\quad \text{for}\quad  b^{\text{SS}}_{\text{max}}<b.
\end{array}\right.
\end{eqnarray}

\item When $r_{\text{pho}}<r^{\text{SS}}_{\text{inn}}<r^{\text{SS}}_{\text{out}}$, there are
\begin{eqnarray}
\label{equ2.28}F_{\text{s}}(b)&=&\left\{
\begin{array}{ll}
\displaystyle\int_{0}^{+\infty}\int_{r^{\text{SS}}_{\text{inn}}}^{r^{\text{SS}}_{\text{out}}}\left(g_{\text{s}}^{(\text{outw})}\right)^{3}\left(-\frac{p_{0}}{p^{r}}\right)^{(\text{outw})}j(\nu_{\text{e}})\text{d}r\text{d}\nu_{\text{e}},& \quad \text{for}\quad  0\leqslant b\leqslant b_{\text{cri}},\medskip\\
\displaystyle2\int_{0}^{+\infty}\int_{r^{\text{SS}}_{\text{inn}}}^{r^{\text{SS}}_{\text{out}}}\left(g_{\text{s}}^{(\text{outw})}\right)^{3}\left(-\frac{p_{0}}{p^{r}}\right)^{(\text{outw})}j(\nu_{\text{e}})\text{d}r\text{d}\nu_{\text{e}},& \quad \text{for}\quad  b_{\text{cri}}< b< b^{\text{SS}}_{\text{min}},\medskip\\
\displaystyle2\int_{0}^{+\infty}\int_{r_{2}}^{r^{\text{SS}}_{\text{out}}}\left(g_{\text{s}}^{(\text{outw})}\right)^{3}\left(-\frac{p_{0}}{p^{r}}\right)^{(\text{outw})}j(\nu_{\text{e}})\text{d}r\text{d}\nu_{\text{e}},& \quad \text{for}\quad b^{\text{SS}}_{\text{min}}\leqslant b\leqslant b^{\text{SS}}_{\text{max}},\medskip\\
0,&\quad \text{for}\quad  b^{\text{SS}}_{\text{max}}<b,
\end{array}\right.\\
\label{equ2.29}F_{\text{f}}(b)&=&\left\{
\begin{array}{ll}
\displaystyle\int_{0}^{+\infty}\int_{r^{\text{SS}}_{\text{inn}}}^{r^{\text{SS}}_{\text{out}}}\left(g_{\text{f}}^{(\text{outw})}\right)^{3}\left(-\frac{p_{0}}{p^{r}}\right)^{(\text{outw})}j(\nu_{\text{e}})\text{d}r\text{d}\nu_{\text{e}},& \quad \text{for}\quad  0\leqslant b\leqslant b_{\text{cri}},\medskip\\
\displaystyle\int_{0}^{+\infty}\int_{r^{\text{SS}}_{\text{inn}}}^{r^{\text{SS}}_{\text{out}}}\left[\left(g_{\text{f}}^{(\text{inw})}\right)^{3}+\left(g_{\text{f}}^{(\text{outw})}\right)^{3}\right]\left(-\frac{p_{0}}{p^{r}}\right)^{(\text{outw})}j(\nu_{\text{e}})\text{d}r\text{d}\nu_{\text{e}},& \quad \text{for}\quad   b_{\text{cri}}< b< b^{\text{SS}}_{\text{min}},\medskip\\
\displaystyle\int_{0}^{+\infty}\int_{r_{2}}^{r^{\text{SS}}_{\text{out}}}\left[\left(g_{\text{f}}^{(\text{inw})}\right)^{3}+\hspace{-0.05cm}\left(g_{\text{f}}^{(\text{outw})}\right)^{3}\right]\left(-\frac{p_{0}}{p^{r}}\right)^{(\text{outw})}j(\nu_{\text{e}})\text{d}r\text{d}\nu_{\text{e}},& \quad \text{for}\quad b^{\text{SS}}_{\text{min}}\leqslant b\leqslant b^{\text{SS}}_{\text{max}},\medskip\\
0,&\quad \text{for}\quad   b^{\text{SS}}_{\text{max}}<b.
\end{array}\right.
\end{eqnarray}
\end{itemize}
\begin{figure}[tbp]
	\centering
	\begin{subfigure}{0.32\linewidth}
		\centering
		\includegraphics[width=1\linewidth]{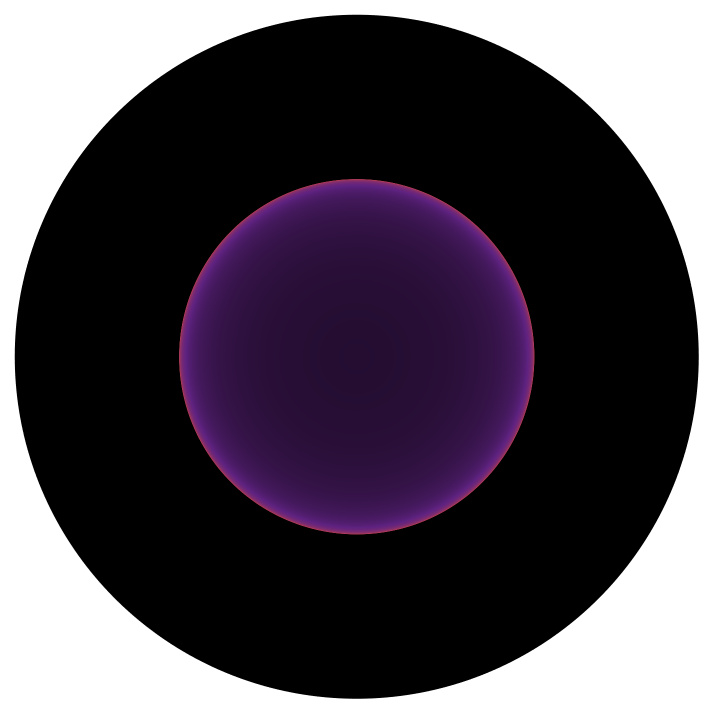}
		\caption{Static model}
	\end{subfigure}
	\centering
	\begin{subfigure}{0.32\linewidth}
		\centering
		\includegraphics[width=1\linewidth]{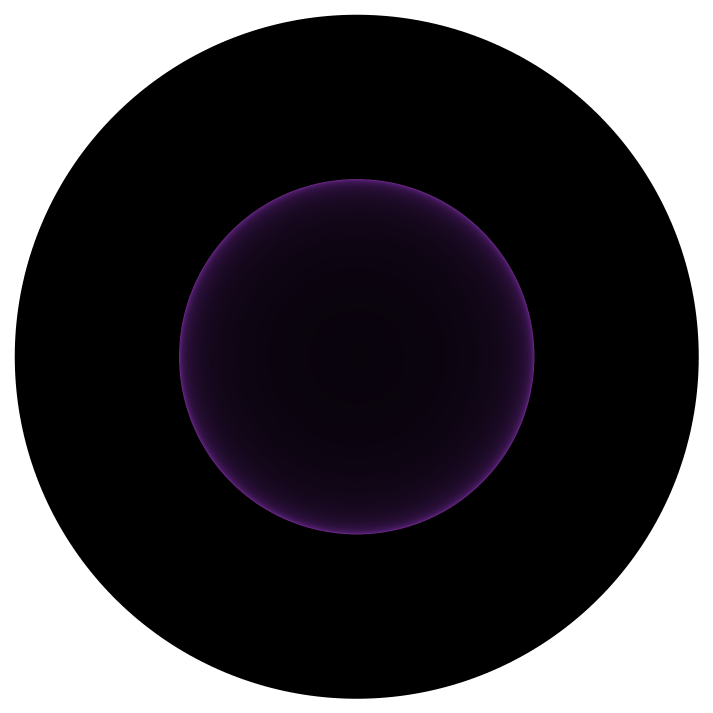}
		\caption{Infalling model with $r^{\text{SS}}_{\text{ini}}=2.9m$}
	\end{subfigure}
    \centering
	\begin{subfigure}{0.32\linewidth}
		\centering
		\includegraphics[width=1\linewidth]{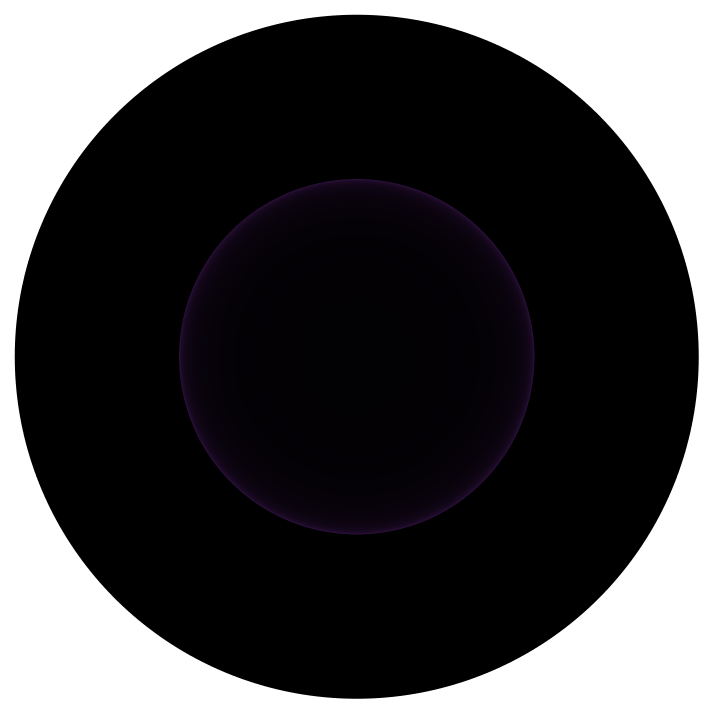}
		\caption{Infalling model with $r^{\text{SS}}_{\text{ini}}=4m$}
	\end{subfigure}
	\caption{The Schwarzschild BH images observed by a distant observer in the SS models of optically thin accretion with  $r^{\text{SS}}_{\text{inn}}=2.1m$ and $r^{\text{SS}}_{\text{out}}=2.9m$. The dark central regions are BH shadows with radius $b_{\text{cri}}\approx5.196m$.$\phantom{111111111111}$}
	\label{fig2}
\end{figure}

\subsection{The features of the Schwarzschild BH images in the static and infalling SS models of optically thin accretion}

\begin{figure}[tbp]
	\centering
	\begin{subfigure}{0.32\linewidth}
		\centering
		\includegraphics[width=1\linewidth]{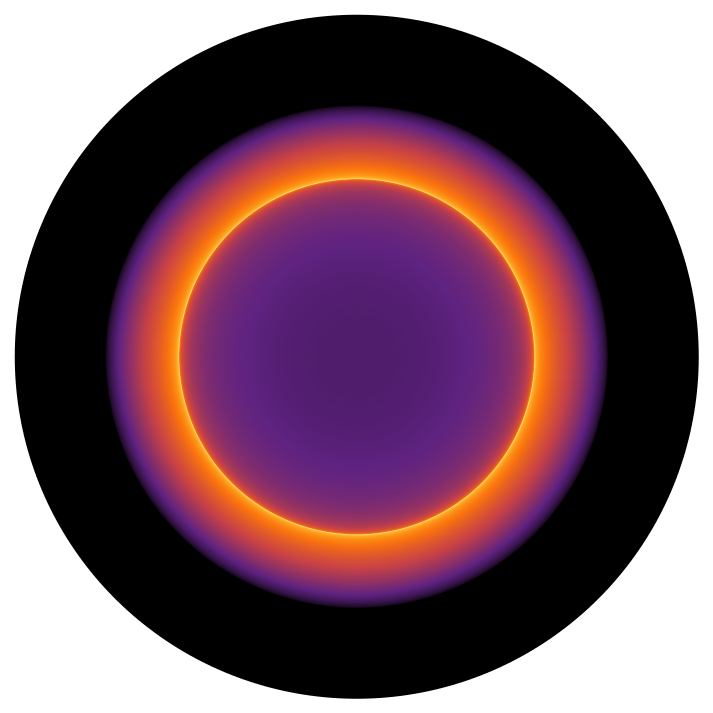}
		\caption{Static model}
	\end{subfigure}
	\centering
	\begin{subfigure}{0.32\linewidth}
		\centering
		\includegraphics[width=1\linewidth]{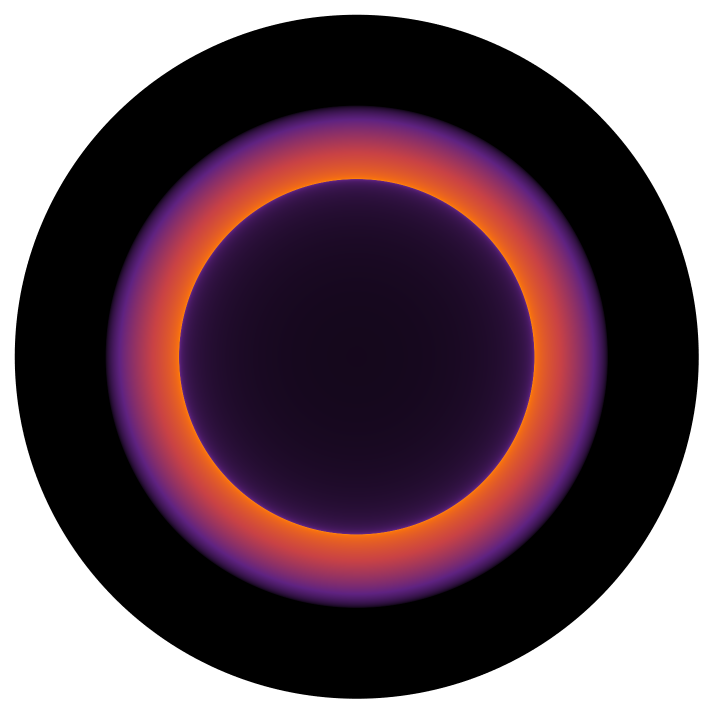}
		\caption{Infalling model with $r^{\text{SS}}_{\text{ini}}=6m$}
	\end{subfigure}
    \centering
	\begin{subfigure}{0.32\linewidth}
		\centering
		\includegraphics[width=1\linewidth]{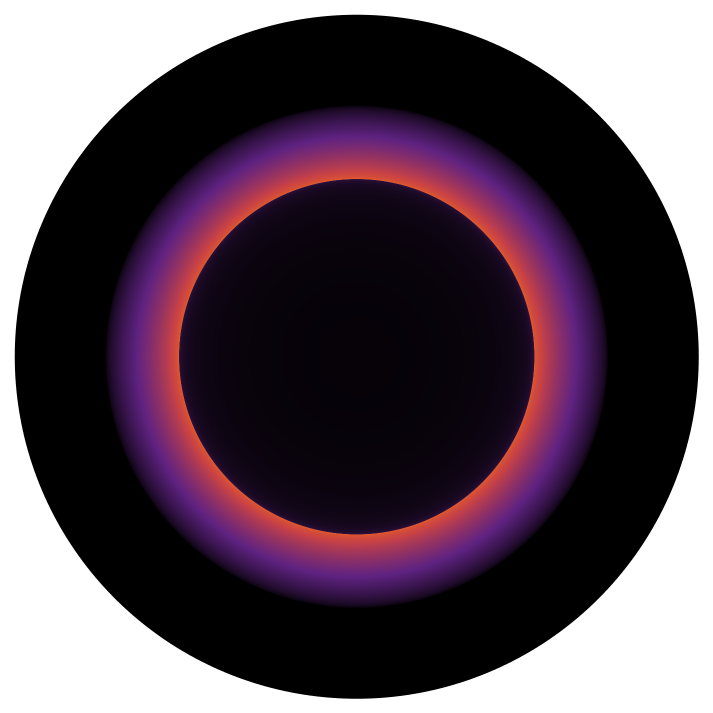}
		\caption{Infalling model with $r^{\text{SS}}_{\text{ini}}=10^{3}m$}
	\end{subfigure}
	\caption{The Schwarzschild BH images observed by a distant observer in the SS models of optically thin accretion with  $r^{\text{SS}}_{\text{inn}}=2.5m$ and $r^{\text{SS}}_{\text{out}}=6m$. The dark central regions are BH shadows with radius $b_{\text{cri}}\approx5.196m$, and the circles enclosing the bright regions have radius $b_{\text{max}}^{\text{SS}}\approx7.348m$. $\phantom{1111111111111111111111111111111111111111111111111111}$}
	\label{fig3}
\end{figure}

\begin{figure}[tbp]
	\centering
	\begin{subfigure}{0.32\linewidth}
		\centering
		\includegraphics[width=1\linewidth]{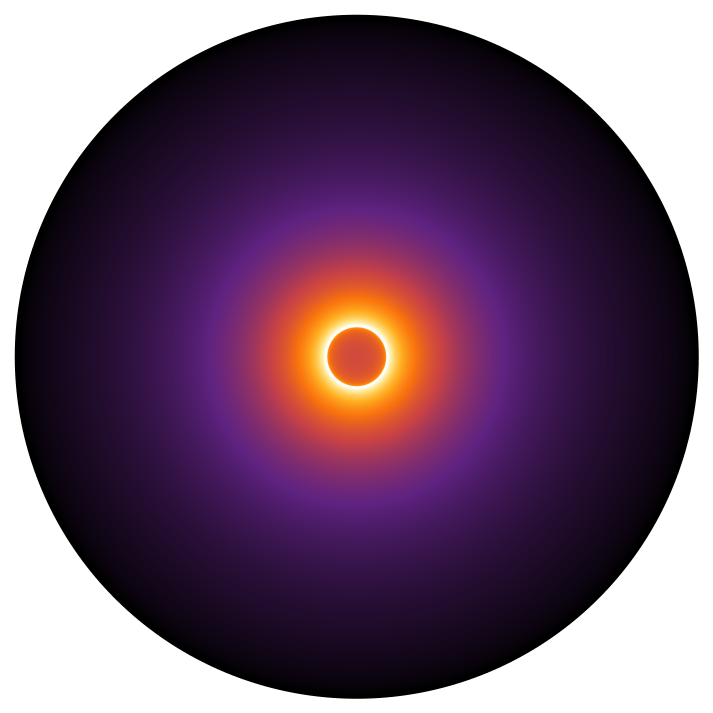}
		\caption{Static model}
	\end{subfigure}
	\centering
	\begin{subfigure}{0.32\linewidth}
		\centering
		\includegraphics[width=1\linewidth]{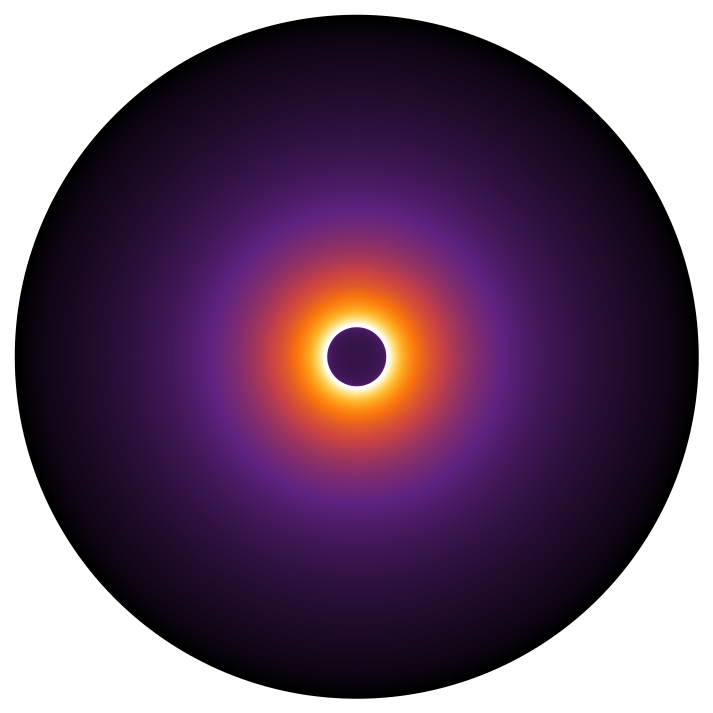}
		\caption{Infalling model with $r^{\text{SS}}_{\text{ini}}=58m$}
	\end{subfigure}
    \centering
	\begin{subfigure}{0.32\linewidth}
		\centering
		\includegraphics[width=1\linewidth]{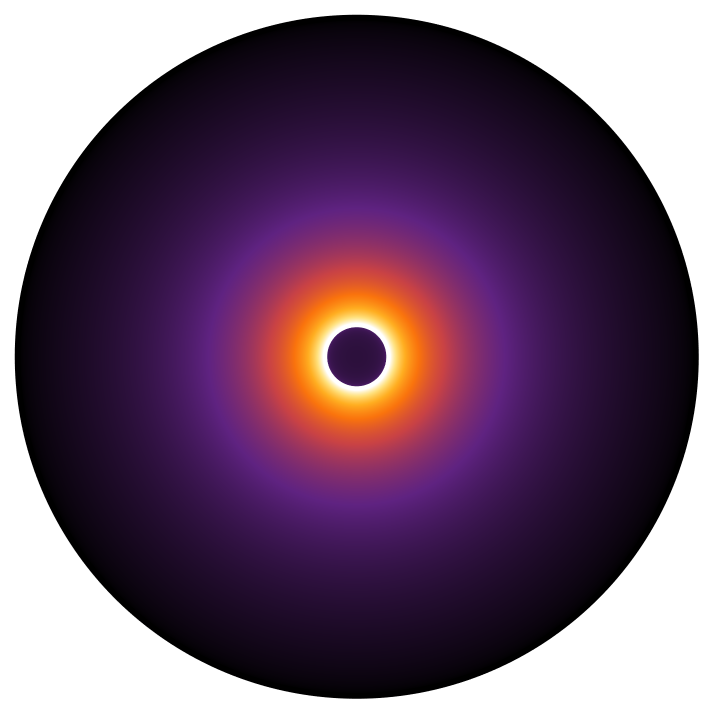}
		\caption{Infalling model with $r^{\text{SS}}_{\text{ini}}=10^{3}m$}
	\end{subfigure}
	\caption{The Schwarzschild BH images observed by a distant observer in the SS models of optically thin accretion with  $r^{\text{SS}}_{\text{inn}}=2.5m$ and $r^{\text{SS}}_{\text{out}}=58m$. The dark central regions are BH shadows with radius $b_{\text{cri}}\approx5.196m$, and the circles enclosing the bright regions have radius $b_{\text{max}}^{\text{SS}}\approx59.027m$. $\phantom{1111111111111111111111111111111111111111111111111111}$}
	\label{fig4}
\end{figure}

\begin{figure}[tbp]
	\centering
	\begin{subfigure}{0.32\linewidth}
		\centering
		\includegraphics[width=1\linewidth]{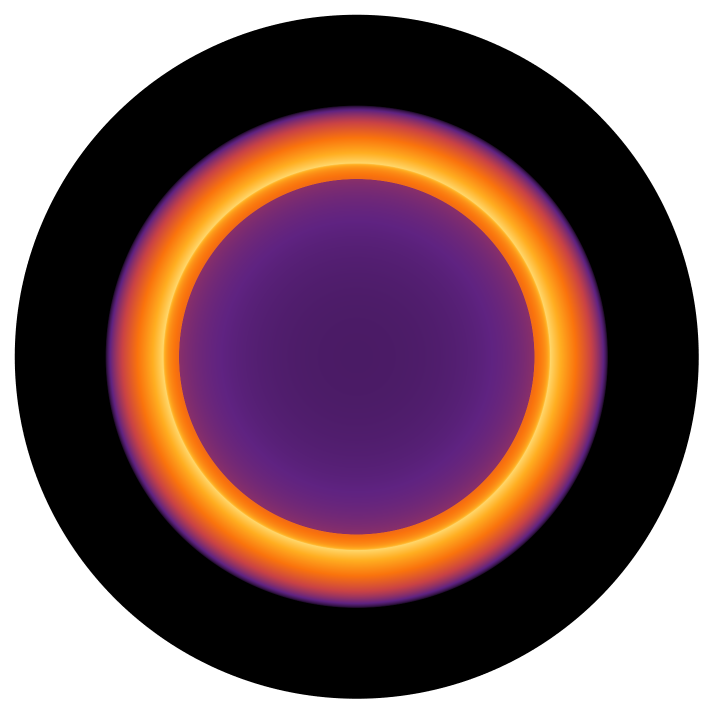}
		\caption{Static model}
	\end{subfigure}
	\centering
	\begin{subfigure}{0.32\linewidth}
		\centering
		\includegraphics[width=1\linewidth]{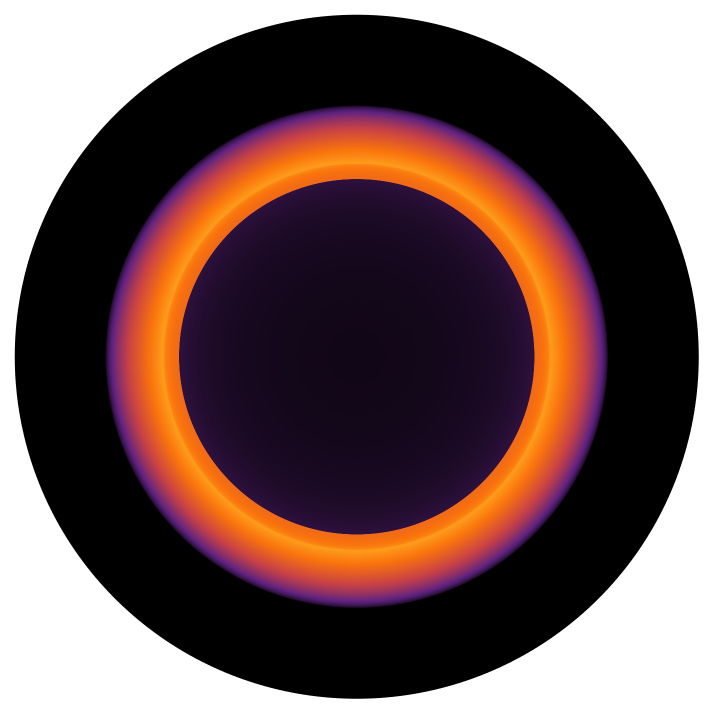}
		\caption{Infalling model with $r^{\text{SS}}_{\text{ini}}=8m$}
	\end{subfigure}
    \centering
	\begin{subfigure}{0.32\linewidth}
		\centering
		\includegraphics[width=1\linewidth]{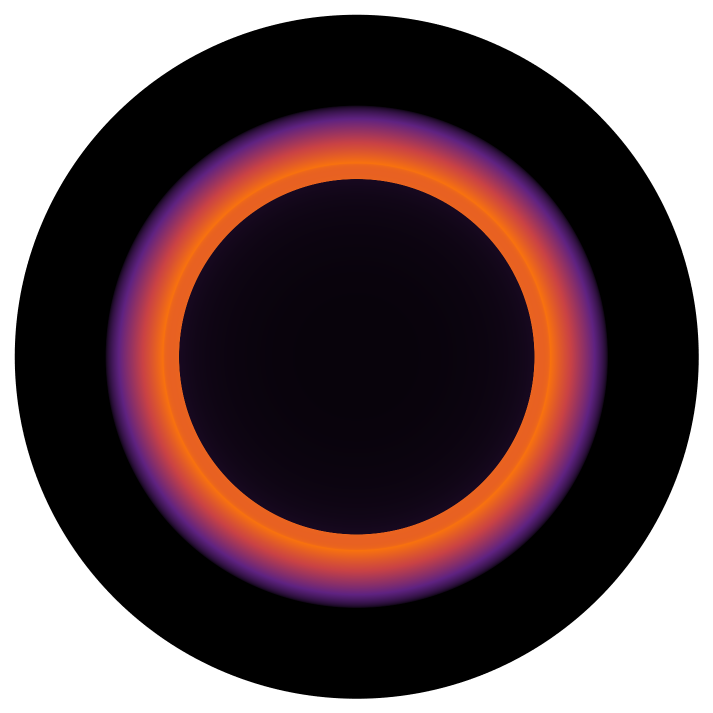}
		\caption{Infalling model with $r^{\text{SS}}_{\text{ini}}=10^{3}m$}
	\end{subfigure}
	\caption{The Schwarzschild BH images observed by a distant observer in the SS models of optically thin accretion with  $r^{\text{SS}}_{\text{inn}}=4m$ and $r^{\text{SS}}_{\text{out}}=6m$. The dark central regions are BH shadows with radius $b_{\text{cri}}\approx5.196m$, the circles enclosing the bright regions have radius $b_{\text{max}}^{\text{SS}}\approx7.348m$, and the radial position of the luminosity peak is at $b_{\text{min}}^{\text{SS}}\approx5.657m$. $\phantom{11111111}$}
	\label{fig5}
\end{figure}

\begin{figure}[tbp]
	\centering
	\begin{subfigure}{0.32\linewidth}
		\centering
		\includegraphics[width=1\linewidth]{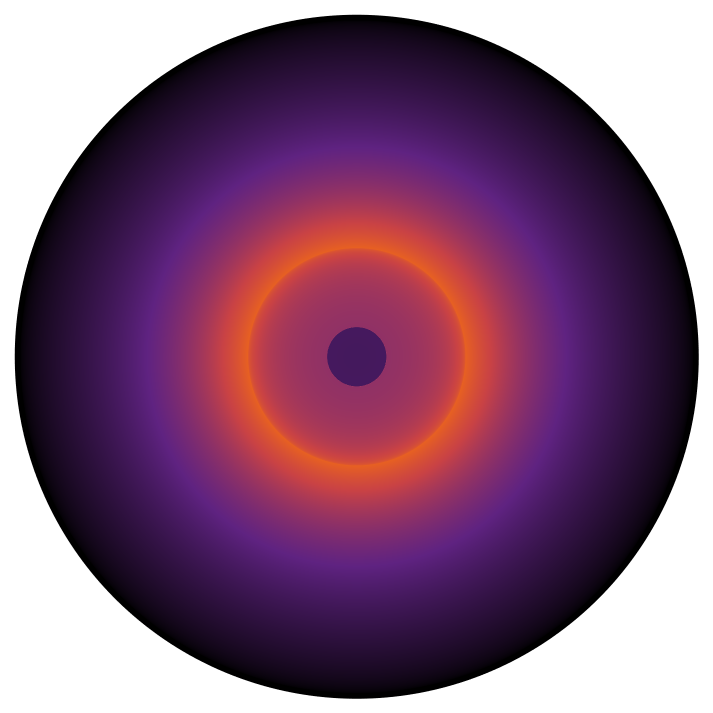}
		\caption{Static model}
	\end{subfigure}
	\centering
	\begin{subfigure}{0.32\linewidth}
		\centering
		\includegraphics[width=1\linewidth]{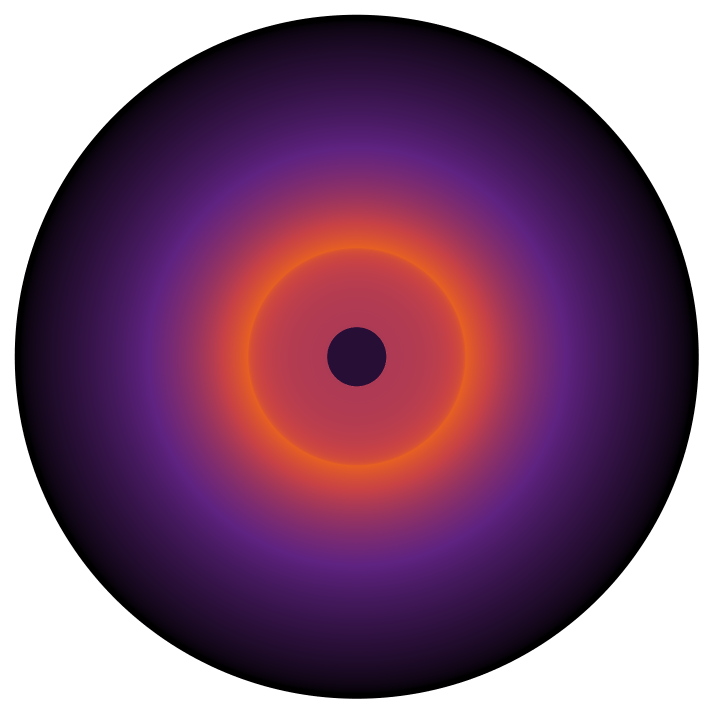}
		\caption{Infalling model with $r^{\text{SS}}_{\text{ini}}=62m$}
	\end{subfigure}
    \centering
	\begin{subfigure}{0.32\linewidth}
		\centering
		\includegraphics[width=1\linewidth]{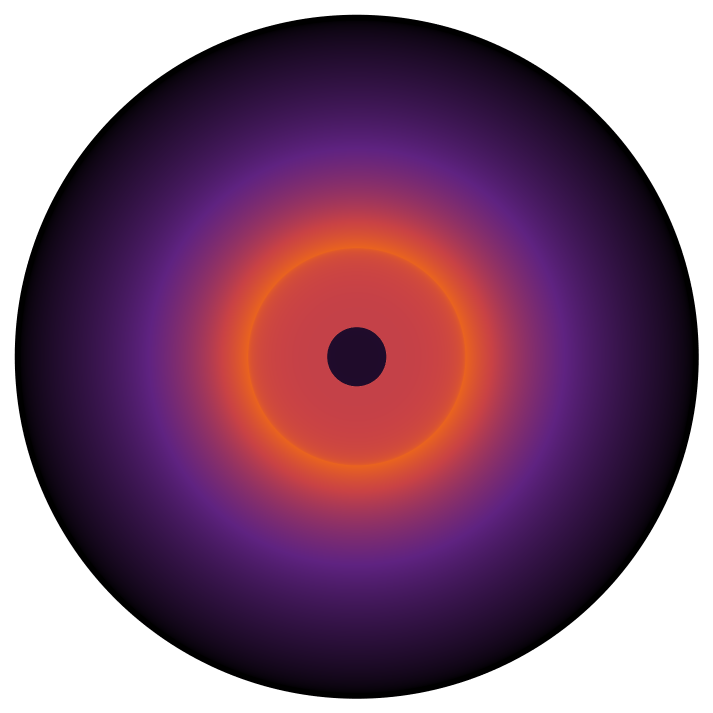}
		\caption{Infalling model with $r^{\text{SS}}_{\text{ini}}=10^{3}m$}
	\end{subfigure}
	\caption{The Schwarzschild BH images observed by a distant observer in the SS models of optically thin accretion with  $r^{\text{SS}}_{\text{inn}}=18m$ and $r^{\text{SS}}_{\text{out}}=58m$. The dark central regions are BH shadows with radius $b_{\text{cri}}\approx5.196m$, the circles enclosing the bright regions have radius $b_{\text{max}}^{\text{SS}}\approx59.027m$, and the radial position of the luminosity peak is at $b_{\text{min}}^{\text{SS}}\approx19.092m$. $\phantom{111111111111111111111111111111111111111111111111111111111111111111111111111111111111111}$}
	\label{fig6}
\end{figure}

\begin{figure}[tbp]
	\centering
	\begin{subfigure}{0.32\linewidth}
		\centering
		\includegraphics[width=1\linewidth]{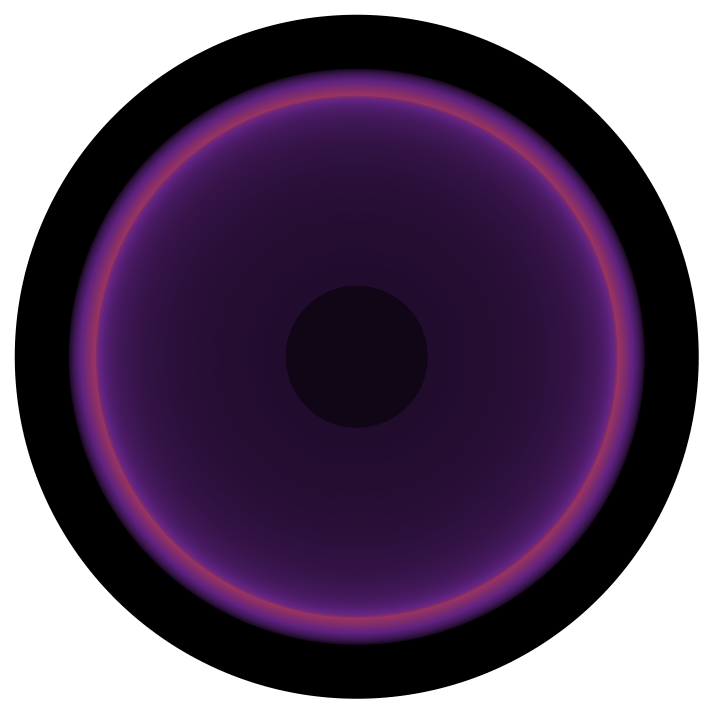}
		\caption{Static model}
	\end{subfigure}
	\centering
	\begin{subfigure}{0.32\linewidth}
		\centering
		\includegraphics[width=1\linewidth]{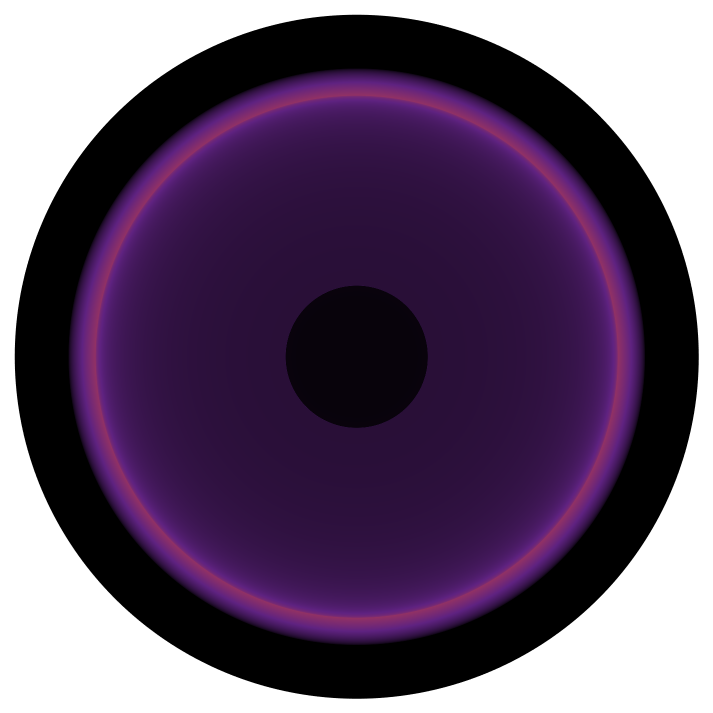}
		\caption{Infalling model with $r^{\text{SS}}_{\text{ini}}=34m$}
	\end{subfigure}
    \centering
	\begin{subfigure}{0.32\linewidth}
		\centering
		\includegraphics[width=1\linewidth]{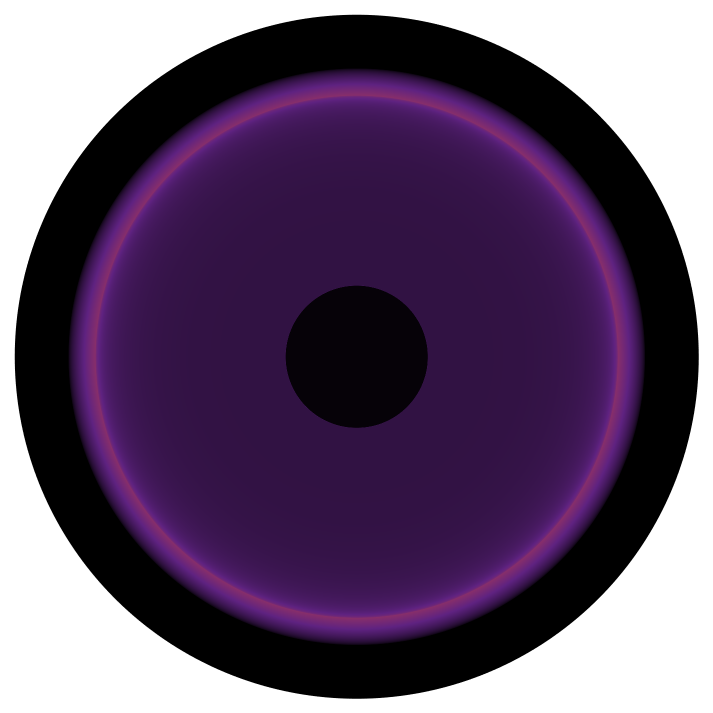}
		\caption{Infalling model with $r^{\text{SS}}_{\text{ini}}=10^{3}m$}
	\end{subfigure}
	\caption{The Schwarzschild BH images observed by a distant observer in the SS models of optically thin accretion with  $r^{\text{SS}}_{\text{inn}}=18m$ and $r^{\text{SS}}_{\text{out}}=20m$. The dark central regions are BH shadows with radius $b_{\text{cri}}\approx5.196m$, the circles enclosing the bright regions have radius $b_{\text{max}}^{\text{SS}}\approx21.082m$, and the radial position of the luminosity peak is at $b_{\text{min}}^{\text{SS}}\approx19.092m$. $\phantom{111111111111111111111111111111111111111111111111111111111111111111111111111111111111111}$}
	\label{fig7}
\end{figure}

Equations~(\ref{equ2.24})--(\ref{equ2.29}) can be used to generate the Schwarzschild BH images in the static and infalling SS models of optically thin accretion. According to these equations, once the emissivity $j(\nu_{\text{e}})$ per unit volume in the rest frame of the emitter is given, one is able to directly evaluate the integrated intensities on the screen of a distant observer when accreting matters are static and radially infalling. Thus, the corresponding BH images could be plotted. In what follows of this section, we will take a common emission pattern~\cite{Zeng:2020dco,Bambi:2013nla} as example to analyze the geometric and luminosity features of the BH images in the SS models. Concerning the emission pattern, it is assumed that the emission is monochromatic with rest-frame frequency $\nu_{0}$ and a $1/r^{2}$ radial profile, namely,
\begin{eqnarray}
\label{equ2.30}j(\nu_{\text{e}})\propto\frac{\delta(\nu_{\text{e}}-\nu_{0})}{r^{2}},
\end{eqnarray}
and with it, on the base of Eqs.~(\ref{equ2.24})--(\ref{equ2.29}), the BH images under the cases of $r^{\text{SS}}_{\text{inn}}<r^{\text{SS}}_{\text{out}}<r_{\text{pho}}$, $r^{\text{SS}}_{\text{inn}}\leqslant r_{\text{pho}}\leqslant r^{\text{SS}}_{\text{out}}$, and $r_{\text{pho}}<r^{\text{SS}}_{\text{inn}}<r^{\text{SS}}_{\text{out}}$ are plotted and shown in Figs.~\ref{fig2}--\ref{fig7}.
As mentioned in the previous section, photons on the lightlike geodesics with $b<b_{\text{cri}}$ are always moving away from the BH, whereas those on the lightlike geodesics with $b>b_{\text{cri}}$ could first move toward the BH and then move outward. Therefore, it can be expected that there is a luminosity jump at $b=b_{\text{cri}}$ in the BH images,
which explains why the edge curve of the shadow coincides with the bound photon orbit. As to the comparison of the shadow luminosities between the static and infalling SS models, by resorting to
the expressions of $F_{\text{s}}(b)$ and $F_{\text{f}}(b)$ under $0\leqslant b< b_{\text{cri}}$ in Eqs.~(\ref{equ2.24})--(\ref{equ2.29}), we could write
\begin{eqnarray}
\label{equ2.31}F_{\text{f}}(b)-F_{\text{s}}(b)\propto\int_{r^{\text{SS}}_{\text{inn}}}^{r^{\text{SS}}_{\text{out}}}\left[\left(g_{\text{f}}^{(\text{outw})}\right)^{3}-\left(g_{\text{s}}^{(\text{outw})}\right)^{3}\right]\left(-\frac{p_{0}}{p^{r}}\right)^{(\text{outw})}\frac{1}{r^2}\text{d}r
\end{eqnarray}
and
\begin{eqnarray}
\label{equ2.32}\frac{\left(g_{\text{f}}^{(\text{outw})}\right)^{3}}{\left(g_{\text{s}}^{(\text{outw})}\right)^{3}}-1=\left(\frac{\displaystyle\sqrt{1-\frac{2m}{r}}}{\displaystyle\sqrt{1-\frac{2m}{r^{\text{SS}}_{\text{ini}}}}+\sqrt{1-\frac{b^2}{r^2}\left(1-\frac{2m}{r}\right)}\sqrt{\frac{2m}{r}-\frac{2m}{r^{\text{SS}}_{\text{ini}}}}}\right)^{3}-1,
\end{eqnarray}
where Eq.~(\ref{equ2.32}) is derived from Eqs.~(\ref{equ2.13}) and (\ref{equ2.14}).
Due to $r^{\text{SS}}_{\text{ini}}\geqslant r^{\text{SS}}_{\text{out}}\geqslant r$, it is not difficult to conclude that when $0\leqslant b< b_{\text{cri}}$, the inequality
\begin{eqnarray}
\label{equ2.33}\left(g_{\text{f}}^{(\text{outw})}\right)^{3}\leqslant\left(g_{\text{s}}^{(\text{outw})}\right)^{3}
\end{eqnarray}
holds, which leads to
\begin{eqnarray}
\label{equ2.34}F_{\text{f}}(b)\leqslant F_{\text{s}}(b).
\end{eqnarray}
This result clearly indicates that the shadow luminosity in the infalling model is always lower than that in the static model. This conclusion manifests that the variation in the shadow luminosity between these two SS models is exactly the same as that between the corresponding spherical models~\cite{Narayan:2019imo}. In addition, from Eqs.~(\ref{equ2.31}) and (\ref{equ2.32}), one could also find that the shadow luminosity in the infalling model will decrease as $r^{\text{SS}}_{\text{ini}}$ increases.

\begin{figure}[tbp]
	\centering
	\begin{subfigure}{0.329\linewidth}
		\centering
		\includegraphics[width=0.95\linewidth]{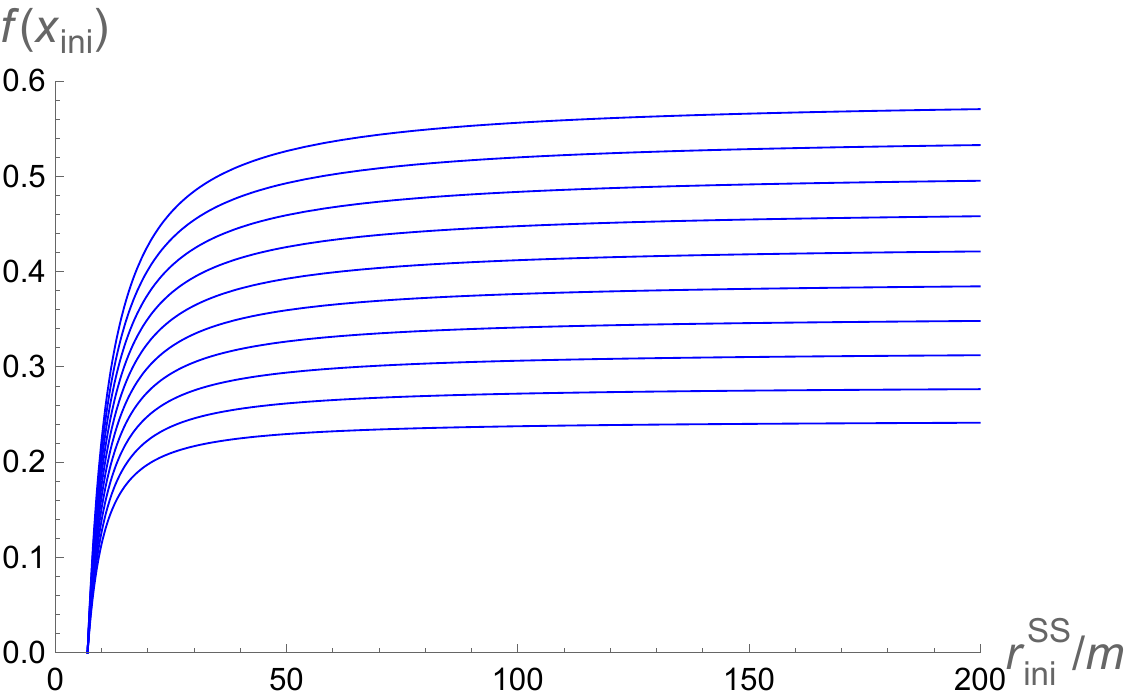}
		\caption{$r=7m$}
	\end{subfigure}
	\centering
	\begin{subfigure}{0.329\linewidth}
		\centering
		\includegraphics[width=0.95\linewidth]{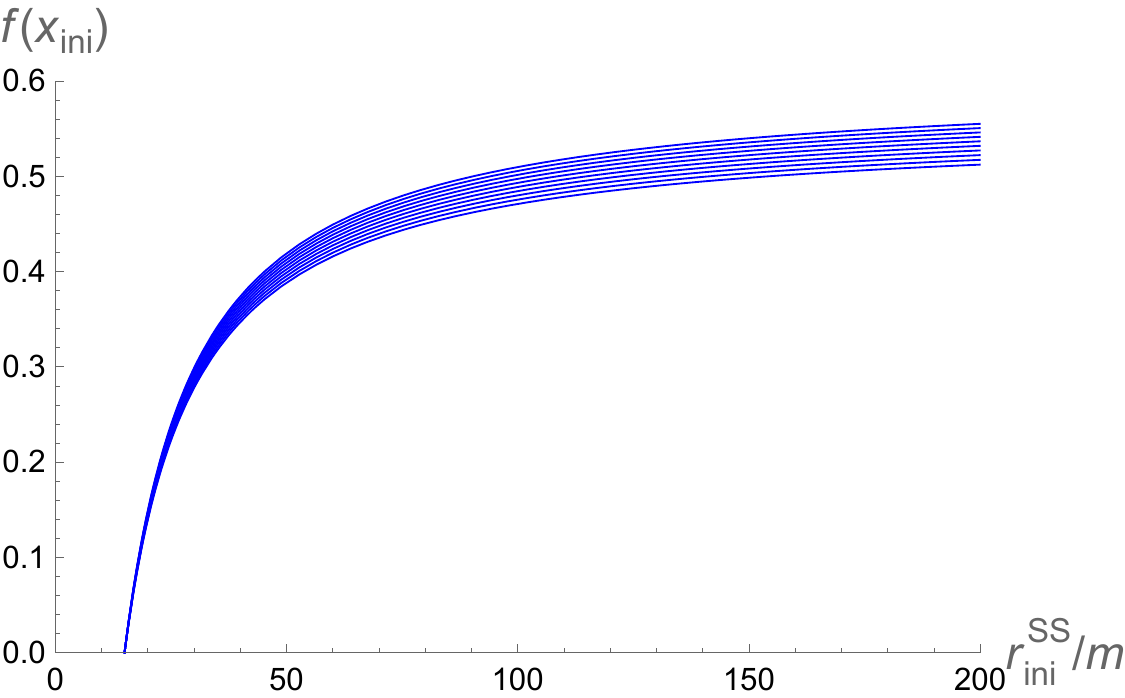}
		\caption{$r=15m$}
	\end{subfigure}
    \centering
	\begin{subfigure}{0.329\linewidth}
		\centering
		\includegraphics[width=0.95\linewidth]{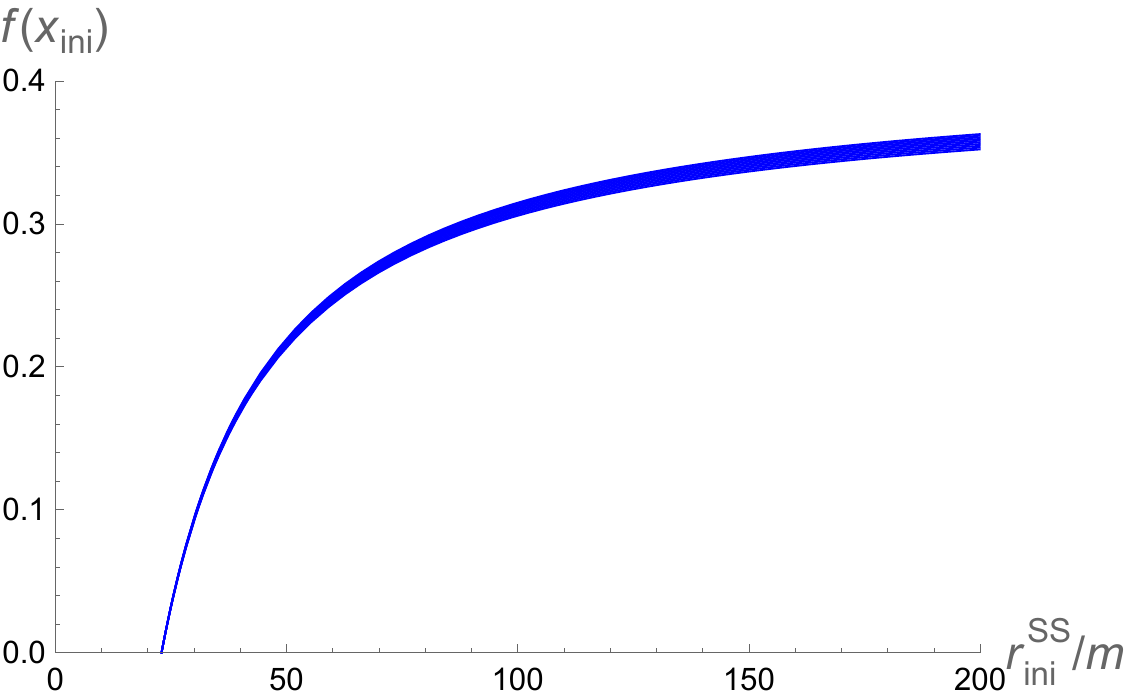}
		\caption{$r=23m$}
	\end{subfigure}
	\caption{Behaviors of the function $f(x_{\text{ini}})$ for $b=b_{\text{cri}}+(0.1i)m\ (i=1,2,\cdots10)$ with $r=7m, 15m$, and $23m$, where $r$ is the integral variable in Eqs.~(\ref{equ2.28}) and (\ref{equ2.29}). Note that the impact parameter $b$ of the lightlike geodesics should be slightly larger than $b_{\text{cri}}$ because only the small region close to the edge of the shadow is considered, and the integral variable $r$ should be larger than $r_{\text{pho}}=3m$ because the integral variable $r$ is confined between the inner and outer boundaries of the SS and $b_{\text{pea}}=b^{\text{SS}}_{\text{min}}$ is sufficiently larger than $b_{\text{cri}}$.  $\phantom{1111111111111111111111111111111111111111111}$}
	\label{fig8}
\end{figure}

For the SS models, the fact that the outer boundary of the SS does not extend to infinity in general means that
in a BH image, there exists a maximal impact parameter so that while $b$ is larger than it, the luminosity goes to zero. According to Eqs.~(\ref{equ2.24})--(\ref{equ2.29}), when $r^{\text{SS}}_{\text{out}}<r_{\text{pho}}$, the maximal impact parameter is $b_{\text{cri}}$, and the luminosity in the entire region outside the shadow is zero. Furthermore, when $r^{\text{SS}}_{\text{out}}>r_{\text{pho}}$, the maximal impact parameter is $b^{\text{SS}}_{\text{max}}$ defined by Eq.~(\ref{equ2.4}), and the luminosity in the region $b>b^{\text{SS}}_{\text{max}}$ is zero. Both the two arguments can easily be validated from the BH images in Figs.~\ref{fig2}--\ref{fig7}. In the region $b_{\text{cri}}\leqslant b\leqslant b^{\text{SS}}_{\text{max}}$ of the BH images, the luminosity variation is a bit complicated. The BH images in Figs.~\ref{fig2}--\ref{fig7} display that under the cases of $r^{\text{SS}}_{\text{inn}}\leqslant r_{\text{pho}}\leqslant r^{\text{SS}}_{\text{out}}$ and $r_{\text{pho}}<r^{\text{SS}}_{\text{inn}}<r^{\text{SS}}_{\text{out}}$,  the observed luminosities reach peak values at $b=b_{\text{cri}}$ and $b=b^{\text{SS}}_{\text{min}}$, respectively. In the former case, Eqs.~(\ref{equ2.26}) and (\ref{equ2.27}) indicate that the peak luminosity should be infinity, but it is due to numerical limitations that the actual calculated results never go to infinity. For the convenience of later use, the radial position of the peak luminosity could be rewritten as
\begin{eqnarray}
\label{equ2.35}b=b_{\text{pea}}:=\left\{
\begin{array}{ll}
b_{\text{cri}},\qquad&\text{for}\ r^{\text{SS}}_{\text{inn}}<r_{\text{pho}},\\
b^{\text{SS}}_{\text{min}},\qquad&\text{for}\  r^{\text{SS}}_{\text{inn}}>r_{\text{pho}}.
\end{array}\right.
\end{eqnarray}
As before, in order to compare the luminosities in the region $b_{\text{cri}}\leqslant b\leqslant b^{\text{SS}}_{\text{max}}$ between the static and infalling SS models,
based on the expressions of $F_{\text{s}}(b)$ and $F_{\text{f}}(b)$ in Eqs.~(\ref{equ2.26})--(\ref{equ2.29}), we need to write
\begin{eqnarray}
\label{equ2.36}F_{\text{f}}(b)-F_{\text{s}}(b)\propto2\hspace{-0.1cm}\int_{r^{\text{SS}}_{\text{inn}}\ \text{or}\ r_{2}}^{r^{\text{SS}}_{\text{out}}}\hspace{-0.1cm}\left[\frac{1}{2}\left(g_{\text{f}}^{(\text{inw})}\right)^{3}+\frac{1}{2}\left(g_{\text{f}}^{(\text{outw})}\right)^{3}-\left(g_{\text{s}}^{(\text{outw})}\right)^{3}\right]\left(-\frac{p_{0}}{p^{r}}\right)^{(\text{outw})}\frac{1}{r^{2}}\text{d}r\qquad\
\end{eqnarray}
and
\begin{eqnarray}
\label{equ2.37}\frac{\left(g_{\text{f}}^{(\text{inw})}\right)^{3}+\left(g_{\text{f}}^{(\text{outw})}\right)^{3}}{2\left(g_{\text{s}}^{(\text{outw})}\right)^{3}}\hspace{-0.1cm}-\hspace{-0.1cm}1=:f(x_{\text{ini}}),
\end{eqnarray}
where the lower limit of the integral in Eq.~(\ref{equ2.36}) is set to be $r^{\text{SS}}_{\text{inn}}$ or $r_{2}$ for different $b$, and by virtue of Eqs.~(\ref{equ2.13})--(\ref{equ2.15}), $f(x_{\text{ini}})$ is defined by
\begin{eqnarray}
\label{equ2.38}f(x_{\text{ini}})=\frac{1}{2}\left(\frac{\sqrt{\alpha}}{\sqrt{x_{\text{ini}}}-\beta\sqrt{x_{\text{ini}}-\alpha}}\right)^{3}\hspace{-0.1cm}+\hspace{-0.1cm}
\frac{1}{2}\left(\frac{\sqrt{\alpha}}{\sqrt{x_{\text{ini}}}+\beta\sqrt{x_{\text{ini}}-\alpha}}\right)^{3}\hspace{-0.1cm}-\hspace{-0.1cm}1
\end{eqnarray}
with
\begin{eqnarray}
\label{equ2.39}x_{\text{ini}}:=1-\frac{2m}{r^{\text{SS}}_{\text{ini}}},\quad \alpha:=1-\frac{2m}{r},\quad\beta:=\sqrt{1-\frac{b^2}{r^2}\left(1-\frac{2m}{r}\right)}.
\end{eqnarray}
Let us first focus on the luminosity near the peak, and there are two situations that need to be addressed.
\begin{itemize}
\item If $b^{\text{SS}}_{\text{max}}-b_{\text{pea}}$ is sufficiently small, Eqs.~(\ref{equ2.4}), (\ref{equ2.6}), and (\ref{equ2.35}) show that $r^{\text{SS}}_{\text{out}}-r_{\text{pho}}$ in the case of $r^{\text{SS}}_{\text{inn}}\leqslant r_{\text{pho}}\leqslant r^{\text{SS}}_{\text{out}}$ and $r^{\text{SS}}_{\text{out}}-r^{\text{SS}}_{\text{inn}}$ in the case of $r_{\text{pho}}<r^{\text{SS}}_{\text{inn}}<r^{\text{SS}}_{\text{out}}$ are also sufficiently small. Thus, for a lightlike geodesic with $b\approx b_{\text{pea}}$, according to the corresponding integral expressions of $F_{\text{s}}(b)$ and $F_{\text{f}}(b)$ in Eqs.~(\ref{equ2.26})--(\ref{equ2.29}), the radial coordinate $r_{2}$ of the periastron is less than or approximately equal to the integral variable $r$. This result, together with Eq.~(\ref{equA30}), leads to $\beta\gtrsim0$, which implies that the behavior of the function $f(x_{\text{ini}})$ in this situation can be analyzed with the help of the Taylor expansion, namely,
     \begin{eqnarray}
     \label{equ2.40}f(x_{\text{ini}})=\frac{\alpha ^{3/2}}{x_{\text{ini}}^{3/2}}-1+o(\beta).
     \end{eqnarray}
     Based on Eqs.~(\ref{equ2.39}), the condition $r^{\text{SS}}_{\text{ini}}\geqslant r^{\text{SS}}_{\text{out}}\geqslant r$ means $x_{\text{ini}}\geqslant \alpha$, and then with Eqs.~(\ref{equ2.36}), (\ref{equ2.37}), and (\ref{equ2.40}), we get
     \begin{eqnarray}
     \label{equ2.41}\frac{1}{2}\left(g_{\text{f}}^{(\text{inw})}\right)^{3}+\frac{1}{2}\left(g_{\text{f}}^{(\text{outw})}\right)^{3}\leqslant\left(g_{\text{s}}^{(\text{outw})}\right)^{3}
     \end{eqnarray}
     and
     \begin{eqnarray}
     \label{equ2.42}F_{\text{f}}(b)\leqslant F_{\text{s}}(b).
     \end{eqnarray}
     Obviously, when $b^{\text{SS}}_{\text{max}}-b_{\text{pea}}$ is sufficiently small, the luminosity near $b=b_{\text{pea}}$ in the infalling model is lower than that in the static model. In addition, from the above equations, one could also find that in the infalling model, the luminosity near $b=b_{\text{pea}}$ will decrease as $r^{\text{SS}}_{\text{ini}}\ (x_{\text{ini}})$ increases. The BH images in Figs.~\ref{fig3},~\ref{fig5}, and~\ref{fig7} verify these discussions.
\item If $b^{\text{SS}}_{\text{max}}-b_{\text{pea}}$ is sufficiently large, the complex integrations involved in Eqs.~(\ref{equ2.26})--(\ref{equ2.29}) make the analytical analysis difficult. But as shown in Figs.~\ref{fig4} and~\ref{fig6}, the numerical calculations indicate that the observed luminosities near $b=b_{\text{pea}}$ in the static and infalling SS models are close to each other, and the luminosity in the infalling model is not sensitive to $r^{\text{SS}}_{\text{ini}}$.
\end{itemize}
We will next study the luminosity near the exterior of the shadow when $r_{\text{pho}}<r^{\text{SS}}_{\text{inn}}<r^{\text{SS}}_{\text{out}}$.
If $b_{\text{pea}}=b^{\text{SS}}_{\text{min}}$ is close to $b_{\text{cri}}$,
the small region $b\gtrsim b_{\text{cri}}$ is near the peak, namely $b=b_{\text{pea}}$, so under this situation, the luminosity variations in the small region $b\gtrsim b_{\text{cri}}$ between the static and infalling SS models have been addressed in the previous two paragraphs. Therefore, the situation that needs to be handled is when $b_{\text{pea}}$ is sufficiently larger than $b_{\text{cri}}$. For this situation, Figs.~\ref{fig8} display that $f(x_{\text{ini}})$ is positive and it will increase as $r^{\text{SS}}_{\text{ini}}$ increases. With these arguments in mind, by Eqs.~(\ref{equ2.36}) and (\ref{equ2.37}), it is concluded that there are
\begin{eqnarray}
\label{equ2.43}\frac{1}{2}\left(g_{\text{f}}^{(\text{inw})}\right)^{3}+\frac{1}{2}\left(g_{\text{f}}^{(\text{outw})}\right)^{3}\geqslant\left(g_{\text{s}}^{(\text{outw})}\right)^{3}
\end{eqnarray}
and
\begin{eqnarray}
\label{equ2.44}F_{\text{f}}(b)\geqslant F_{\text{s}}(b),
\end{eqnarray}
which clearly shows that in the case of $r_{\text{pho}}<r^{\text{SS}}_{\text{inn}}<r^{\text{SS}}_{\text{out}}$, if $b_{\text{pea}}=b^{\text{SS}}_{\text{min}}$ is sufficiently larger than $b_{\text{cri}}$, the luminosity near the exterior of the shadow
in the infalling model is higher than that in the static model. Besides, Figs.~\ref{fig8} also imply that as $r^{\text{SS}}_{\text{ini}}$ increases, the luminosity in the infalling model will increase accordingly. These two conclusions can be confirmed from the BH images in Figs.~\ref{fig6} and~\ref{fig7}.

As shown in Ref.~\cite{Narayan:2019imo}, in the static and infalling spherical models, the features of the BH images are simple, whereas when the spherical models are extended to the SS models, the features of the BH images are rich and varied. It is the nature of the lightlike geodesics passing through the SS that gives rise to the rich and diverse features of the BH images in the SS models. The calculation of the integrated intensity will always involve the integral along a lightlike geodesic within the boundaries of the SS, so the luminosities of the different regions in the BH images will vary with the change of the boundaries of the SS. Now, let us present a comprehensive analysis and discussion.
\begin{enumerate}
\item The features of the shadow: For the lightlike geodesics received by the observer, those with $b<b_{\text{cri}}$ can pass through the bound photon orbit and only have the outward segments, but those with $b>b_{\text{cri}}$ lie outside the bound photon orbit and have both inward and outward segments. The distinct behaviors of these two types of the lightlike geodesics result in that there is a luminosity jump at $b=b_{\text{cri}}$. As stated in Ref.~\cite{Narayan:2019imo}, the Doppler beaming causes the lightlike geodesics with $b>b_{\text{cri}}$ to be brighter than those with $b<b_{\text{cri}}$. As a result, the lightlike geodesics with $b<b_{\text{cri}}$ create a relatively dark central region in BH image, and this is the BH shadow. Obviously, the shadow is always enclosed by the circle $b=b_{\text{cri}}$, so its size is fixed. As to the luminosity of the shadow, the fact that the lightlike geodesics with $b<b_{\text{cri}}$ only have the outward segment means photons on these geodesics are always moving away from the BH. However, in the infalling SS model, the emitters are always moving toward the BH, so the gravitational and Doppler redshift effects of the lightlike geodesics lead to the luminosity of the shadow being reduced. Obviously, with the increase of the initial radial position of accreting matters, due to an increase in radial velocities of the emitters, the Doppler redshift effects are strengthened, so that the luminosity of the shadow is further reduced.
\item The outer edge of the bright region: When both the boundaries of the SS lie inside the bound photon orbit, only the lightlike geodesics with $b<b_{\text{cri}}$ can pass through the SS, and the photons are always moving away from the BH. Therefore, the luminosity of the shadow is non-zero, whereas the luminosity outside the shadow is zero. When the outer boundary of the SS lies outside the bound photon orbit, among all the lightlike geodesics satisfying $b>b_{\text{cri}}$, those with  $b>b^{\text{SS}}_{\text{max}}$ (cf.~Eq.~(\ref{equ2.4})) can not pass through the SS because their periastrons are located outside the outer boundary of the SS, which results in that the luminosity of the region $b>b^{\text{SS}}_{\text{max}}$ is zero. Under this circumstance, the outer boundary of the SS determines the outer edge of the bright region in a BH image.
\item Radial position of the luminosity peak: When the inner and outer boundaries of the SS lie inside the bound photon orbit, the luminosity of a BH image will reach peak at  $b=b_{\text{cri}}$. The reason is that for a lightlike geodesic, when $b\rightarrow b_{\text{cri}}$,  it can bend at a greater angle around the BH, which means that the photons emitted from more emitters will be received by the observer. Furthermore, if the outer boundary of the SS lies outside the bound photon orbit, the luminosity of a BH image will still reach peak at  $b=b_{\text{cri}}$, and the peak luminosity can reach infinity because the lightlike geodesic can travel around the BH an infinite number of times. When the inner boundary of the SS lies outside the bound photon orbit, among all the lightlike geodesics satisfying $b>b_{\text{cri}}$, those with  $b<b^{\text{SS}}_{\text{min}}$ (cf.~Eq.~(\ref{equ2.6})) can pass through the inner boundary of the SS whereas those with  $b>b^{\text{SS}}_{\text{min}}$ can not, so the circle $b=b^{\text{SS}}_{\text{min}}$ also represents a critical curve. Under this circumstance, the BH images suggest that the inner boundary of the SS determines the radial position of the luminosity peak.
\item Luminosity variations near the peak between the static and infalling SS models: As mentioned above, the position of the luminosity peak is at $b=b_{\text{pea}}$ defined in Eq.~(\ref{equ2.35}). When the SS is thin enough and its outer boundary lies outside the bound photon orbit, for a considered lightlike geodesic with $b\approx b_{\text{pea}}$, the integral expressions of $F_{\text{s}}(b)$ and $F_{\text{f}}(b)$ in Eqs.~(\ref{equ2.26})--(\ref{equ2.29}) imply that the radial coordinate $r_{2}$ of the periastron is less than or approximately equal to the integral variable $r$. This result indicates that for the lightlike geodesics with $b\approx b_{\text{pea}}$, newly emitted photons have low radial velocities, which means that the Doppler blueshift effects contributed by the inward segments of these geodesics are very weak. So, although the lightlike geodesics with $b\approx b_{\text{pea}}$ have both inward and outward segments, the Doppler redshift effects contributed by the outward segments play a dominant role in the infalling SS model. Consequently, like the case of shadow, the gravitational and Doppler redshift effects of the lightlike geodesics lead to the observed luminosity near $b=b_{\text{cri}}$ in the infalling model being reduced, and the luminosity will be further reduced as the initial radial position of accreting matters increases. As a contrast, when the SS is thick enough, for the lightlike geodesics with $b\approx b_{\text{pea}}$, the Doppler shift effects contributed by the outward and inward segments approximately cancel each other out because the numerical calculations indicate that the observed luminosity near $b=b_{\text{pea}}$ in the infalling model shows negligible variation compared to that in the static model, and the luminosity is not sensitive to the initial radial position of accreting matters.
\item Luminosity variations near the exterior of the shadow between the static and infalling SS models: When the inner boundary of the SS lies near the exterior of the bound photon orbit, the position of luminosity peak is in the small region outside the shadow, so
    this small region is also around the peak, and thus it is not difficult to infer that under this case, the luminosities near the exterior of the shadow in the infalling SS model are identical to those addressed above. When the inner boundary of the SS is far from the bound photon orbit, the numerical calculations indicate that in the infalling model, the observed luminosity near the exterior of the shadow is enhanced, and the luminosity will be further enhanced as the initial radial position of accreting matters increases. This conclusion explicitly shows that for the lightlike geodesics with $b\gtrsim b_{\text{cri}}$, the Doppler blueshift effects contributed by the inward segments play a dominant role. In fact, while performing the integral along such a geodesic in Eqs.~(\ref{equ2.28}) and (\ref{equ2.29}), one will find that the radial coordinate $r_{2}$ of the periastron is much smaller than the integral variable $r$ because $r$ is confined between the inner and outer boundaries of the SS, which results in that newly emitted photons exhibit sufficiently large radial velocity. Under this case, the Doppler blueshift effects contributed by the inward segments of the considered lightlike geodesics in the infalling model could become strong enough, which leads to the observed luminosity near the exterior of the shadow being enhanced. In addition, with the increase of the initial radial position of accreting matters, the radial velocities of the emitters within the SS will increase, which strengthens the Doppler blueshift effects, so that the observed luminosity in the small region is further enhanced accordingly.
\end{enumerate}

The above summarized features of the BH images in the SS accretion models are generalizations of those in the spherical accretion models, and they strongly depend on the boundary positions of the SS. Firstly, the boundary positions of the SS play crucial roles in defining the geometric features of the BH images. For example, when the outer boundary of the SS lies outside the bound photon orbit, it determines the outer edge of the bright region in a BH image, and when the inner boundary of the SS lies outside the bound photon orbit, it determines the radial position of the luminosity peak. Secondly, the luminosity variations within specific regions between the static and infalling models heavily rely on the nature of the lightlike geodesics passing through the SS. To be specific, between these two models, the relative luminosity difference is determined by the comparison of the redshift factors. For the lightlike geodesics with $b<b_{\text{cri}}$, they only have the outward segments, so the gravitational and Doppler redshift effects in the infalling model lead to the luminosity of the shadow being reduced. However, for the lightlike geodesics with $b>b_{\text{cri}}$, they have both inward and outward segments, and the relative magnitude of the Doppler shift effects contributed by the inward and outward segments varies as the positions of the SS boundaries change. According to the above summary, for most situations in the infalling model, the gravitational and Doppler redshift effects lead to the observed luminosities being reduced, whereas when the inner boundary of the SS is far from the bound photon orbit, the Doppler blueshift effects of the lightlike geodesics with $b\gtrsim b_{\text{cri}}$ are strengthened, so that the observed luminosity near the exterior of the shadow is enhanced. Given that the unusual luminosity enhancement phenomenon is rare, it could be viewed as a notable luminosity feature of the BH images in the infalling SS model.

In view that the peak luminosity of a BH image is very important in observations, we also summarize the radial position of the luminosity peak and the luminosity variations near the peak between the static and infalling SS models. It is shown that once the radial position of the luminosity peak does not coincide with the edge of the shadow, one can conclude that the inner boundary of the SS lies outside the bound photon orbit. In addition, the above summaries also indicate that in the infalling model, when the SS is thin enough, the observed luminosity near the peak will be reduced, but when the SS is thick enough, the luminosity around the peak is almost unchanged. Finally, it should be emphasized that Eqs.~(\ref{equ2.24})--(\ref{equ2.29}) constitute the core ingredients for generating and analyzing the Schwarzschild BH images in the static and infalling SS models of optically thin accretion. As mentioned earlier, the previous results are obtained on the basis of the monochromatic emission pattern, and they constitute an application example of Eqs.~(\ref{equ2.24})--(\ref{equ2.29}).  By further employing them, the corresponding results for more realistic emission patterns can readily be achieved, and in reality, one only needs to insert the new expression of the emissivity per unit volume into these equations so that the integrated intensities observed by a distant observer in the static and infalling SS models of optically thin accretion on a Schwarzschild BH can be evaluated, and then the BH images can be plotted.
\section{The BH images in the static, infalling, and rotating CA models of optically and geometrically thin accretion~\label{Sec:third}}
In addition to the spherical accretion models, the disk accretion models are also significant in the study of BH physics~\cite{Hu:2023pyd,Luminet:1979nyg,Gyulchev:2019tvk,Gyulchev:2021dvt,Paul:2019trt,Rahaman:2021kge,Liu:2021lvk,Guo:2023grt,Hu:2023bzy}. For the static disk model of optically and geometrically thin accretion on Schwarzschild BH, a basic method to evaluate the integrated intensity observed by a distant observer at the face-on orientation is provided in Ref.~\cite{Gralla:2019xty}, and the BH images for three emission patterns are presented. These results illustrate that the size of the shadow is very much dependent on the emission pattern, and the luminosities of BH images are mainly contributed by the first and second order emissions. As mentioned in Ref.~\cite{Gralla:2019xty}, these results need to be generalized because of  the highly idealized nature of the physical scenarios. In more realistic situations, accreting matters should be distributed in a limited area within the infinite disk, and may radially move towards or orbit around the BH. Moreover, in general, the observer should view the accretion disk at a definite inclination angle. Thus, it is indeed necessary to further generalize the static disk accretion model in Ref.~\cite{Gralla:2019xty} to more realistic accretion models. For the rotating disk accretion model, the BH images are plotted by means of analytical methods in Ref.~\cite{Luminet:1979nyg}. In this section, motivated by the techniques in Ref.~\cite{Luminet:1979nyg}, we will make use of the equations of the lightlike geodesics to generate
the Schwarzschild BH images in the static, infalling, and rotating circular-annulus (CA) models of optically and geometrically thin accretion and analyze their features, where accreting matters located within a CA centered at the center of the BH are considered.

\subsection{The analytical forms of the transfer functions working for all impact parameter values}

\begin{figure}[tbp]
\centering
\includegraphics[width=.75\textwidth]{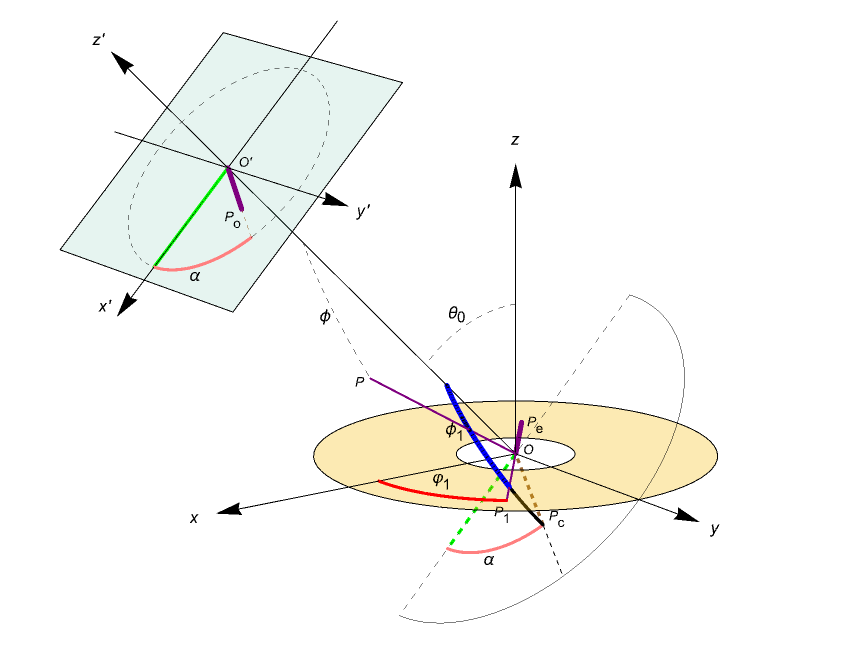}
\caption{\label{fig9} The coordinate systems (see text).}
\end{figure}

Let $r^{\text{CA}}_{\text{inn}}$ and $r^{\text{CA}}_{\text{out}}$ be the radial coordinates of the inner and outer boundaries of the CA. Suppose that the CA with a negligible thickness is on the equatorial plane of the BH, and the radiation is emitted isotropically in the rest frame of the accreting matter. In Fig.~\ref{fig9}, the yellow annulus is employed to represent the CA. As in this figure, the right-handed rectangular coordinate systems $Oxyz$ and $O'x'y'z'$  are chosen. For the former, the origin $O$ is at the center of the BH, and the $Oxy$ coordinate plane coincides with the equatorial plane. For convenience, here, we denote the positive unit vectors of $x$-axis, $y$-axis, and $z$-axis by $\boldsymbol{e}_{x}$, $\boldsymbol{e}_{y}$, and $\boldsymbol{e}_{z}$, respectively.
As to the latter, the origin $O'$ is the position of the observer, and since the observer is practically fixed at infinity in the gravitational field of the BH, the Schwarzschild radial coordinate of $O'$ ought to be $+\infty$. The direction of $O'$ could be set to be along
\begin{eqnarray}
\label{equ3.1}\boldsymbol{e}_{\scriptstyle{OO'}}=\sin{\theta_{0}}\boldsymbol{e}_{x}+\cos{\theta_{0}}\boldsymbol{e}_{z},
\end{eqnarray}
where $\theta_{0}$ is the inclination angle of the observer relative to the normal of the equatorial plane. With $\theta_{0}$, the positive unit vectors of $x'$-axis, $y'$-axis, and $z'$-axis are defined by
\begin{eqnarray}
\label{equ3.2}\boldsymbol{e}_{x'}=\cos{\theta_{0}}\boldsymbol{e}_{x}-\sin{\theta_{0}}\boldsymbol{e}_{z},\quad \boldsymbol{e}_{y'}=\boldsymbol{e}_{y},\quad \boldsymbol{e}_{z'}=\boldsymbol{e}_{\scriptstyle{OO'}}.
\end{eqnarray}
In general, the observer should view the BH (not the CA) at the face-on orientation, and therefore, the plane where the screen of the observer is located is just $O'x'y'$ plane (the cyan plane in Fig.~\ref{fig9}).

Among all the lightlike geodesics emitted from the CA, we only consider those whose asymptotic directions are along $\boldsymbol{e}_{\scriptstyle{OO'}}$ because they could be received by the observer. Assume that starting from an emitting point $P_{\text{e}}$ on the CA, the trajectory of such a lightlike geodesic reaches the observational screen at a receiving point $P_{\text{o}}$. As mentioned in the Appendix, the entire path of the geodesic is in the plane $P_{\text{e}}OO'$, and in order to obtain its equation, we need to treat this plane as a new equatorial plane of the BH, where under this circumstance, for a point $P$ on the geodesic, the angle $\phi$ shown in Fig.~\ref{fig9} is the corresponding azimuthal angle. Since the observer is located at infinity, the azimuthal angle coordinate of the receiving point $P_{\text{o}}$ in the plane $P_{\text{e}}OO'$ should be zero. As a consequence, in Eqs.~(\ref{equA54}), (\ref{equA56}), and (\ref{equA57}), after we replace the original azimuthal angle $\varphi$ by the new one $\phi$ and then set $r_{0}=+\infty$ and $\varphi_{0}=0$, the equation of a lightlight geodesic emitted from the CA intersecting the observational screen is obtained,
\begin{eqnarray}
\label{equ3.3}r=\left\{
\begin{array}{ll}
\displaystyle m\left[X_{1}+Y(X_{1})\left(\frac{1+\scn\hspace{-0.05cm}{[+\phi\sqrt{2Y(X_{1})}}-\F{(\vartheta_{\text{les}}(0),k_{\text{les}})]}}{1-\scn\hspace{-0.05cm}{[+\phi\sqrt{2Y(X_{1})}}-\F{(\vartheta_{\text{les}}(0),k_{\text{les}})]}}\right)\right]^{-1},&\quad \text{for}\quad  0\leqslant b<b_{\text{cri}},\medskip\\
\displaystyle m\left[\frac{1}{2}\tanh^{2}{\left(+\frac{1}{2}\phi+\arctanh{\bigg(\hspace{-0.12cm}\sqrt{\frac{1}{3}}\,\bigg)}\right)}-\frac{1}{6}\right]^{-1},&\quad \text{for}\quad b=b_{\text{cri}},\medskip\\
\displaystyle m\left[X_{1}+(X_{2}-X_{1})\ssn\hspace{-0.1cm}^{2}{\left(+\phi\sqrt{\frac{X_{3}-X_{1}}{2}}+\F{\bigg(\vartheta_{\text{larI}}(0),k_{\text{lar}}\bigg)}\right)}\right]^{-1},&\quad \text{for}\quad b>b_{\text{cri}},
\end{array}\right.
\end{eqnarray}
where in Eqs.~(\ref{equA54}), (\ref{equA56}), and (\ref{equA57}), only the positive signs before $(\varphi-\varphi_{0})$ are kept, and due to the symmetry of the geometrical configuration, this sign selection does not affect the final BH images. From Eqs.~(\ref{equA16})--(\ref{equA24}), $X_{1}$, $X_{2}$, and $X_{3}$ appearing in Eq.~(\ref{equ3.3}) all depend on the impact parameter $b$ of the lightlike geodesic, which results in that the radial coordinate of a point on the geodesic is an explicit function of $\phi$ and $b$, and therefore, Eq.~(\ref{equ3.3}) can be written as the following form,
\begin{eqnarray}
\label{equ3.4}r=h(\phi,b).
\end{eqnarray}
It should be pointed out that in the derivation of Eq.~(\ref{equ3.3}), we did not use the original equations of the lightlike geodesics, namely~(\ref{equA17}), (\ref{equA31}), and (\ref{equA36}) because in such a formulation of the lightlike geodesic equations,
two branches of the lightlike geodesic with $b>b_{\text{cri}}$ are described by different equations, which could cause ambiguities in applications~\cite{Cadez:2004cg}. In the Appendix, the original equations of the lightlike geodesics are rewritten in a form suitable for
practical use so that the equation of a lightlight geodesic emitted from the CA intersecting the observational screen can be presented in a neat form.

When a lightlike geodesic is traced from the receiving point $P_{\text{o}}$ on the observational screen backwards towards the emitting point $P_{\text{e}}$ on the CA, it could intersect the equatorial plane $Oxy$ many times. If the emitting point $P_{\text{e}}$ is the $k$th $(k=1,2,3,\cdots)$ intersection point and the geodesic only picks up luminosity from the emission of the point $P_{\text{e}}$, the geodesic is referred to as the $k$th order lightlike geodesic and the corresponding emission is called the $k$th order emission. Now, we denote the first intersection point of the lightlike geodesic with the equatorial plane as $P_{1}$, and as implied in Fig.~\ref{fig9},  the coordinates of $P_{1}$ in the plane $P_{\text{e}}OO'$ are $(r_{1},\phi_{1})$, where it can be easily seen that as the angle $\varphi_{1}$ varies from $0$ to $2\pi$, the angle $\phi_{1}$ satisfies
\begin{eqnarray}
\label{equ3.5}\frac{\pi}{2}-\theta_{0}\leqslant\phi_{1}\leqslant\frac{\pi}{2}+\theta_{0}.
\end{eqnarray}
With the angle $\phi_{1}$, the coordinates $(r_{\text{e}},\phi_{\text{e}})$ of the emitting point $P_{\text{e}}$ in the plane $P_{\text{e}}OO'$ can be completely determined. One could infer that when $P_{\text{e}}$ is the $k$th intersection point of the geodesic with the equatorial plane, the angle $\phi_{\text{e}}$ should be
\begin{eqnarray}
\label{equ3.6}\phi_{\text{e}}=\phi_{1}+(k-1)\pi.
\end{eqnarray}
Given the angle $\phi_{\text{e}}$, by directly plugging it into the equation~(\ref{equ3.4}) of the geodesic, the radial coordinate of $P_{\text{e}}$ is also obtained,
\begin{eqnarray}
\label{equ3.7}r_{\text{e}}=h(\phi_{\text{e}},b).
\end{eqnarray}
On the observational screen, the polar coordinates of the receiving point $P_{\text{o}}$ are $(b,\alpha)$. The first task of this section is to find the relationship between the angles $\phi_{\text{e}}$ and $\alpha$, so that the radial coordinate $r_{\text{e}}$ of the emitting point could be expressed in terms of $b$ and $\alpha$.
This task is not difficult. In the rectangular coordinate system $O'x'y'z'$, the direction of $P_{\text{o}}$ relative to the origin $O'$ is along
\begin{eqnarray}
\label{equ3.8}\boldsymbol{e}_{\scriptstyle{O'P_{\text{o}}}}&=&\cos{\alpha}\boldsymbol{e}_{x'}+\sin{\alpha}\boldsymbol{e}_{y'},
\end{eqnarray}
and by inserting the definitions of $\boldsymbol{e}_{x'}$ and $\boldsymbol{e}_{y'}$ in Eqs.~(\ref{equ3.2}), the expansion of $\boldsymbol{e}_{\scriptstyle{O'P_{\text{o}}}}$ in the coordinate system $Oxyz$ is offered,
\begin{eqnarray}
\label{equ3.9}&&\boldsymbol{e}_{\scriptstyle{O'P_{\text{o}}}}=\cos{\theta_{0}}\cos{\alpha}\boldsymbol{e}_{x}+\sin{\alpha}\boldsymbol{e}_{y}-\sin{\theta_{0}}\cos{\alpha}\boldsymbol{e}_{z}.
\end{eqnarray}
From Eq.~(\ref{equ3.1}), it can be proved that $\boldsymbol{e}_{\scriptstyle{O'P_{\text{o}}}}$ satisfies the property
\begin{eqnarray}
\label{equ3.10}&&\boldsymbol{e}_{\scriptstyle{O'P_{\text{o}}}}\cdot\boldsymbol{e}_{\scriptstyle{OO'}}=0.
\end{eqnarray}
For the convenience of later derivation, a point $P_{\text{c}}$ whose direction relative to the origin $O$ is along $\boldsymbol{e}_{\scriptstyle{OP_{\text{c}}}}=\boldsymbol{e}_{\scriptstyle{O'P_{\text{o}}}}$ is found, and thus, there are
\begin{eqnarray}
\label{equ3.11}&&\boldsymbol{e}_{\scriptstyle{OP_{\text{c}}}}=\cos{\theta_{0}}\cos{\alpha}\boldsymbol{e}_{x}+\sin{\alpha}\boldsymbol{e}_{y}-\sin{\theta_{0}}\cos{\alpha}\boldsymbol{e}_{z},\\
\label{equ3.12}&&\boldsymbol{e}_{\scriptstyle{OP_{\text{c}}}}\cdot\boldsymbol{e}_{\scriptstyle{OO'}}=0.
\end{eqnarray}
The fact that $P_{\text{o}}$ is on the plane $P_{\text{e}}OO'$ leads that $P_{\text{c}}$ is also on this plane, and the conclusion, together with $\boldsymbol{e}_{\scriptstyle{OP_{\text{c}}}}\cdot\boldsymbol{e}_{\scriptstyle{OO'}}=0$, implies that the unit vector $\boldsymbol{e}_{\scriptstyle{OP_{1}}}$  representing the direction of the point $P_{1}$ relative to the origin $O$ could be expanded as
\begin{eqnarray}
\label{equ3.13}\boldsymbol{e}_{\scriptstyle{OP_{1}}}&=&\cos{\phi_{1}}\boldsymbol{e}_{\scriptstyle{OO'}}+\sin{\phi_{1}}\boldsymbol{e}_{\scriptstyle{OP_{\text{c}}}}.
\end{eqnarray}
Then, after substituting Eqs.~(\ref{equ3.1}) and (\ref{equ3.11}) in the above expansion, the final expansion of $\boldsymbol{e}_{\scriptstyle{OP_{1}}}$ is achieved,
\begin{eqnarray}
\label{equ3.14}\boldsymbol{e}_{\scriptstyle{OP_{1}}}&=&(\sin{\theta_{0}}\cos{\phi_{1}}+\cos{\theta_{0}}\cos{\alpha}\sin{\phi_{1}})\boldsymbol{e}_{x}+\sin{\alpha}\sin{\phi_{1}}\boldsymbol{e}_{y}\nonumber\\
&&+(\cos{\theta_{0}}\cos{\phi_{1}}-\sin{\theta_{0}}\cos{\alpha}\sin{\phi_{1}})\boldsymbol{e}_{z}.
\end{eqnarray}
As mentioned before, the point $P_{1}$ is on the equatorial plane (plane $Oxy$), which means that in Eq.~(\ref{equ3.14}), the $z$-component should disappear, namely,
\begin{eqnarray}
\label{equ3.15}\cos{\theta_{0}}\cos{\phi_{1}}-\sin{\theta_{0}}\cos{\alpha}\sin{\phi_{1}}=0.
\end{eqnarray}
In this section, the inclination angle $\theta_{0}$ of the observer is restricted to $[0,\pi/2)$, and thus, by solving Eq.~(\ref{equ3.15}) under the condition~(\ref{equ3.5}), a preliminary expression of $\phi_{1}$ in terms of $\alpha$ is given,
\begin{eqnarray}
\label{equ3.16}\phi_{1}=\left\{
\begin{array}{ll}
\displaystyle\arctan{(\cot{\theta_{0}}\sec{\alpha})},&\displaystyle\ \ \text{for}\ \   0\leqslant \alpha\leqslant\frac{\pi}{2}\ \ \text{or}\ \ \frac{3\pi}{2}\leqslant \alpha<2\pi,\medskip\\
\displaystyle\pi+\arctan{(\cot{\theta_{0}}\sec{\alpha})},&\displaystyle\ \ \text{for}\ \   \frac{\pi}{2}< \alpha<\frac{3\pi}{2}.
\end{array}\right.
\end{eqnarray}
The above expression of $\phi_{1}$ can be further simplified, and by means of the property of the function $\arccos$, it could be rewritten in the following concise form,
\begin{eqnarray}
\label{equ3.17}\phi_{1}=\arccos{\left(\frac{\tan{\theta_{0}}\cos{\alpha}}{\sqrt{1+\tan^2{\theta_{0}}\cos^2{\alpha}}}\right)}.
\end{eqnarray}
In view that the angle $\phi_{\text{e}}$ has been given in terms of $\phi_{1}$ in Eq.~(\ref{equ3.6}),  the formal relationship between the angles $\phi_{\text{e}}$ and $\alpha$ is directly provided by using Eq.~(\ref{equ3.17}), namely,
\begin{eqnarray}
\label{equ3.18}\phi_{\text{e}}=\iota_{k}(\alpha,\theta_{0}):=(k-1)\pi+\arccos{\left(\frac{\tan{\theta_{0}}\cos{\alpha}}{\sqrt{1+\tan^2{\theta_{0}}\cos^2{\alpha}}}\right)}.
\end{eqnarray}
In the above process, we adopt an alternative approach to deriving the relationship between the angles $\phi_{\text{e}}$ and $\alpha$, and in Refs.~\cite{Luminet:1979nyg,muller2009analytic,Gyulchev:2021dvt,gyulchev2020observational}, the equivalent versions of this relationship can also be evaluated with the aid of spherical trigonometry, and the reader wishing to go into more details may consult these references.
To conclude the task of this part, we need to insert the above relationship between the angles $\phi_{\text{e}}$ and $\alpha$ into Eq.~(\ref{equ3.7}) to acquire
\begin{eqnarray}
\label{equ3.19}r_{\text{e}}=h\left(\iota_{k}(\alpha,\theta_{0}),b\right)
\end{eqnarray}
which is the radial coordinate $r_{\text{e}}$ of the emitting point $P_{\text{e}}$ in terms of $b$ and $\alpha$. The implication of Eq.~(\ref{equ3.19}) is that for a $k$th order lightlike geodesic emitted from the CA, once the polar coordinates of its intersection point with the observational screen are given, the radial coordinate of the emitting point is determined, and therefore, $h\left(\iota_{k}(\alpha,\theta_{0}),b\right)$ is named as the $k$th order transfer function.
\begin{figure}[tbp]
	\centering
	\begin{subfigure}{0.235\linewidth}
		\centering
		\includegraphics[width=1\linewidth]{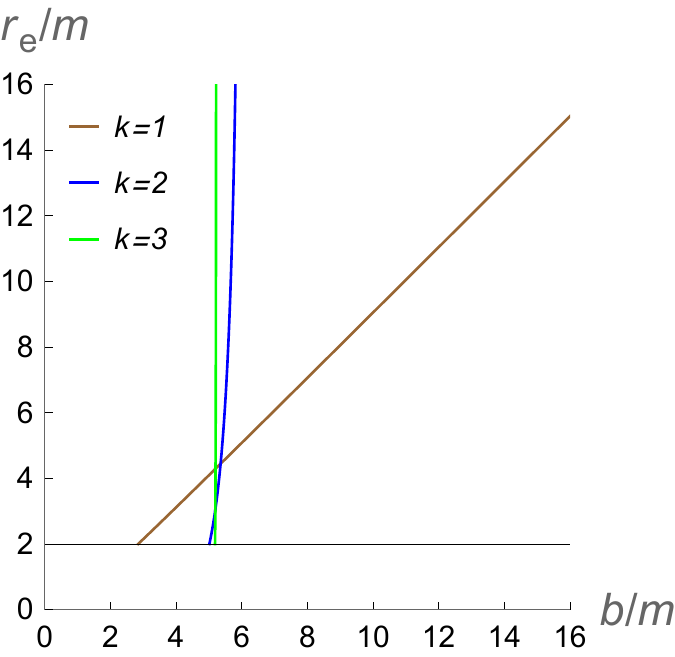}
		\caption{$h\left(\iota_{k}(\alpha,0),b\right)$}
		\label{figg10.1}
	\end{subfigure}
	\centering
	\begin{subfigure}{0.235\linewidth}
		\centering
		\includegraphics[width=1\linewidth]{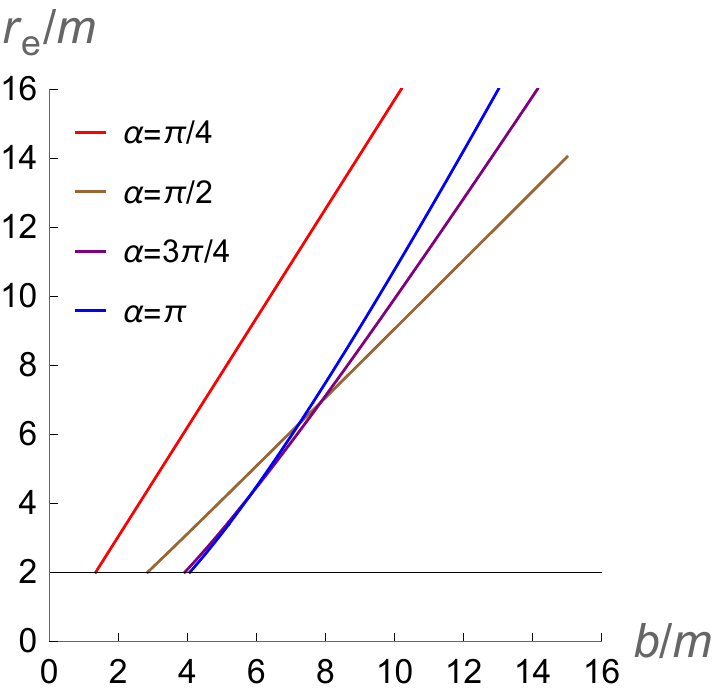}
		\caption{$h\left(\iota_{1}(\alpha,\pi/3),b\right)$}
		\label{figg10.2}
	\end{subfigure}
    \centering
	\begin{subfigure}{0.235\linewidth}
		\centering
		\includegraphics[width=1\linewidth]{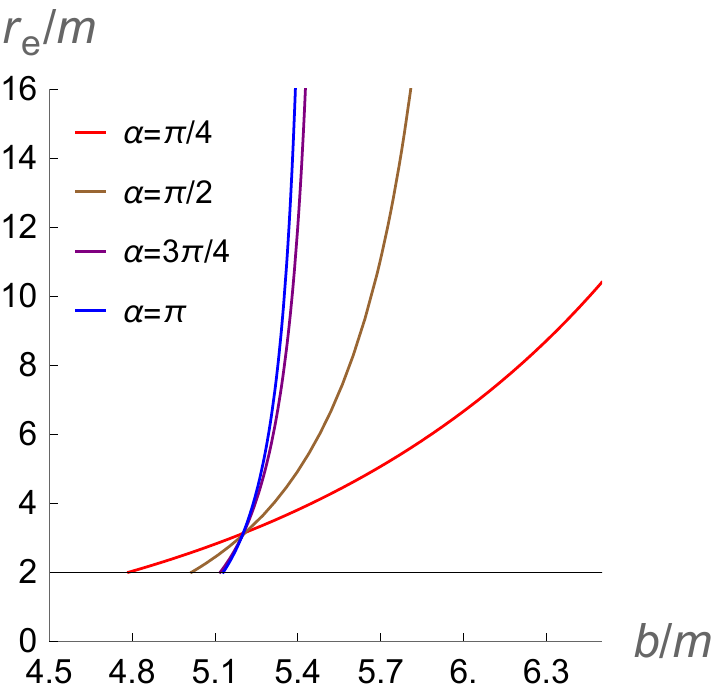}
		\caption{$h\left(\iota_{2}(\alpha,\pi/3),b\right)$}
		\label{figg10.3}
	\end{subfigure}
    \centering
	\begin{subfigure}{0.235\linewidth}
		\centering
		\includegraphics[width=1\linewidth]{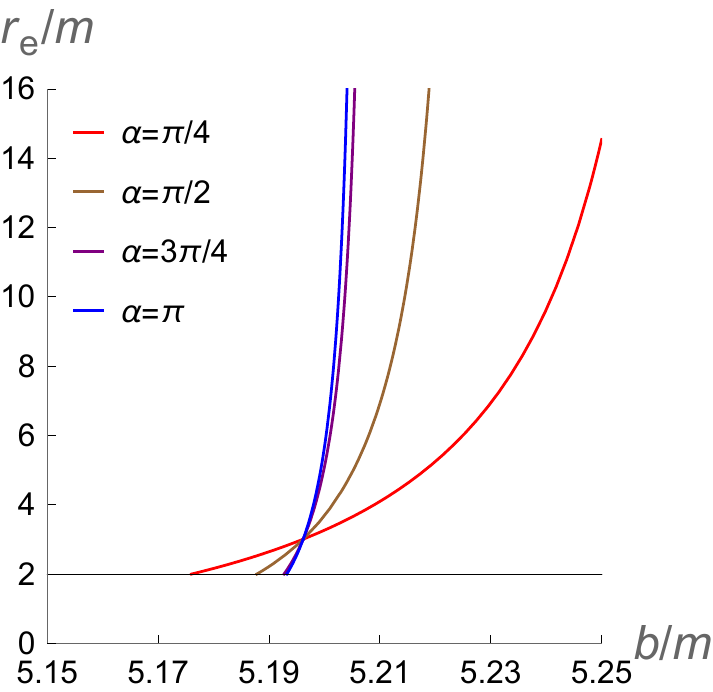}
		\caption{$h\left(\iota_{3}(\alpha,\pi/3),b\right)$}
		\label{figg10.4}
	\end{subfigure}
	\caption{Behaviors of the first three order transfer functions for $\theta_{0}=0$ and $\theta_{0}=\pi/3$.}
	\label{fig10}
\end{figure}
When $\theta_{0}=0$, one could verify that $\iota_{k}(\alpha,0)=(k-1/2)\pi$, so in this case, the $k$th order transfer function $h\left(\iota_{k}(\alpha,0),b\right)$ actually does not depend on the angle $\alpha$. The behaviors of the first three order transfer functions for $\theta_{0}=0$ are shown in Fig.~\ref{figg10.1}, and it is not difficult to recognize that they are identical to those in Ref.~\cite{Gralla:2019xty}, which means that the analytical forms of the transfer functions presented in Eq.~(\ref{equ3.19})  generalize the results in Ref.~\cite{Gralla:2019xty}. On the basis of the analytical results presented in the section, the behaviors of the first three order transfer functions for $\theta_{0}=\pi/3$ are displayed  in Figs.~\ref{figg10.2}--\ref{figg10.4}, from which it is concluded that when the angle $\alpha$ is fixed, the $k$th order transfer function is a monotonically increasing function of the impact parameter $b$. One should note that the analytical forms of the transfer functions given in Ref.~\cite{Luminet:1979nyg} only hold under the situation of $b>b_{\text{cri}}$, but the results derived in this section work for all impact parameter values.

The transfer functions play central roles in determining the geometric features of the BH images. As we initially envisioned, accreting matters are distributed within a CA, and the radial coordinates of the boundaries of the CA are $r^{\text{CA}}_{\text{inn}}$ and $r^{\text{CA}}_{\text{out}}$. The $k$th order transfer function can be employed to determine the inner and outer edge curves of the bright region in the $k$th order BH image, the BH image plotted only based on the $k$th order lightlike geodesics emitted from the CA. In fact, based on the solutions $b=b^{\text{CA}}_{\text{min}}(\alpha,\theta_{0},k)$ and $b=b^{\text{CA}}_{\text{max}}(\alpha,\theta_{0},k)$ to the following two equations
\begin{eqnarray}
\label{equ3.20}h\left(\iota_{k}(\alpha,\theta_{0}),b\right)=r^{\text{CA}}_{\text{inn}},\qquad h\left(\iota_{k}(\alpha,\theta_{0}),b\right)=r^{\text{CA}}_{\text{out}},
\end{eqnarray}
one could plot two curves on the observational screen. Since the $k$th order transfer function is monotonically increasing as $b$ increases, the fact that the radial coordinate $r_{\text{e}}$ of the emitting points satisfies $r^{\text{CA}}_{\text{inn}}\leqslant r_{\text{e}}\leqslant r^{\text{CA}}_{\text{out}}$ means that the radial coordinate $b$ of the receiving points on the observational screen should satisfy
\begin{eqnarray}
\label{equ3.21}b^{\text{CA}}_{\text{min}}(\alpha,\theta_{0},k)\leqslant b\leqslant b^{\text{CA}}_{\text{max}}(\alpha,\theta_{0},k).
\end{eqnarray}
Consequently, for the $k$th order BH image, the luminosities in the regions $b<b^{\text{CA}}_{\text{min}}(\alpha,\theta_{0},k)$ and $b>b^{\text{CA}}_{\text{max}}(\alpha,\theta_{0},k)$ ought to be zero, which means that the above two curves bound the bright region of the image. In the same manner, the isoradial curve for a fixed $r_{\text{e}}$ can also be plotted based on Eq.~(\ref{equ3.19}), and obviously, the two edge curves of the bright region in the $k$th order BH image are just two particular isoradial curves.
\begin{figure}[tbp]
	\centering
	\begin{subfigure}{0.32\linewidth}
		\centering
		\includegraphics[width=1\linewidth]{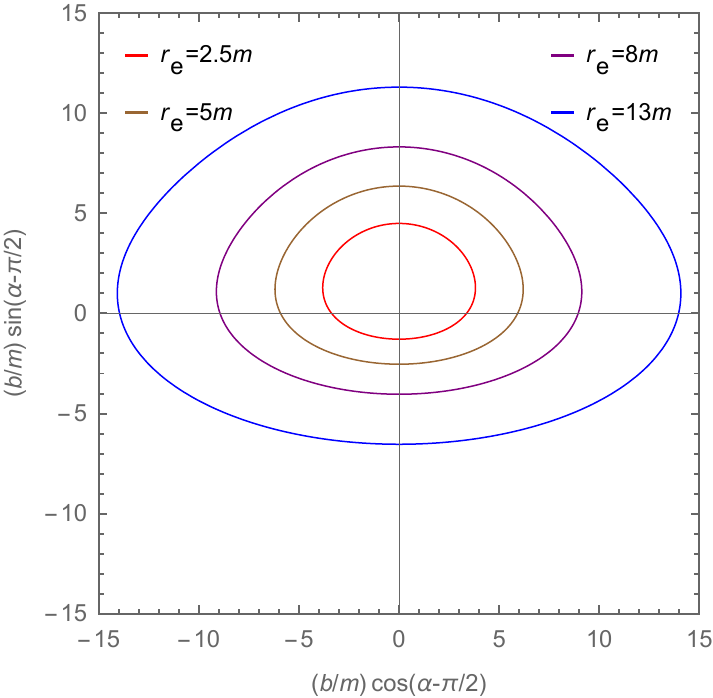}
		\caption{The first order, $\theta_{0}=\pi/3$}
		\label{figg11.1}
	\end{subfigure}
	\centering
	\begin{subfigure}{0.32\linewidth}
		\centering
		\includegraphics[width=1\linewidth]{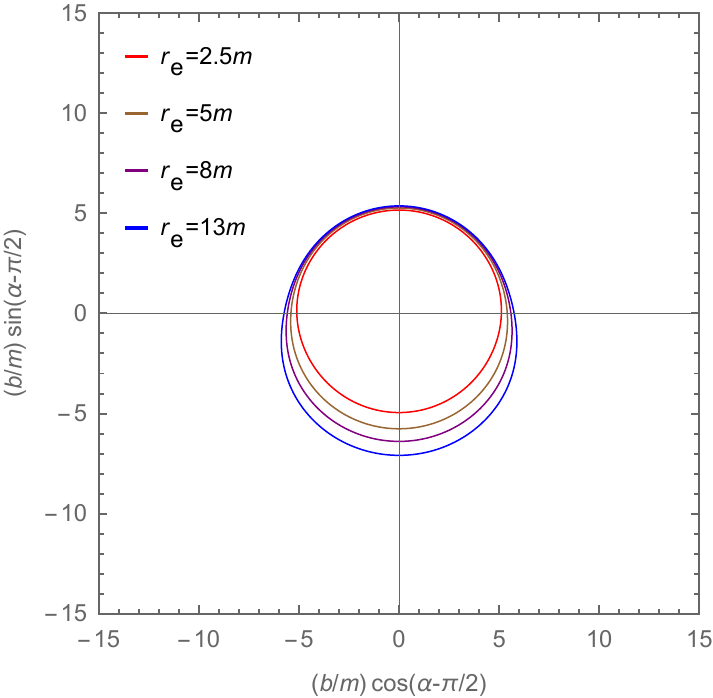}
		\caption{The second order, $\theta_{0}=\pi/3$}
		\label{figg11.2}
	\end{subfigure}
    \centering
	\begin{subfigure}{0.32\linewidth}
		\centering
		\includegraphics[width=1\linewidth]{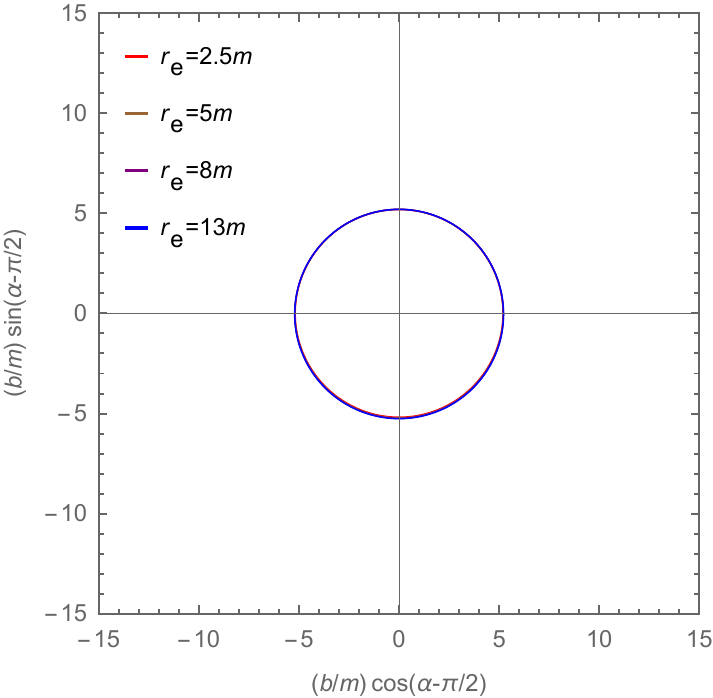}
		\caption{The third order, $\theta_{0}=\pi/3$}
		\label{figg11.3}
	\end{subfigure}
	\centering
	\begin{subfigure}{0.32\linewidth}
		\centering
		\includegraphics[width=1\linewidth]{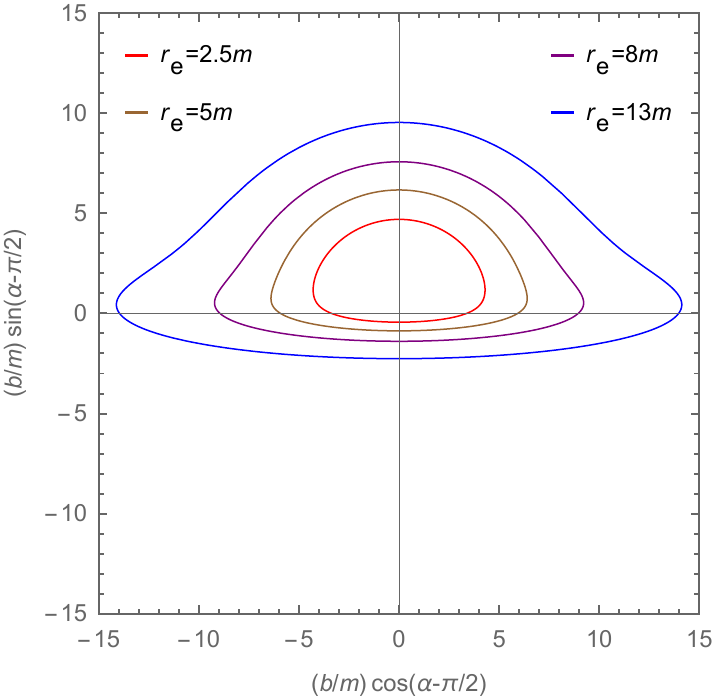}
		\caption{The first order, $\theta_{0}=4\pi/9$}
		\label{figg11.4}
	\end{subfigure}
	\centering
	\begin{subfigure}{0.32\linewidth}
		\centering
		\includegraphics[width=1\linewidth]{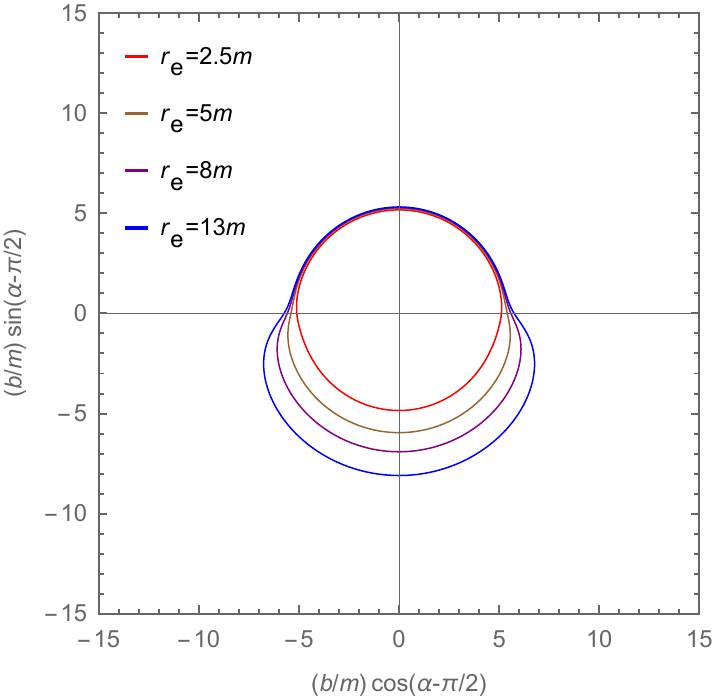}
		\caption{The second order, $\theta_{0}=4\pi/9$}
		\label{figg11.5}
	\end{subfigure}
    \centering
	\begin{subfigure}{0.32\linewidth}
		\centering
		\includegraphics[width=1\linewidth]{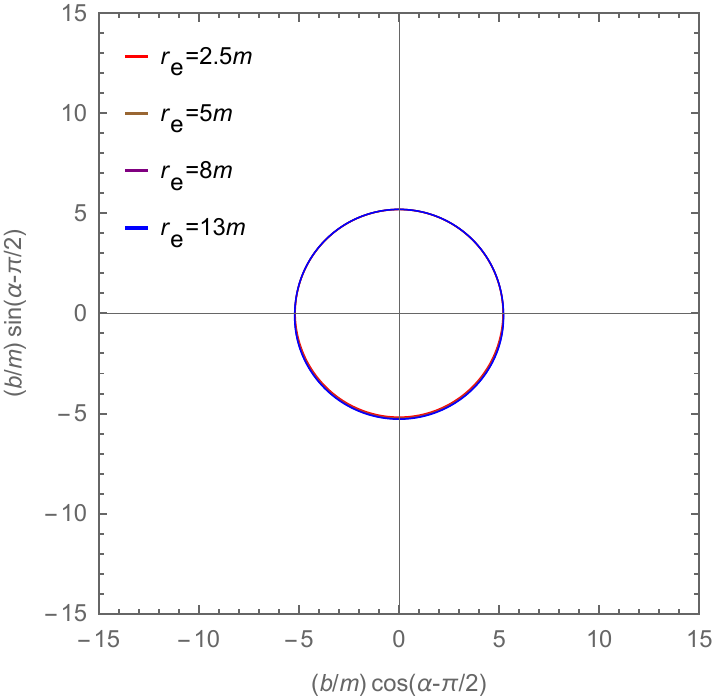}
		\caption{The third order, $\theta_{0}=4\pi/9$}
		\label{figg11.6}
	\end{subfigure}
	\caption{The isoradial curves of the first three order BH images for different $r_{\text{e}}$ in the CA models of optically and geometrically thin accretion with  $r^{\text{CA}}_{\text{inn}}=2.5m$ and $r^{\text{CA}}_{\text{out}}=13m$. The red and blue lines represent the two edge curves of the BH images. In each of these images, the horizontal axis corresponds to the $y'$-axis on the screen of the observer (with positive direction to the right), and the vertical axis corresponds to the $x'$-axis on the screen (with positive direction downward), where the polar angle $\alpha$ is measured from the $x'$-axis (cf.~Fig.~\ref{fig9}).
$\phantom{1111111111111111111111111}$}
	\label{fig11}
\end{figure}
The isoradial curves of the first three order BH images for different $r_{\text{e}}$ are shown in Fig.~\ref{fig11}, from which, it is known that the shapes of isoradial curves in a BH image depend on the order of the image and the inclination angle $\theta_{0}$ of the observer, and it has nothing to do with the emission pattern of the CA. In addition, as described later, for the third and higher order BH images, the bright regions bounded by the edge curves are greatly demagnified, and all their isoradial curves are concentrated near $b=b_{\text{cri}}$. Hence, in general, the shadow of a BH image in the CA accretion models should refer to the union of the central dark areas in the first and second order BH images. Thus, in view that the inner edge curve of the bright region for a given order BH image is dependent on $r^{\text{CA}}_{\text{inn}}$, one could conclude that the size and shape of the shadow will vary with the change of the inner boundary of the CA.

\subsection{ The redshift factors in the static, infalling, and rotating CA models and the formulas for the integrated intensity observed by a distant observer viewing the CA at an inclination angle}

After the transfer functions in the CA accretion models are presented, we will next embark on the derivation of the observed integrated intensity. Starting from the formulas in Refs.~\cite{Wen:2022hkv,Gralla:2019xty}, the evaluation of the observed specific intensity $I_{\text{o}}(\nu_{\text{o}})$ at the frequency $\nu_{\text{o}}$ along a lightlike geodesic is not difficult. Since accreting matters are assumed to be optically and geometrically thin in the CA accretion models, the emitted specific intensity could be denoted as
\begin{eqnarray}
\label{equ3.22}I_{\text{e}}(r_{\text{\text{e}}},\nu_{\text{e}})\quad\text{with}\quad r^{\text{CA}}_{\text{inn}}\leqslant r_{\text{e}}\leqslant r^{\text{CA}}_{\text{out}},
\end{eqnarray}
where $\nu_{\text{e}}$ is the emission frequency in the rest frame of the emitter. Along a lightlike geodesic, based on  identity~\cite{Gralla:2019xty}
\begin{eqnarray}
\label{equ3.23}\frac{I_{\text{e}}(r_{\text{e}},\nu_{\text{e}})}{\nu_{\text{e}}^{3}}=\frac{I_{\text{o}}(\nu_{\text{o}})}{\nu_{\text{o}}^{3}},
\end{eqnarray}
the observed specific intensity $I_{\text{o}}(\nu_{\text{o}})$ is given as
\begin{eqnarray}
\label{equ3.24}I_{\text{o}}(\nu_{\text{o}})=g^{3}I_{\text{e}}(r_{\text{e}},\nu_{\text{e}})
\end{eqnarray}
with $g:=\nu_{\text{o}}/\nu_{\text{e}}$ as the redshift factor. Thus, similarly to the formula~(\ref{equ2.2}), by integrating $I_{\text{o}}(\nu_{\text{o}})$ with respect to the frequency $\nu_{\text{o}}$, the integrated intensity of radiation on the observational screen is derived,
\begin{eqnarray}
\label{equ3.25}F_{\text{o}}=\int_{0}^{+\infty}I_{\text{o}}(\nu_{\text{o}})\text{d}\nu_{\text{o}}=g^{4}\int_{0}^{+\infty}I_{\text{e}}(r_{\text{e}},\nu_{\text{e}})\text{d}\nu_{\text{e}}.
\end{eqnarray}
In the above equation, $r_{\text{e}}$ is the radial coordinate of the emitting points, and since the observed integrated intensity should be a sum of the intensities contributed by the lightlike geodesics of all orders, it can be evaluated by inserting Eqs.~(\ref{equ3.19}) and (\ref{equ3.21}), namely,
\begin{eqnarray}
\label{equ3.26}F_{\text{o}}=\sum_{k=1}^{+\infty}g^{4}\int_{0}^{+\infty}I_{\text{e}}(r_{\text{e}},\nu_{\text{e}})\text{d}\nu_{\text{e}}\bigg|_{r_{\text{e}}=h\left(\iota_{k}(\alpha,\theta_{0}),b\right)},\quad b^{\text{CA}}_{\text{min}}(\alpha,\theta_{0},k)\leqslant b\leqslant b^{\text{CA}}_{\text{max}}(\alpha,\theta_{0},k),\quad
\end{eqnarray}
where one needs to note that the redshift factor $g$ is also a function of $r_{\text{e}}$. As seen from Eqs.~(\ref{equ3.4})  and~(\ref{equ3.18}), the above result actually provides the semianalytical method for evaluating the observed integrated intensity, and it is the generalization of the corresponding one in Ref.~\cite{Gralla:2019xty}. Differently from the case in the SS accretion models, the above expression of $F_{\text{o}}$ does not involve $b^{\text{CA}}_{\text{min}}(\alpha,\theta_{0},k)$ and $b^{\text{CA}}_{\text{max}}(\alpha,\theta_{0},k)$, which implies that the radial positions of the boundaries of the CA do not influence the luminosity of the bright region in a BH image.

Next, we will derive the redshift factors in the static, infalling, and rotating CA models of optically and geometrically thin accretion.
In what follows, by following the similar rule in Sec.~\ref{Sec:second}, the subscripts ``s'' ,  ``f'', and ``r''  represent that the corresponding quantities are defined in the static, infalling, and rotating models, respectively. Simple calculations indicate that the four-velocities of the static, infalling, and roatating emitters are~\cite{Luminet:1979nyg,Bambi:2011jq,Bambi:2012tg,Tian:2019yhn,Liu:2021lvk}, respectively,
\begin{eqnarray}
\label{equ3.27}u^{\mu}_{\text{es}}&=&(u^{0}_{\text{es}},0,0,0),\\
\label{equ3.28}u^{\mu}_{\text{ef}}&=&(u^{0}_{\text{ef}},u^{r}_{\text{ef}},0,0),\\
\label{equ3.29}u^{\mu}_{\text{er}}&=&(u^{0}_{\text{er}},0,0,u^{\varphi}_{\text{er}})
\end{eqnarray}
with
\begin{eqnarray}
\label{equ3.30}u^{0}_{\text{es}}&=&\frac{c}{\displaystyle\sqrt{1-\frac{2m}{r_{\text{e}}}}},\quad u^{0}_{\text{ef}}=\frac{\mathcal{E}^{\text{CA}}}{\displaystyle c\left(1-\frac{2m}{r_{\text{e}}}\right)},\quad u^{r}_{\text{ef}}= -c\sqrt{\frac{(\mathcal{E}^{\text{CA}})^{2}}{c^4}-\left(1-\frac{2m}{r_{\text{e}}}\right)},\\
\label{equ3.31}u^{0}_{\text{er}}&=&\frac{c}{\displaystyle\sqrt{1-\frac{3m}{r_{\text{e}}}}},\quad u^{\varphi}_{\text{er}}=\frac{\mathit{\Omega}}{\displaystyle \sqrt{1-\frac{3m}{r_{\text{e}}}}}=\frac{c\sqrt{\displaystyle\frac{m}{r^{3}_{\text{e}}}}}{\displaystyle \sqrt{1-\frac{3m}{r_{\text{e}}}}},
\end{eqnarray}
where in the infalling model, if $r^{\text{CA}}_{\text{ini}}$ is the coordinate of the initial radial position of accretion matters, there is
\begin{eqnarray}
\label{equ3.32}\mathcal{E}^{\text{CA}}:=c^{2}\sqrt{1-\frac{2m}{r^{\text{CA}}_{\text{ini}}}},
\end{eqnarray}
and in the rotating model, $\mathit{\Omega}:=\text{d}\varphi_{\text{er}}/\text{d}t_{\text{er}}=c\sqrt{m/r^{3}_{\text{e}}}$ is the angular velocity of the rotating emitters. For the rotating model, if accreting matters are rotating in opposite direction, the angular velocity should be $\mathit{\Omega}=-c\sqrt{m/r^{3}_{\text{e}}}$.  Before the formal evaluation of the redshift factor, the equation of the lightlike geodesic $x^{\mu}(\lambda)$ needs to be reconsidered because its entire path is not in the equatorial plane (plane $Oxy$). Thus, as a result, equations~(\ref{equA6})--(\ref{equA9}) should be slightly modified. In the Schwarzschild coordinate system displayed in Fig.~\ref{fig9}, the spherical polar coordinates are $(ct,r,\theta,\varphi)$. Based on the fact that the four-momentum $p^{\mu}=\text{d}x^{\mu}/\text{d}\lambda$ of photons is a lightlike vector and there are four killing vector fields (cf.~Eqs.~(\ref{equA2})--(\ref{equA5})) in the Schwarzschild spacetime, the equations satisfied by $p^{\mu}$ are
\begin{eqnarray}
\label{equ3.33} 0&=&\displaystyle g_{\alpha\beta}\frac{\text{d}x^{\alpha}}{\text{d}\lambda}\frac{\text{d}x^{\beta}}{\text{d}\lambda}=-\left(1-\frac{2m}{r}\right)c^2\left(\frac{\text{d}t}{\text{d}\lambda}\right)^2+\frac{\displaystyle\left(\frac{\text{d}r}{\text{d}\lambda}\right)^2}{\displaystyle 1-\frac{2m}{r}}+r^2\left[\left(\frac{\displaystyle\text{d}\theta}{\text{d}\lambda}\right)^2+\sin^{2}{\theta}\left(\frac{\displaystyle\text{d}\varphi}{\text{d}\lambda}\right)^2\right],\qquad\\
\label{equ3.34}-E&=&\displaystyle g_{\alpha\beta}\varepsilon_{0}^{\alpha}p^{\beta}=-c^2\left(1-\frac{2m}{r}\right)\frac{\text{d}t}{\text{d}\lambda},\\
\label{equ3.35}L_{1}&=&\displaystyle  g_{\alpha\beta}\varepsilon_{1}^{\alpha}p^{\beta}=-r^2\sin{\varphi}\frac{\text{d}\theta}{\text{d}\lambda}-r^2\sin{\theta}\cos{\theta}\cos{\varphi}\frac{\text{d}\varphi}{\text{d}\lambda},\\
\label{equ3.36}L_{2}&=&\displaystyle  g_{\alpha\beta}\varepsilon_{2}^{\alpha}p^{\beta}=r^2\cos{\varphi}\frac{\text{d}\theta}{\text{d}\lambda}-r^2\sin{\theta}\cos{\theta}\sin{\varphi}\frac{\text{d}\varphi}{\text{d}\lambda},\\
\label{equ3.37}L_{3}&=&\displaystyle  g_{\alpha\beta}\varepsilon_{3}^{\alpha}p^{\beta}=r^2\sin^{2}{\theta}\frac{\text{d}\varphi}{\text{d}\lambda},
\end{eqnarray}
where $E$, $L_{1}$, $L_{2}$, and $L_{3}$ are the conserved quantities related to the energy and the components of the angular momentum of photons. At infinity, if the coordinate time is represented by $\tilde{t}$, from Eq.~(\ref{equ3.34}), there is $E/c^2=\text{d}\tilde{t}/\text{d}\lambda$.
A straightforward check suggests that
\begin{eqnarray}
\label{equ3.38}\frac{L_{1}}{E}c^{2}&=&-r^2\sin{\varphi}\frac{\text{d}\theta}{\text{d}\tilde{t}}-r^2\sin{\theta}\cos{\theta}\cos{\varphi}\frac{\text{d}\varphi}{\text{d}\tilde{t}},\\
\label{equ3.39}\frac{L_{2}}{E}c^{2}&=&r^2\cos{\varphi}\frac{\text{d}\theta}{\text{d}\tilde{t}}-r^2\sin{\theta}\cos{\theta}\sin{\varphi}\frac{\text{d}\varphi}{\text{d}\tilde{t}},\\
\label{equ3.40}\frac{L_{3}}{E}c^{2}&=&r^2\sin^{2}{\theta}\frac{\text{d}\varphi}{\text{d}\tilde{t}}
\end{eqnarray}
could be exactly identified as the components of the angular momentum of the photons at infinity, and thus, the magnitude of the angular momentum ought to be
\begin{eqnarray}
\label{equ3.41}\frac{L}{E}c^{2}&=&r^2\left[\left(\frac{\displaystyle\text{d}\theta}{\text{d}\tilde{t}}\right)^2+\sin^{2}{\theta}\left(\frac{\displaystyle\text{d}\varphi}{\text{d}\tilde{t}}\right)^2\right]^{1/2}.\qquad
\end{eqnarray}
By applying $E/c^2=\text{d}\tilde{t}/\text{d}\lambda$ again, the desired result is achieved,
\begin{eqnarray}
\label{equ3.42}\frac{L^{2}}{r^{2}}=r^2\left[\left(\frac{\displaystyle\text{d}\theta}{\text{d}\lambda}\right)^2+\sin^{2}{\theta}\left(\frac{\displaystyle\text{d}\varphi}{\text{d}\lambda}\right)^2\right].
\end{eqnarray}
With this result, after plugging it into Eq.~(\ref{equ3.33}) and then employing Eq.~(\ref{equ3.34}),  the formulas similar to those in Eqs.~(\ref{equA9}) are derived, namely,
\begin{equation}\label{equ3.43}
\left\{\begin{array}{ll}
\displaystyle p^{0}&=\displaystyle \frac{E}{\displaystyle c\left(1-\frac{2m}{r}\right)},\smallskip\\
\displaystyle p^{r}&=\displaystyle \pm\frac{L}{b}\sqrt{1-\frac{b^2}{r^2}\left(1-\frac{2m}{r}\right)}
\end{array}\right.
\end{equation}
with $b=cL/E$ as the impact parameter of the geodesics, where the sign $+(-)$ indicates that photons go away from (approach) the BH. Now, we are in a position to derive the expressions of the redshift factors. As implied earlier, the spatial coordinates of the observer are  $(+\infty,\theta_{0},0)$, and then, his four-velocity is
\begin{eqnarray}
\label{equ3.44}u^{\mu}_{\text{o}}=(c,0,0,0).
\end{eqnarray}
With the four-velocities of the observer and emitter, by virtue of Eqs.~(\ref{equ3.27})--(\ref{equ3.32}) and (\ref{equ3.44}), the redshift factors in the three models are given by
\begin{eqnarray}
\label{equ3.45}&&g_{\text{s}}=\frac{-p_{\mu}u^{\mu}_{\text{o}}}{-p_{\nu}u^{\nu}_{\text{es}}}=\frac{1}{u^{0}_{\text{es}}/c}=\sqrt{1-\frac{2m}{r_{\text{e}}}},\\
\label{equ3.46}&&g^{(\text{outw})}_{\text{f}}=\frac{-p_{\mu}u^{\mu}_{\text{o}}}{-p_{\nu}u^{\nu}_{\text{ef}}}=\frac{1}{\displaystyle\frac{u^{0}_{\text{ef}}}{c}-\left(-\frac{p_{r}}{p_{0}}\right)^{(\text{outw})}\frac{u^{r}_{\text{ef}}}{c}}\nonumber\\
&&\qquad=\left(1-\frac{2m}{r_{\text{e}}}\right)\left(\sqrt{1-\frac{2m}{r^{\text{CA}}_{\text{ini}}}}+\sqrt{1-\frac{b^2}{r_{\text{e}}^2}\left(1-\frac{2m}{r_{\text{e}}}\right)}\sqrt{\frac{2m}{r_{\text{e}}}-\frac{2m}{r^{\text{CA}}_{\text{ini}}}}\right)^{-1},\\
\label{equ3.47}&&g^{(\text{inw})}_{\text{f}}=\frac{-p_{\mu}u^{\mu}_{\text{o}}}{-p_{\nu}u^{\nu}_{\text{ef}}}=\frac{1}{\displaystyle\frac{u^{0}_{\text{ef}}}{c}-\left(-\frac{p_{r}}{p_{0}}\right)^{(\text{inw})}\frac{u^{r}_{\text{ef}}}{c}}\nonumber\\
&&\qquad=\left(1-\frac{2m}{r_{\text{e}}}\right)\left(\sqrt{1-\frac{2m}{r^{\text{CA}}_{\text{ini}}}}-\sqrt{1-\frac{b^2}{r_{\text{e}}^2}\left(1-\frac{2m}{r_{\text{e}}}\right)}\sqrt{\frac{2m}{r_{\text{e}}}-\frac{2m}{r^{\text{CA}}_{\text{ini}}}}\right)^{-1},
\end{eqnarray}
\begin{eqnarray}
\label{equ3.48}&&g_{\text{r}}=\frac{-p_{\mu}u^{\mu}_{\text{o}}}{-p_{\nu}u^{\nu}_{\text{er}}}=\frac{1}{\displaystyle\frac{u^{0}_{\text{er}}}{c}+\frac{p_{\varphi}}{p_{0}}\frac{u^{\varphi}_{\text{er}}}{c}}=\sqrt{1-\frac{3m}{r_{\text{e}}}}\left(1+\sqrt{\frac{m}{r_{\text{e}}^{3}}}\,b\sin{\alpha}\sin{\theta_{0}}\right)^{-1},\qquad
\end{eqnarray}
where once again we adopt the rule that for a physical quantity $A$ defined on the lightlike geodesic, $(A)^{(\text{outw})}$  or $(A)^{(\text{inw})}$ represents that it takes values at points on the outward or inward segment of the geodesic, respectively. In the above derivations, the following identities
\begin{eqnarray}
\label{equ3.49}p_{0}&=&-\frac{E}{c}=-\frac{L}{b},\\
\label{equ3.50}\left(-\frac{p_{0}}{p^{r}}\right)^{(\text{outw})}&=&-\left(-\frac{p_{0}}{p^{r}}\right)^{(\text{inw})}=\frac{1}{\displaystyle\sqrt{1-\frac{b^2}{r_{\text{e}}^2}\left(1-\frac{2m}{r_{\text{e}}}\right)}},\\
\label{equ3.51}\left(-\frac{p_{r}}{p_{0}}\right)^{(\text{outw})}&=&-\left(-\frac{p_{r}}{p_{0}}\right)^{(\text{inw})}=\frac{\displaystyle\sqrt{1-\frac{b^2}{r_{\text{e}}^2}\left(1-\frac{2m}{r_{\text{e}}}\right)}}{\displaystyle1-\frac{2m}{r_{\text{e}}}},\\
\label{equ3.52}\frac{p_{\varphi}}{p_{0}}&=&b\sin{\alpha}\sin{\theta_{0}}
\end{eqnarray}
have been utilized. The first three identities can be directly obtained from Eqs.~(\ref{equ3.43}) and the definition of the impact parameter $b$, and the last one needs to be deduced with the aid of the geometric configuration presented in Fig.~\ref{fig9}.
Here, for convenience, we restore the use of $r_{0}$ to represent the radial coordinate of the observer, and in the previous statements, because he is so far away from the BH, we have set it to be $+\infty$. As displayed in Fig.~\ref{fig9}, at the receiving point $P_{\text{o}}$, the position and velocity vectors of photons are
\begin{eqnarray}
\label{equ3.53}\boldsymbol{r}&=&r_{0}\boldsymbol{e}_{\scriptstyle{OO'}}+b\boldsymbol{e}_{\scriptstyle{O'P_{\text{o}}}},\\
\label{equ3.54}\boldsymbol{v}&=&c\boldsymbol{e}_{\scriptstyle{OO'}},
\end{eqnarray}
and then, by Eqs.~(\ref{equ3.1}) and (\ref{equ3.9}), the angular momentum of the photons is given by
\begin{eqnarray}
\label{equ3.55}\boldsymbol{r}\times\boldsymbol{v}=bc(\cos{\theta_{0}}\sin{\alpha}\boldsymbol{e}_{x}-\cos{\alpha}\boldsymbol{e}_{y}-\sin{\theta_{0}}\sin{\alpha}\boldsymbol{e}_{z}).
\end{eqnarray}
On the basis of the previous argument, $L_{3}c^2/E$ is just the $z$-component of the angular momentum of the photons at infinity, so there is
\begin{eqnarray}
\label{equ3.56}\frac{L_{3}}{E}c^{2}=-bc\sin{\theta_{0}}\sin{\alpha}\Rightarrow\frac{p_{\varphi}}{p_{0}}=-\frac{L_{3}}{E}c=b\sin{\theta_{0}}\sin{\alpha}.
\end{eqnarray}

According to formulas~(\ref{equ3.45})--(\ref{equ3.48}), the redshift factors in the static and rotating CA models are easily handled, and the direct substitution of the redshift factors~(\ref{equ3.45}) and (\ref{equ3.48}) in Eq.~(\ref{equ3.26}) gives rise to the observed integrated intensities in these two models,
\begin{eqnarray}
\label{equ3.57}F_{\text{os}}&=&\sum_{k=1}^{+\infty}g_{\text{s}}^{4}\int_{0}^{+\infty}I_{\text{e}}(r_{\text{e}},\nu_{\text{e}})\text{d}\nu_{\text{e}}\bigg|_{r_{\text{e}}=h\left(\iota_{k}(\alpha,\theta_{0}),b\right)},\quad b^{\text{CA}}_{\text{min}}(\alpha,\theta_{0},k)\leqslant b\leqslant b^{\text{CA}}_{\text{max}}(\alpha,\theta_{0},k),\qquad\\
\label{equ3.58}F_{\text{or}}&=&\sum_{k=1}^{+\infty}g_{\text{r}}^{4}\int_{0}^{+\infty}I_{\text{e}}(r_{\text{e}},\nu_{\text{e}})\text{d}\nu_{\text{e}}\bigg|_{r_{\text{e}}=h\left(\iota_{k}(\alpha,\theta_{0}),b\right)},\quad b^{\text{CA}}_{\text{min}}(\alpha,\theta_{0},k)\leqslant b\leqslant b^{\text{CA}}_{\text{max}}(\alpha,\theta_{0},k).
\end{eqnarray}
As a contrast, the calculation of the redshift factor in the infalling model is a little subtle. When accreting matters radially move in towards the BH, the evaluation of the redshift factor along a lightlike geodesic is dependent on whether the emitting point is on the inward or outward segment. Obviously, for a lightlike geodesic with $b<b_{\text{cri}}$, photons are always moving away from the BH, which means that one only needs to choose formula~(\ref{equ3.46}) in this case. However,  when $b>b_{\text{cri}}$, the situation becomes complicated because the emitting point of the geodesic may be on the pre- or post-periastron branch. Thus, while the periastron of a lightlike geodesic coincides with its emitting point, how to determine the value of its impact parameter is crucial. As shown in Fig.~\ref{fig9}, for a $k$th order lightlike geodesic with $b>b_{\text{cri}}$ emitted from the point $P_{\text{e}}$ to the point $P_{\text{o}}$, according to Eq.~(\ref{equA40}),  the total change of the azimuthal angle $\phi$ in the plane  $P_{\text{e}}OO'$ is
\begin{eqnarray}
\label{equ3.59}|\Delta\phi|=2\sqrt{\frac{2}{X_{3}-X_{1}}}\left[\K{(k_{\text{lar}})}-\F{\left(\vartheta_{\text{larI}}(0),k_{\text{lar}}\right)}\right],
\end{eqnarray}
and then, the azimuthal angle coordinate of the periastron of the geodesic should be $\phi_{2}=|\Delta\phi|/2$. According to Fig.~2 in Ref.~\cite{Gralla:2019xty} and Eq.~(\ref{equA42}), it is known that as $b$ increases from $b_{\text{cri}}$ to $+\infty$, $\phi_{2}$ will decrease from $+\infty$ to $\pi/2$.
For a particular $k$th order lightlike geodesic whose periastron coincides with the emitting point, the azimuthal angle coordinate $\phi_{\text{e}}$ of the emitting point $P_{\text{e}}$ ought to satisfy
\begin{figure}[tbp]
\centering
	\begin{subfigure}{0.3\linewidth}
		\centering
		\includegraphics[width=0.8\linewidth]{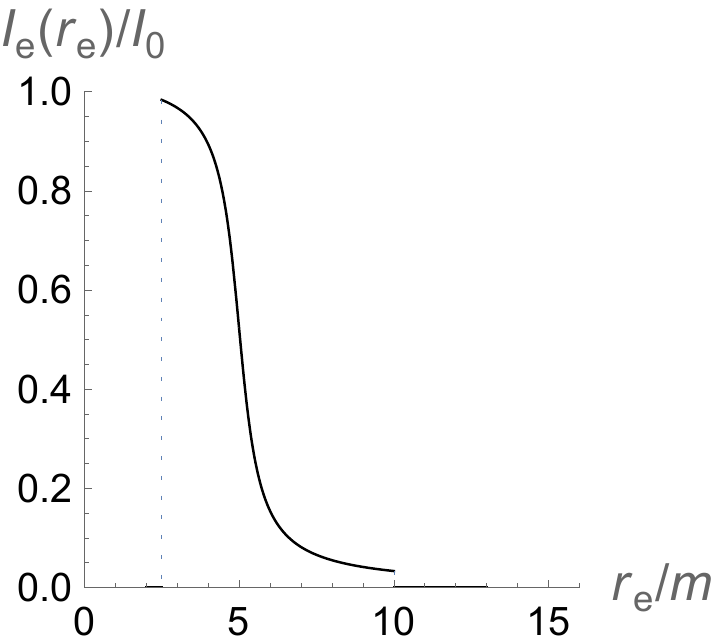}
		\caption{Source profile}
		\label{fig12.0}
	\end{subfigure}
	\centering
	\centering
	\begin{subfigure}{0.3\linewidth}
		\centering
		\includegraphics[width=0.8\linewidth]{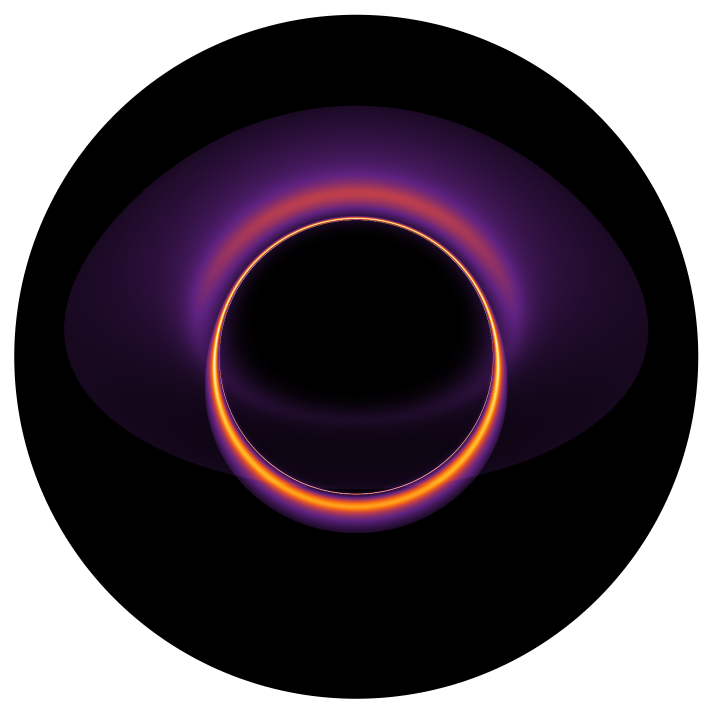}
		\caption{Infalling model, $\theta_{0}=\pi/3$}
		\label{fig12.1}
	\end{subfigure}
	\centering
	\begin{subfigure}{0.3\linewidth}
		\centering
		\includegraphics[width=0.8\linewidth]{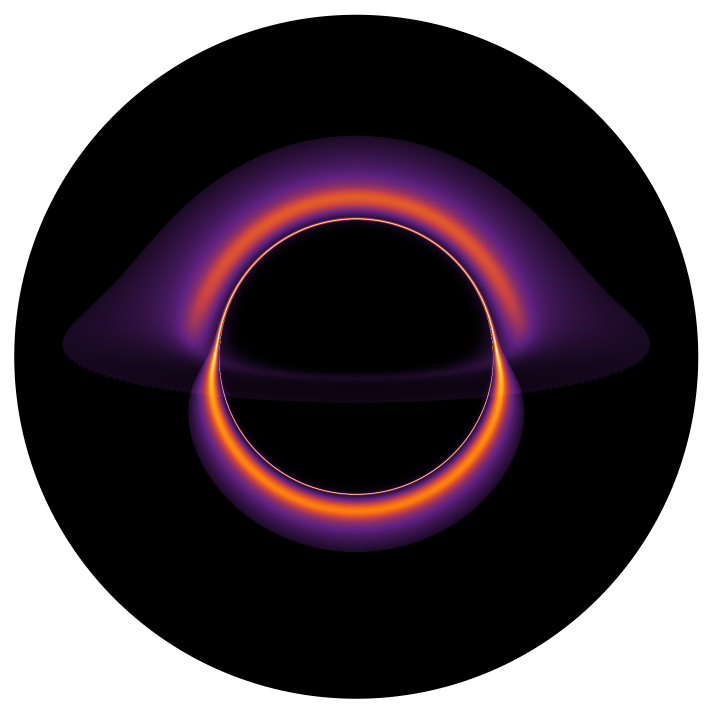}
		\caption{Infalling model, $\theta_{0}=4\pi/9$}
		\label{fig12.2}
	\end{subfigure}
    \centering
	\begin{subfigure}{0.3\linewidth}
		\centering
		\includegraphics[width=0.8\linewidth]{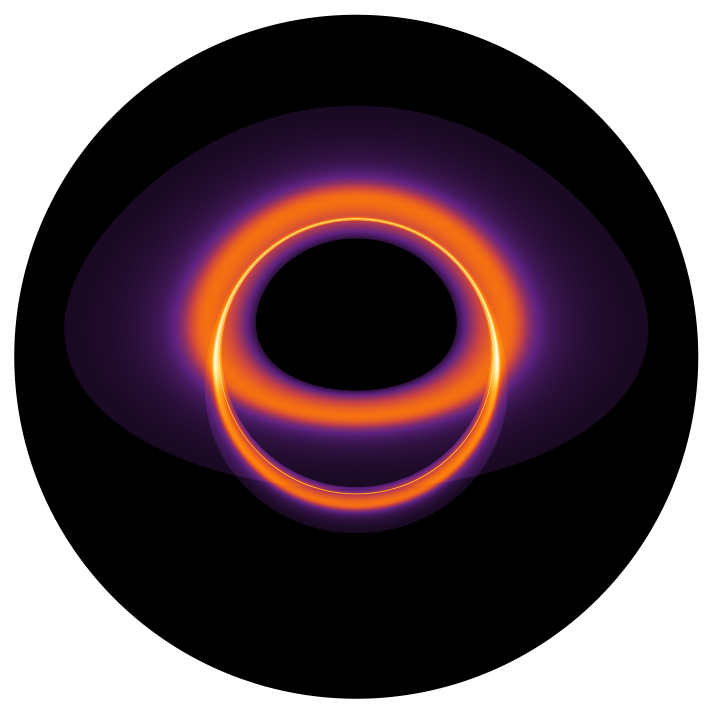}
		\caption{Static model, $\theta_{0}=\pi/3$}
		\label{fig12.3}
	\end{subfigure}
    \centering
	\begin{subfigure}{0.3\linewidth}
		\centering
		\includegraphics[width=0.8\linewidth]{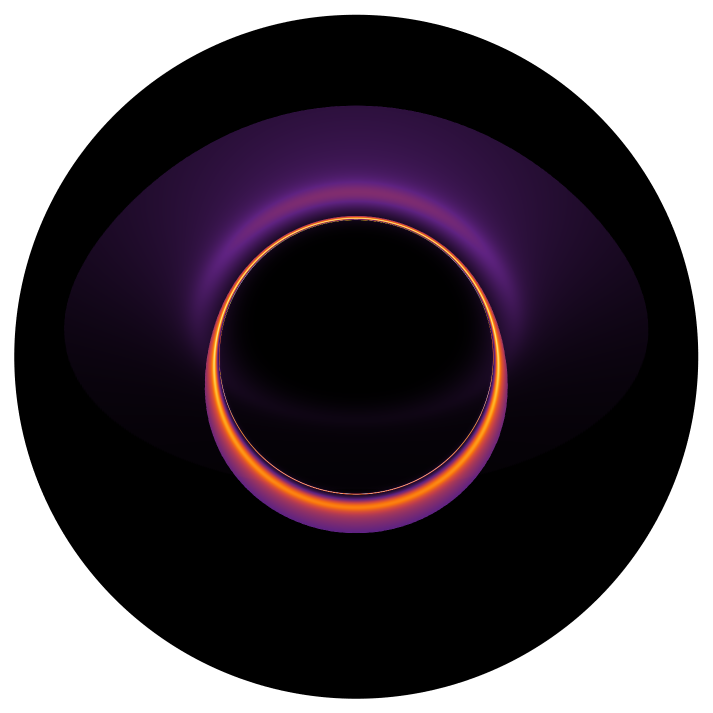}
		\caption{Infalling model, $\theta_{0}=\pi/3$}
		\label{fig12.4}
	\end{subfigure}
\centering
	\begin{subfigure}{0.3\linewidth}
		\centering
		\includegraphics[width=0.8\linewidth]{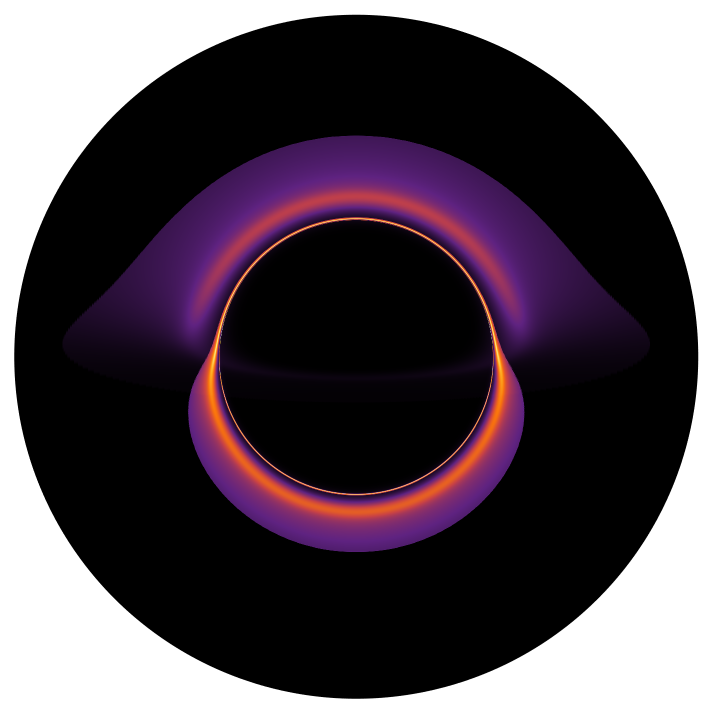}
		\caption{Infalling model, $\theta_{0}=4\pi/9$}
		\label{fig12.5}
	\end{subfigure}
	\centering
	\begin{subfigure}{0.3\linewidth}
		\centering
		\includegraphics[width=0.8\linewidth]{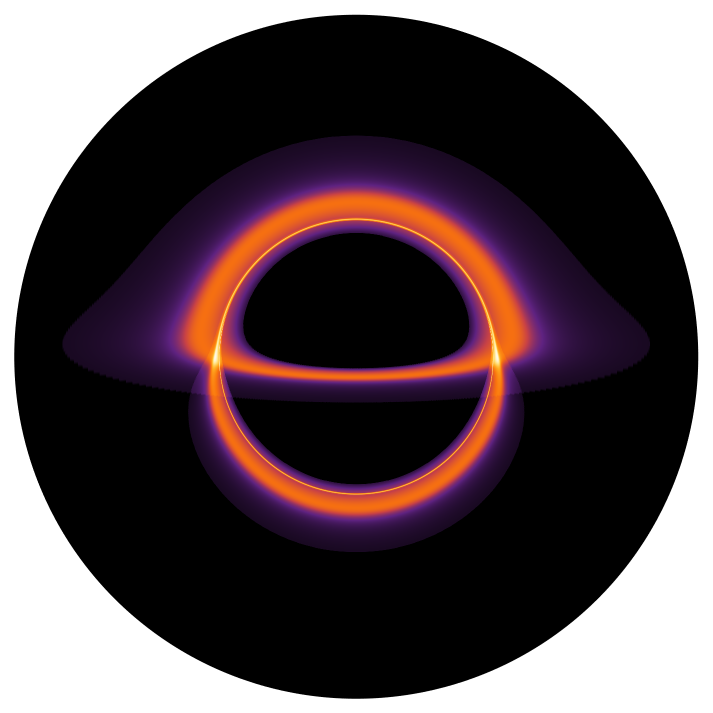}
		\caption{Static model, $\theta_{0}=4\pi/9$}
		\label{fig12.6}
	\end{subfigure}
    \centering
	\begin{subfigure}{0.3\linewidth}
		\centering
		\includegraphics[width=0.8\linewidth]{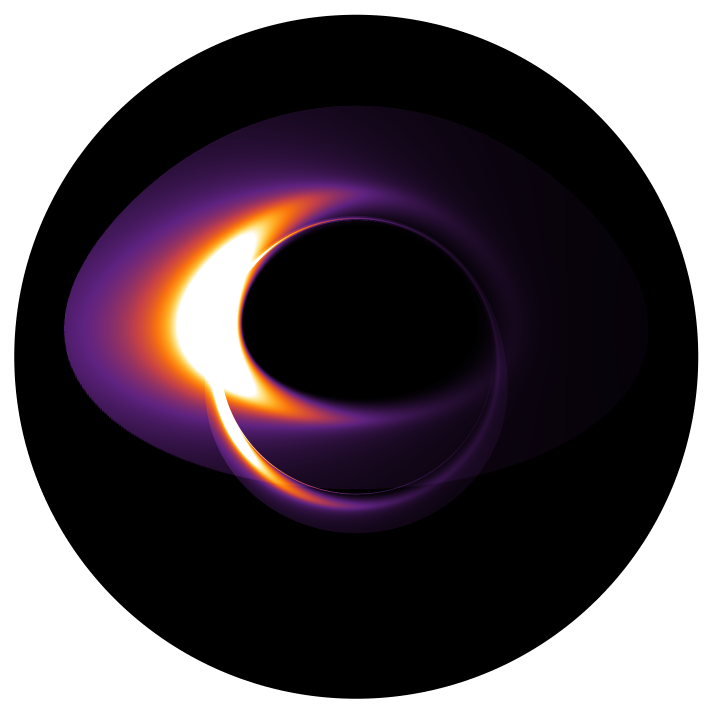}
		\caption{Rotating model, $\theta_{0}=\pi/3$}
		\label{fig12.7}
	\end{subfigure}
    \centering
	\begin{subfigure}{0.3\linewidth}
		\centering
		\includegraphics[width=0.8\linewidth]{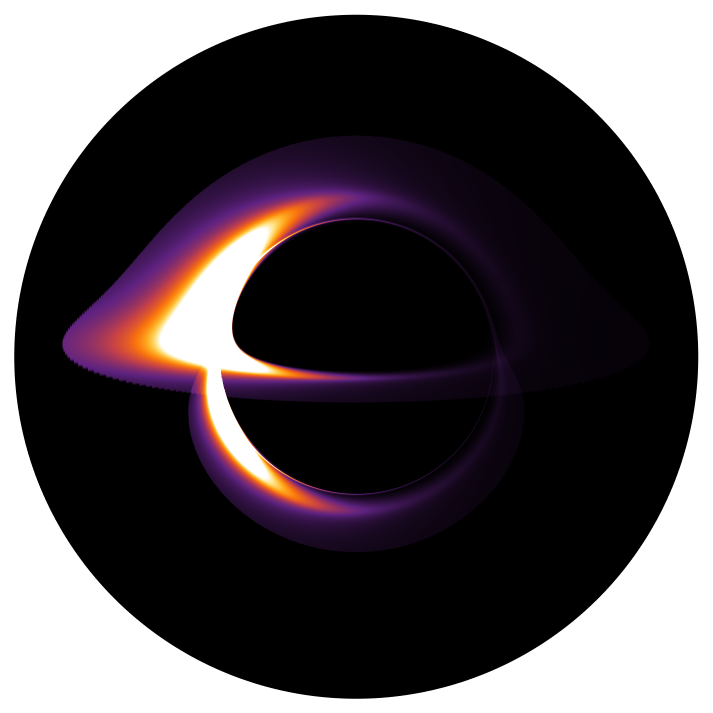}
		\caption{Rotating model, $\theta_{0}=4\pi/9$}
		\label{fig12.8}
	\end{subfigure}
    \caption{The Schwarzschild BH images observed by a distant observer viewing the CA at the inclination angle $\theta_{0}$ in the static, infalling, and rotating CA models of optically and geometrically thin accretion. The emitted and observed intensities are normalized to the maximum value $I_{0}$ of the emitted intensity. The outer boundary of the CA is set to be $r_{\text{out}}^{\text{CA}}=10m$, and the inner boundary of the CA is set to be $r_{\text{inn}}^{\text{CA}}=2.5m$ for the static and infalling models and $r_{\text{inn}}^{\text{CA}}=3m$ for the rotating model. The CA emission mainly arises from $r_{\text{e}}<6m$ but extends all the way down to $r_{\text{e}}=r_{\text{inn}}^{\text{CA}}$. The four images in the infalling model are plotted for $r^{\text{CA}}_{\text{ini}}=10m$ (first row)  and $r^{\text{CA}}_{\text{ini}}=+\infty$  (second row). $\phantom{11111111111111111111111111111111111111111111111111111111111111111111111111111111111111111}$}
	\label{fig12}
\end{figure}
\begin{eqnarray}
\label{equ3.60}\phi_{\text{e}}=\iota_{k}(\alpha,\theta_{0})=\phi_{2}=\frac{|\Delta\phi|}{2},
\end{eqnarray}
and through solving this equation, the impact parameter $b_{\text{hal}}(\alpha,\theta_{0},k)$ of this particular lightlike geodesic is obtained. Based on these discussions, we can now write the redshift factor in the infalling model. Let us first express
the observed integrated intensity in the following form,
\begin{eqnarray}
\label{equ3.61}F_{\text{of}}&=&\sum_{k=1}^{+\infty}g_{\text{f}}^{4}\int_{0}^{+\infty}I_{\text{e}}(r_{\text{e}},\nu_{\text{e}})\text{d}\nu_{\text{e}}\bigg|_{r_{\text{e}}=h\left(\iota_{k}(\alpha,\theta_{0}),b\right)},\quad b^{\text{CA}}_{\text{min}}(\alpha,\theta_{0},k)\leqslant b\leqslant b^{\text{CA}}_{\text{max}}(\alpha,\theta_{0},k).\quad
\end{eqnarray}
For a $k$th order lightlike geodesic with $b>b_{\text{cri}}$, the fact that the azimuthal angle coordinate $\phi_{2}$ of its periastron is a monotonically decreasing function of the impact parameter $b$ (cf.~Fig.~2 in Ref.~\cite{Gralla:2019xty}) means that
\begin{eqnarray}
\label{equ3.62}\left\{
\begin{array}{ll}
\displaystyle\phi_{2}\geqslant\phi_{\text{e}}\Rightarrow g_{\text{f}}=g^{(\text{outw})}_{\text{f}},&\quad\displaystyle\ \ \text{for}\ \   b\leqslant b_{\text{hal}}(\alpha,\theta_{0},k),\medskip\\
\displaystyle\phi_{2}<\phi_{\text{e}}\Rightarrow g_{\text{f}}=g^{(\text{inw})}_{\text{f}},&\quad\displaystyle\ \ \text{for}\ \   b>b_{\text{hal}}(\alpha,\theta_{0},k),
\end{array}\right.
\end{eqnarray}
and more specifically, the redshift factor in the infalling model could be presented as follows.
\begin{itemize}
\item When $b^{\text{CA}}_{\text{max}}(\alpha,\theta_{0},k)<b_{\text{hal}}(\alpha,\theta_{0},k)$, there is
\begin{eqnarray}
\label{equ3.63}g_{\text{f}}=g^{(\text{outw})}_{\text{f}},\qquad\displaystyle b^{\text{CA}}_{\text{min}}(\alpha,\theta_{0},k)\leqslant b\leqslant b^{\text{CA}}_{\text{max}}(\alpha,\theta_{0},k).
\end{eqnarray}
\item When $b^{\text{CA}}_{\text{min}}(\alpha,\theta_{0},k)\leqslant b_{\text{hal}}(\alpha,\theta_{0},k)\leqslant b^{\text{CA}}_{\text{max}}(\alpha,\theta_{0},k)$, there is
\begin{eqnarray}
\label{equ3.64}g_{\text{f}}=\left\{
\begin{array}{ll}
\displaystyle g^{(\text{outw})}_{\text{f}},&\quad\displaystyle\ \ \text{for}\ \   b^{\text{CA}}_{\text{min}}(\alpha,\theta_{0},k)\leqslant b\leqslant b_{\text{hal}}(\alpha,\theta_{0},k),\qquad\medskip\\
\displaystyle g^{(\text{inw})}_{\text{f}},&\quad\displaystyle\ \ \text{for}\ \   b_{\text{hal}}(\alpha,\theta_{0},k)<b\leqslant b^{\text{CA}}_{\text{max}}(\alpha,\theta_{0},k).\qquad
\end{array}\right.
\end{eqnarray}
\begin{figure}[tbp]
\centering
	\begin{subfigure}{0.3\linewidth}
		\centering
		\includegraphics[width=0.8\linewidth]{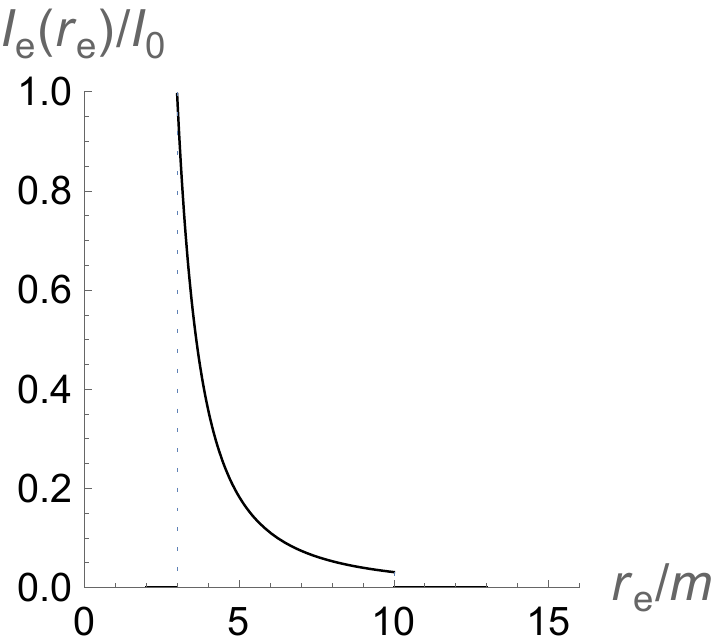}
		\caption{Source profile}
		\label{fig13.0}
	\end{subfigure}
	\centering
	\centering
	\begin{subfigure}{0.3\linewidth}
		\centering
		\includegraphics[width=0.8\linewidth]{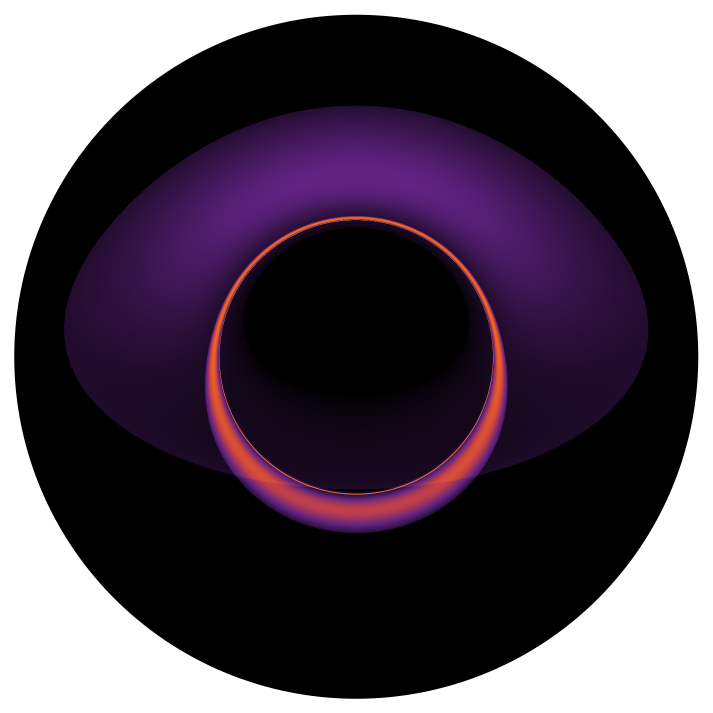}
		\caption{Infalling model, $\theta_{0}=\pi/3$}
		\label{fig13.1}
	\end{subfigure}
	\centering
	\begin{subfigure}{0.3\linewidth}
		\centering
		\includegraphics[width=0.8\linewidth]{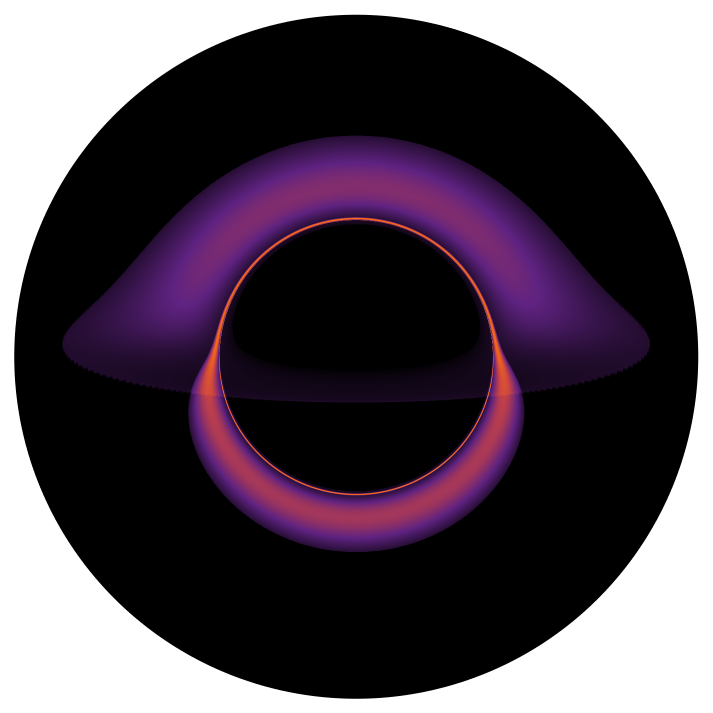}
		\caption{Infalling model, $\theta_{0}=4\pi/9$}
		\label{fig13.2}
	\end{subfigure}
    \centering
	\begin{subfigure}{0.3\linewidth}
		\centering
		\includegraphics[width=0.8\linewidth]{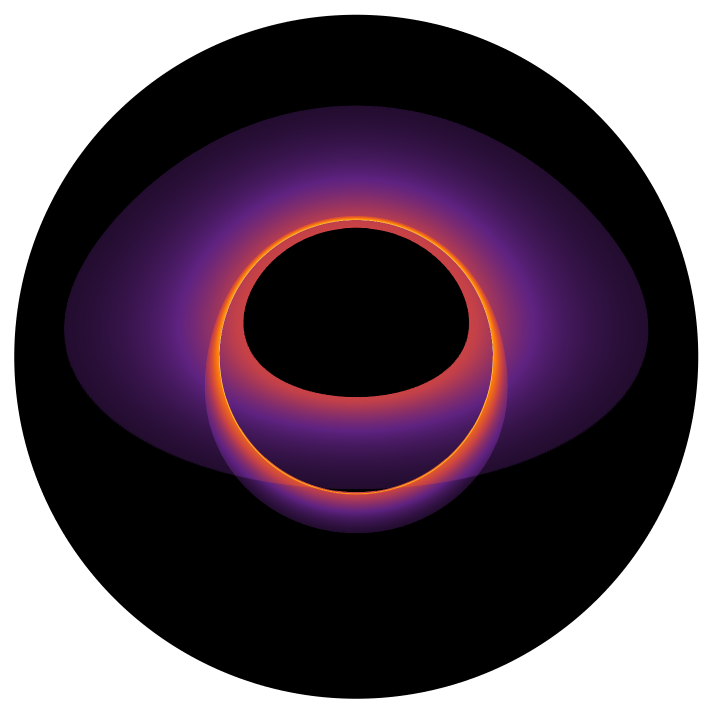}
		\caption{Static model, $\theta_{0}=\pi/3$}
		\label{fig13.3}
	\end{subfigure}
    \centering
	\begin{subfigure}{0.3\linewidth}
		\centering
		\includegraphics[width=0.8\linewidth]{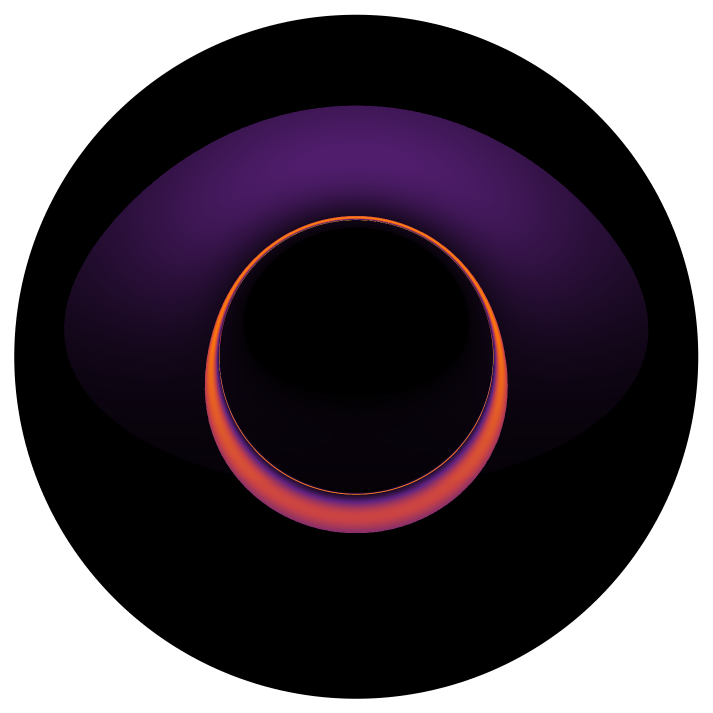}
		\caption{Infalling model, $\theta_{0}=\pi/3$}
		\label{fig13.4}
	\end{subfigure}
\centering
	\begin{subfigure}{0.3\linewidth}
		\centering
		\includegraphics[width=0.8\linewidth]{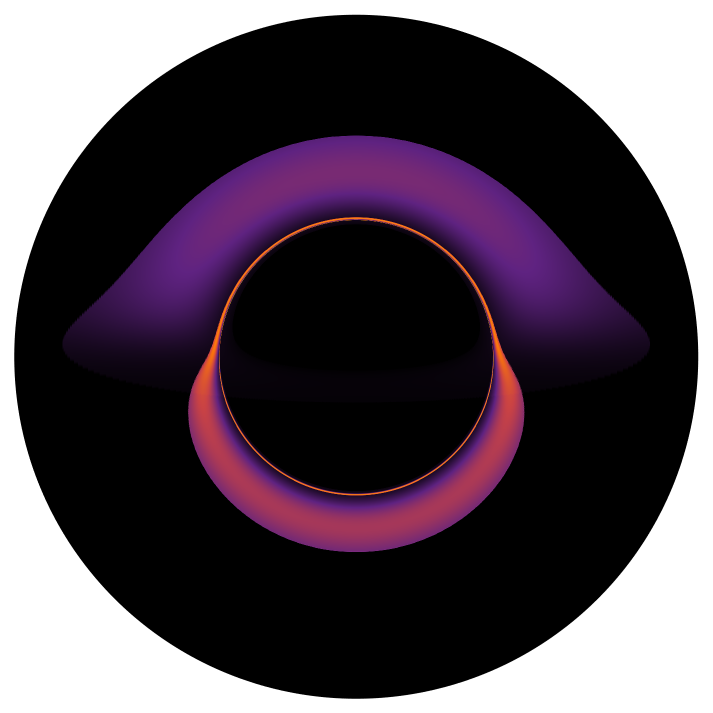}
		\caption{Infalling model, $\theta_{0}=4\pi/9$}
		\label{fig13.5}
	\end{subfigure}
	\centering
	\begin{subfigure}{0.3\linewidth}
		\centering
		\includegraphics[width=0.8\linewidth]{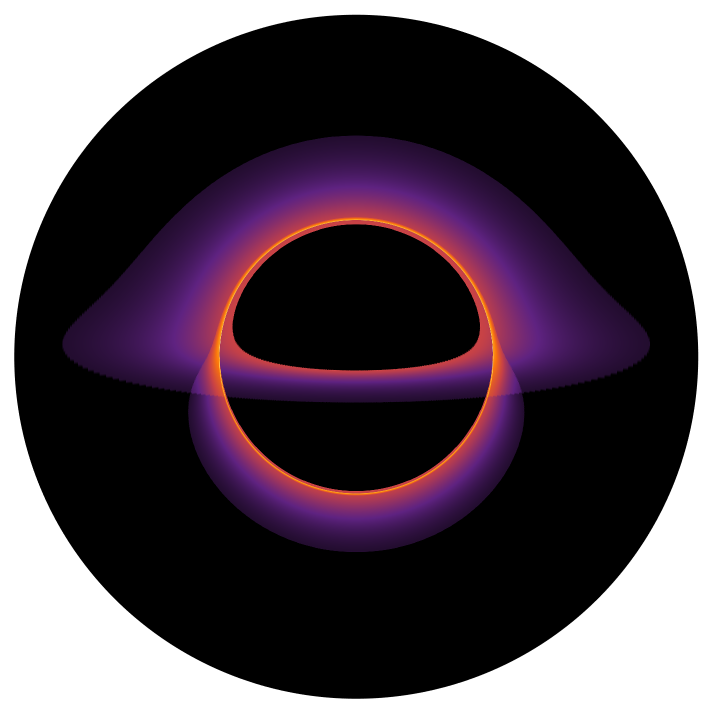}
		\caption{Static model, $\theta_{0}=4\pi/9$}
		\label{fig13.6}
	\end{subfigure}
    \centering
	\begin{subfigure}{0.3\linewidth}
		\centering
		\includegraphics[width=0.8\linewidth]{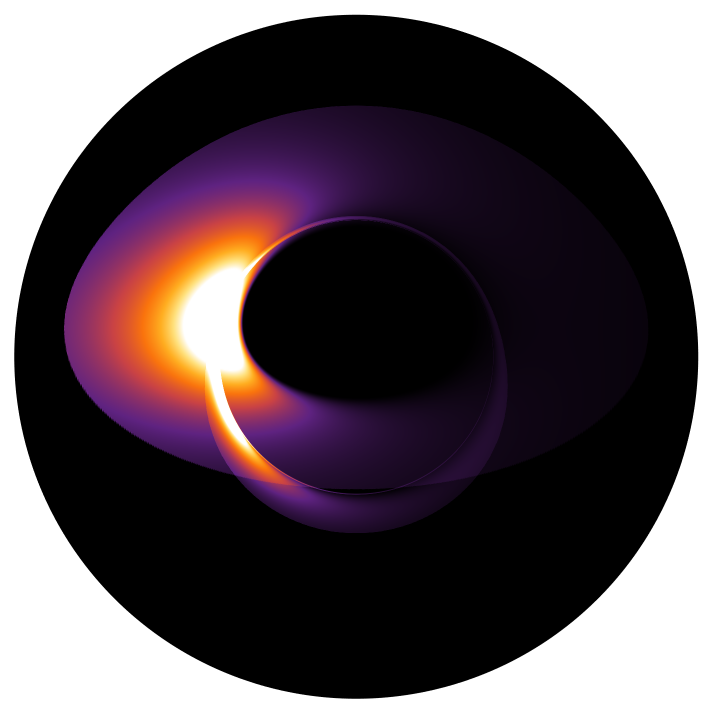}
		\caption{Rotating model, $\theta_{0}=\pi/3$}
		\label{fig13.7}
	\end{subfigure}
    \centering
	\begin{subfigure}{0.3\linewidth}
		\centering
		\includegraphics[width=0.8\linewidth]{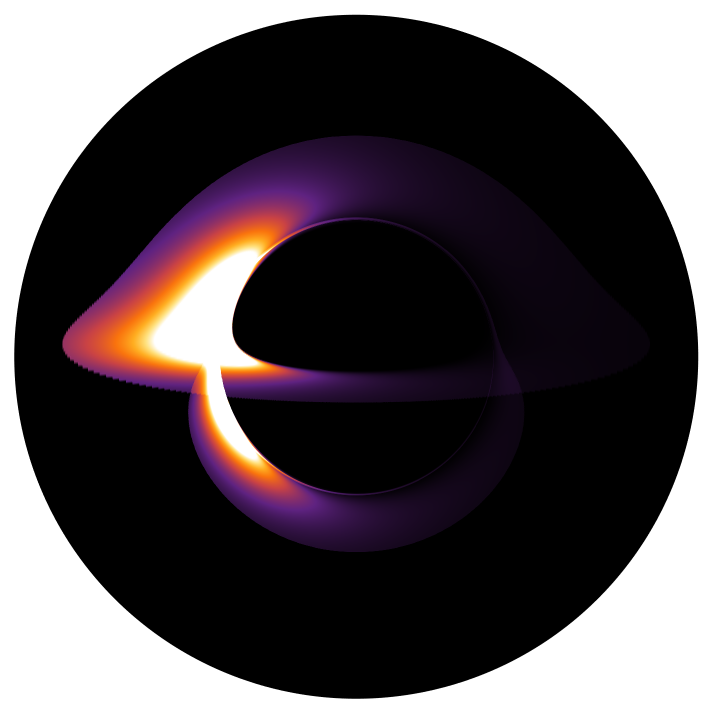}
		\caption{Rotating model, $\theta_{0}=4\pi/9$}
		\label{fig13.8}
	\end{subfigure}
    \caption{The Schwarzschild BH images observed by a distant observer viewing the CA at the inclination angle $\theta_{0}$ in the static, infalling, and rotating CA models of optically and geometrically thin accretion. The emitted and observed intensities are normalized to the maximum value $I_{0}$ of the emitted intensity.  The outer boundary of the CA is set to be $r_{\text{out}}^{\text{CA}}=10m$, and the inner boundary of the CA is set to be $r_{\text{inn}}^{\text{CA}}=3m$. The CA emission sharply reaches a peak at the bound photon orbit $r=3m$ before abruptly dropping. The four images in the infalling model are plotted for $r^{\text{CA}}_{\text{ini}}=10m$ (first row)  and $r^{\text{CA}}_{\text{ini}}=+\infty$  (second row). $\phantom{1111111111111111111111111111111111111111111111111111111}$}
	\label{fig13}
\end{figure}
\item When $b_{\text{hal}}(\alpha,\theta_{0},k)<b^{\text{CA}}_{\text{min}}(\alpha,\theta_{0},k)$, there is
\begin{eqnarray}
\label{equ3.65}g_{\text{f}}=g^{(\text{inw})}_{\text{f}},\qquad\displaystyle b^{\text{CA}}_{\text{min}}(\alpha,\theta_{0},k)\leqslant b\leqslant b^{\text{CA}}_{\text{max}}(\alpha,\theta_{0},k).
\end{eqnarray}
\end{itemize}
What needs to be emphasized is that under the case of $0\leqslant \alpha\leqslant\pi/2$ or $3\pi/2\leqslant \alpha<2\pi$ for $k=1$, equations~(\ref{equA42}), (\ref{equ3.5}),  (\ref{equ3.6}), (\ref{equ3.17}), and (\ref{equ3.62}) imply
\begin{eqnarray}
\label{equ3.66}\frac{\pi}{2}-\theta_{0}\leqslant\phi_{\text{e}}\leqslant\frac{\pi}{2}\Rightarrow\phi_{2}=\frac{|\Delta\phi|}{2}\geqslant\frac{\pi}{2}\geqslant\phi_{\text{e}}\Rightarrow g_{\text{f}}=g^{(\text{outw})}_{\text{f}},
\end{eqnarray}
and since the numerical calculation suggests that $b_{\text{hal}}(\alpha,\theta_{0},1)$ tends to infinity, Eq.~(\ref{equ3.63}) is actually applicable to this case. In addition, we will further prove that
\begin{eqnarray}
\label{equ3.67}b_{\text{cri}}\lesssim b_{\text{hal}}(\alpha,\theta_{0},k)<6m ,\quad \text{for}\quad k\geqslant2.
\end{eqnarray}
The proof is easy because for a particular $k$th $(k\geqslant2)$ order  lightlike geodesic whose periastron coincides with the emitting point, Eqs.~(\ref{equ3.18}) and (\ref{equ3.60}) indicate that
\begin{eqnarray}
\label{equ3.68}\phi_{\text{e}}\geqslant(k-1)\pi\Rightarrow|\Delta\phi|\geqslant2(k-1)\pi\geqslant2\pi,
\end{eqnarray}
and then, (\ref{equ3.67}) can immediately be read off from Fig.~2 in Ref.~\cite{Gralla:2019xty}.
\begin{figure}[tbp]
\centering
	\begin{subfigure}{0.3\linewidth}
		\centering
		\includegraphics[width=0.8\linewidth]{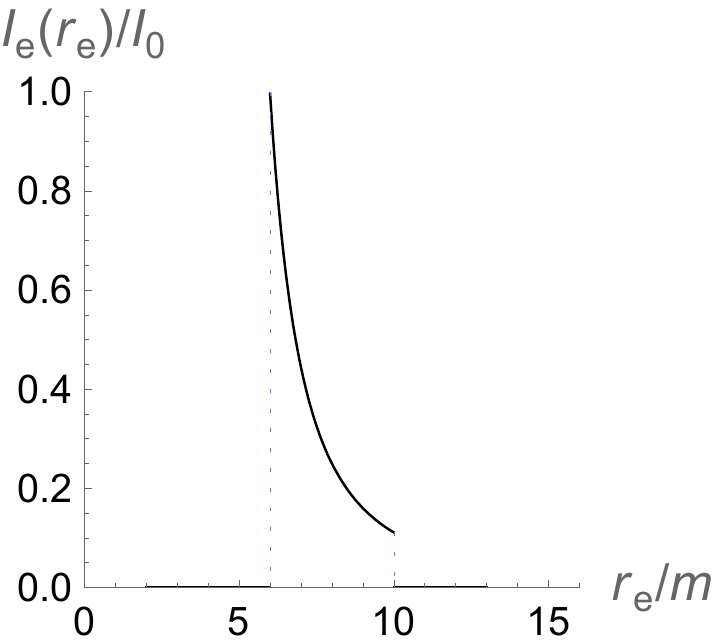}
		\caption{Source profile}
		\label{fig14.0}
	\end{subfigure}
	\centering
	\centering
	\begin{subfigure}{0.3\linewidth}
		\centering
		\includegraphics[width=0.8\linewidth]{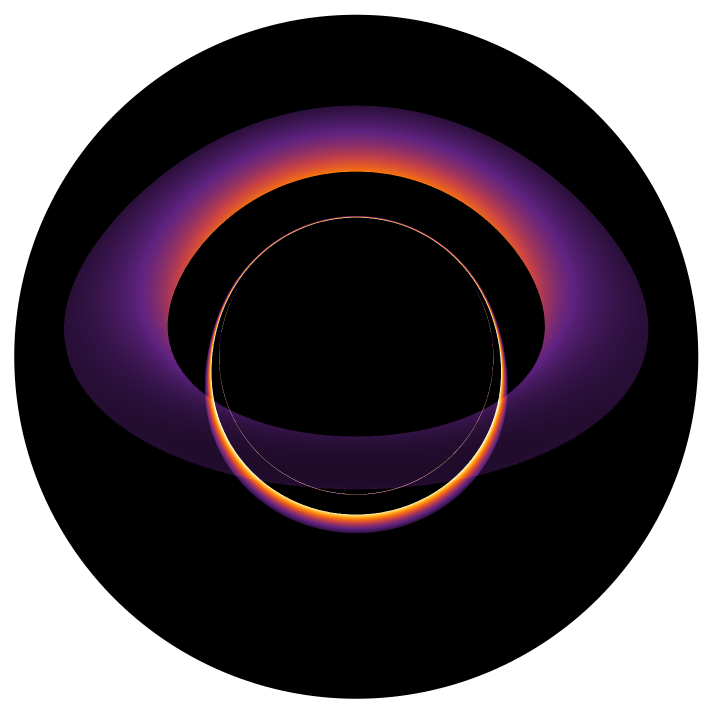}
		\caption{Infalling model, $\theta_{0}=\pi/3$}
		\label{fig14.1}
	\end{subfigure}
	\centering
	\begin{subfigure}{0.3\linewidth}
		\centering
		\includegraphics[width=0.8\linewidth]{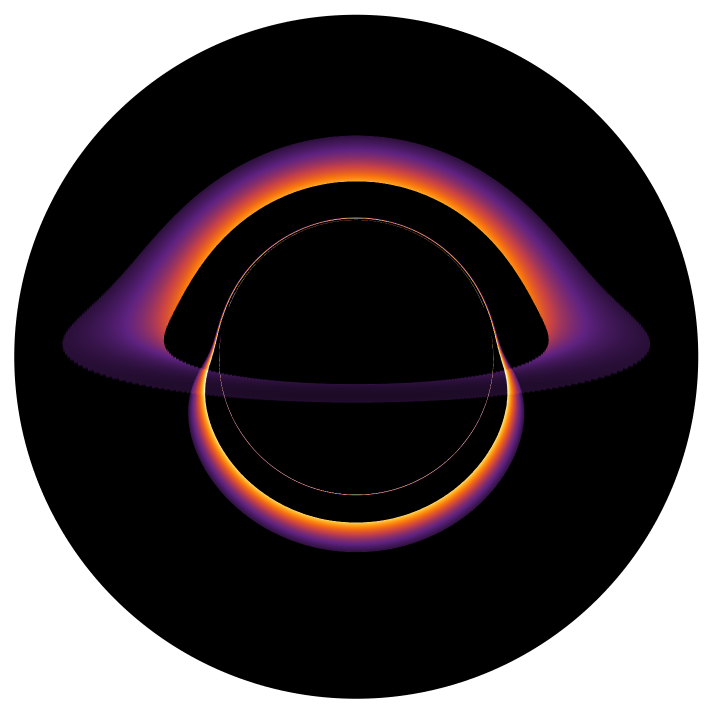}
		\caption{Infalling model, $\theta_{0}=4\pi/9$}
		\label{fig14.2}
	\end{subfigure}
    \centering
	\begin{subfigure}{0.3\linewidth}
		\centering
		\includegraphics[width=0.8\linewidth]{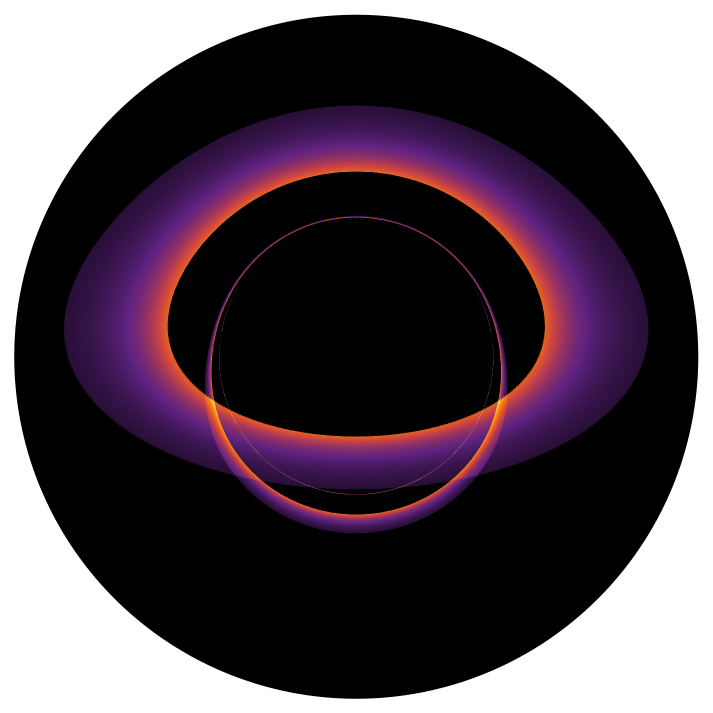}
		\caption{Static model, $\theta_{0}=\pi/3$}
		\label{fig14.3}
	\end{subfigure}
    \centering
	\begin{subfigure}{0.3\linewidth}
		\centering
		\includegraphics[width=0.8\linewidth]{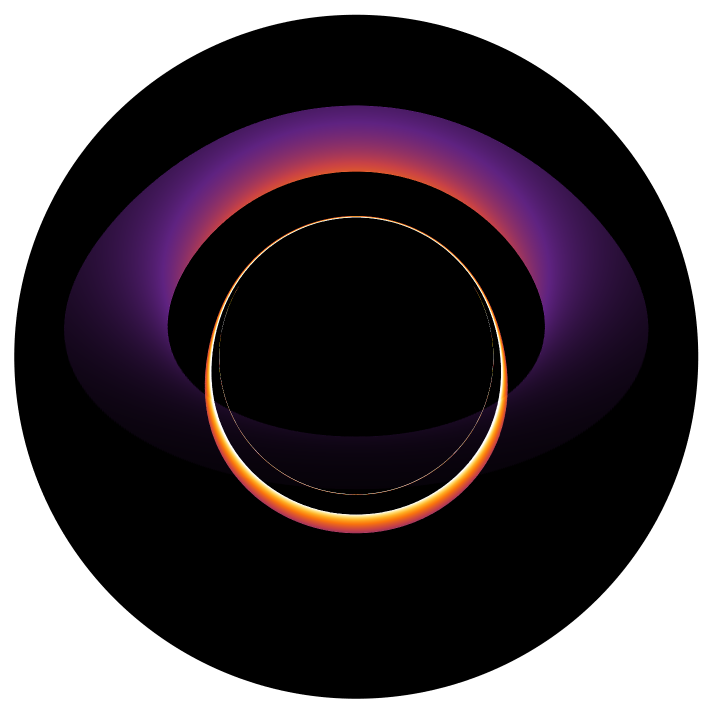}
		\caption{Infalling model, $\theta_{0}=\pi/3$}
		\label{fig14.4}
	\end{subfigure}
\centering
	\begin{subfigure}{0.3\linewidth}
		\centering
		\includegraphics[width=0.8\linewidth]{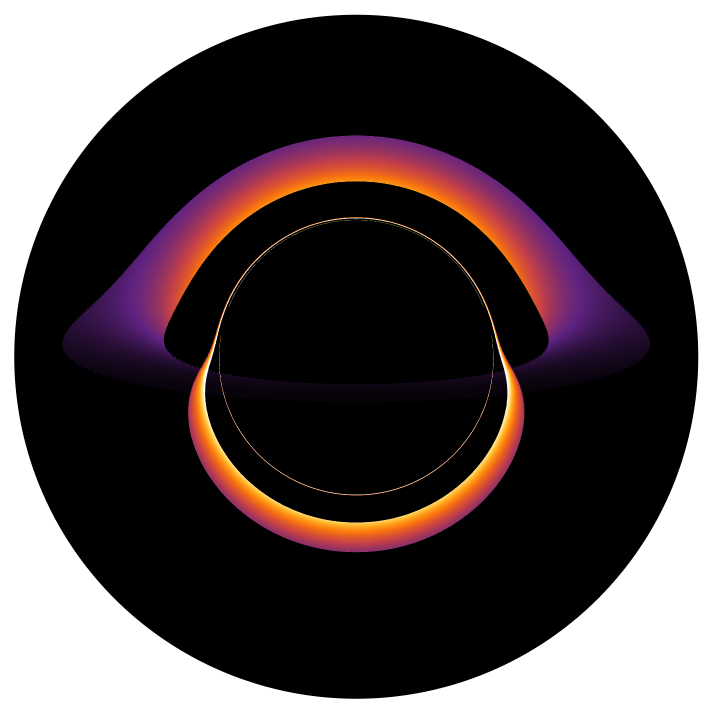}
		\caption{Infalling model, $\theta_{0}=4\pi/9$}
		\label{fig14.5}
	\end{subfigure}
	\centering
	\begin{subfigure}{0.3\linewidth}
		\centering
		\includegraphics[width=0.8\linewidth]{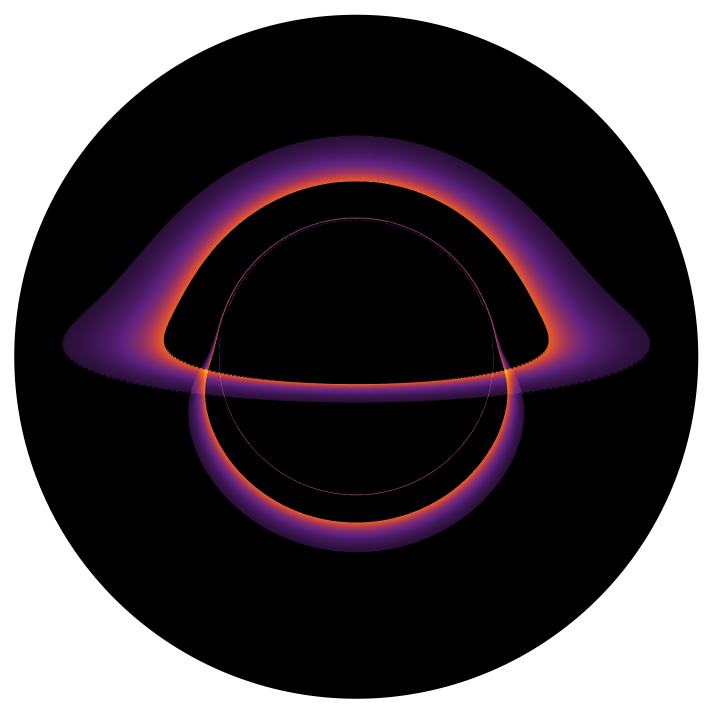}
		\caption{Static model, $\theta_{0}=4\pi/9$}
		\label{fig14.6}
	\end{subfigure}
    \centering
	\begin{subfigure}{0.3\linewidth}
		\centering
		\includegraphics[width=0.8\linewidth]{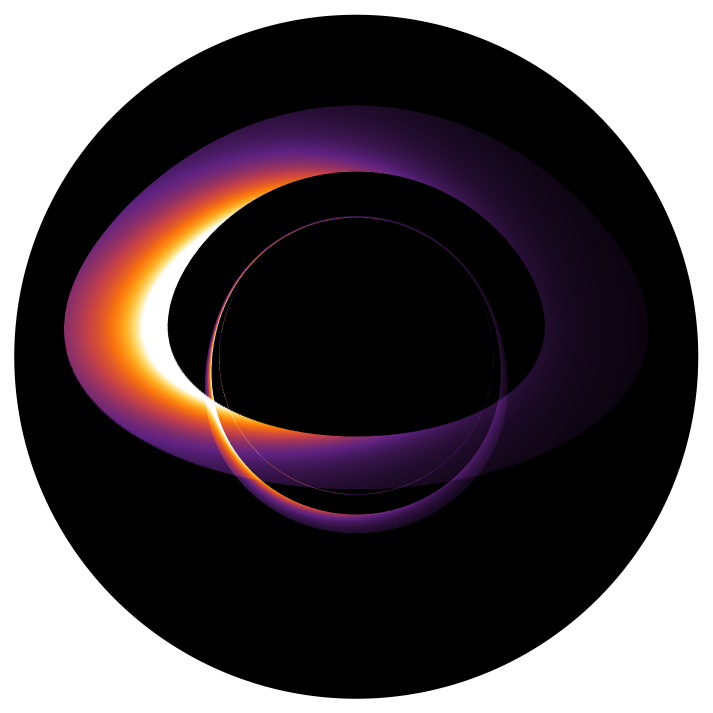}
		\caption{Rotating model, $\theta_{0}=\pi/3$}
		\label{fig14.7}
	\end{subfigure}
    \centering
	\begin{subfigure}{0.3\linewidth}
		\centering
		\includegraphics[width=0.8\linewidth]{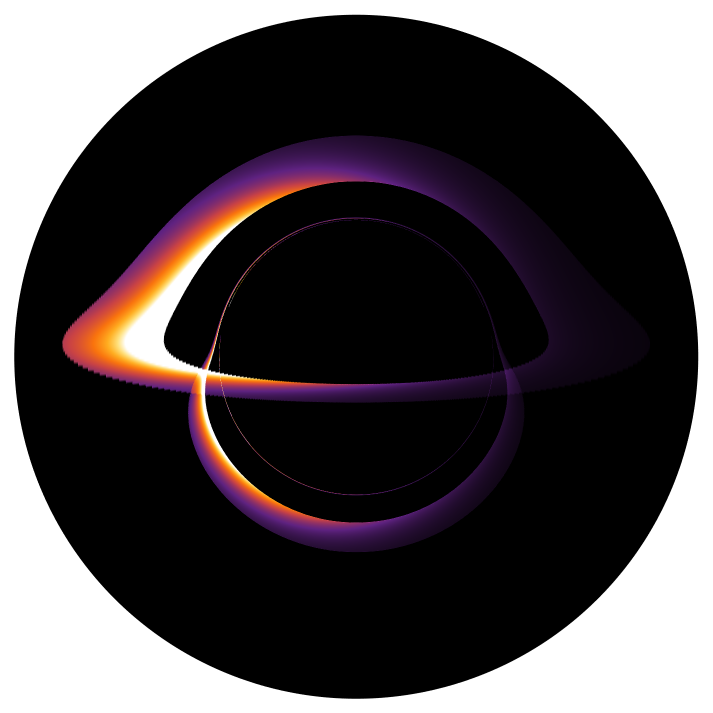}
		\caption{Rotating model, $\theta_{0}=4\pi/9$}
		\label{fig14.8}
	\end{subfigure}
    \caption{The Schwarzschild BH images observed by a distant observer viewing the CA at the inclination angle $\theta_{0}$ in the static, infalling, and rotating CA models of optically and geometrically thin accretion. The emitted and observed intensities are normalized to the maximum value $I_{0}$ of the emitted intensity. The outer boundary of the CA is set to be $r_{\text{out}}^{\text{CA}}=10m$, and the inner boundary of the CA is set to be $r_{\text{inn}}^{\text{CA}}=6m$. The CA emission sharply reaches a peak at $r=6m$ before abruptly dropping. The four images in the infalling model are plotted for $r^{\text{CA}}_{\text{ini}}=10m$ (first row)  and $r^{\text{CA}}_{\text{ini}}=+\infty$  (second row). $\phantom{1111111111111111111111111111111111111111111111111111111111111111111111111111111}$}
	\label{fig14}
\end{figure}

\subsection{The features of the Schwarzschild BH images in the static, infalling, and rotating models of optically and geometrically thin accretion}

Given the formulas~(\ref{equ3.57}), (\ref{equ3.58}), and (\ref{equ3.61}), once the emitted specific intensity $I_{\text{e}}(r_{\text{e}},\nu_{\text{e}})$ in the rest frame of the emitter is known, one is able to make use of them to generate the  Schwarzschild  BH images in the static, infalling, and rotating CA models of optically and geometrically thin accretion. In general, for the fourth and higher order BH images, the bright regions bounded by the two edge curves are extremely demagnified and their luminosities are extremely faint~\cite{Gralla:2019xty}, which implies that the contributions from the emissions of these orders to the complete BH image are negligible. Consequently, when plotting a BH image in the CA models, one only needs to take the first three order emissions into account. Like the case in the SS models, we also assume that the emission is monochromatic with rest-frame frequency $\nu_{0}$, namely,
\begin{eqnarray}
\label{equ3.69}I_{\text{e}}(r_{\text{e}},\nu_{\text{e}})=I_{\text{e}}(r_{\text{e}})\delta(\nu_{\text{e}}-\nu_{0}),
\end{eqnarray}
and then, by inserting it into Eqs.~(\ref{equ3.57}), (\ref{equ3.58}), (\ref{equ3.61}), and (\ref{equ3.63})--(\ref{equ3.65}), the corresponding integrated intensities observed by the observer in the static, infalling, and rotating situations are achieved. The next task of this section is to take the three emission patterns provided in Ref.~\cite{Gralla:2019xty} as examples to generate the corresponding BH images in the CA models and analyze their geometric and luminosity features. Figures~\ref{fig12}--\ref{fig14} are the BH images plotted based on the source profiles in the three emission patterns. Since the outer boundary of the CA is set to be $r_{\text{out}}^{\text{CA}}=10m$, for these source profiles, the outermost emissions occur at $r_{\text{e}}=10m$. In Fig.~\ref{fig12.0}, the inner boundary of the CA is set to be $r_{\text{inn}}^{\text{CA}}=2.5m$ for the static and infalling models and $r_{\text{inn}}^{\text{CA}}=3m$ for the rotating model, so that the innermost emissions of the source profile are observed at $r_{\text{e}}=r_{\text{inn}}^{\text{CA}}$, and it is shown that the emission profile mainly arising from $r_{\text{e}}<6m$ extends all the way down to the interior of the bound photon orbit. For the source profile in Fig.~\ref{fig13.0}, the innermost emissions occur exactly at the bound photon orbit because the inner boundary of the CA is set to be $r_{\text{inn}}^{\text{CA}}=3m$, and the emission profile sharply reaches a peak at $r_{\text{e}}=r_{\text{inn}}^{\text{CA}}$ before abruptly dropping, which is different from that in Fig.~\ref{fig12.0}. In Fig.~\ref{fig14.0}, the shape of the source profile is analogous to that in Fig.~\ref{fig13.0}, but since the inner boundary of the CA is set to be $r_{\text{inn}}^{\text{CA}}=6m$, the innermost emissions and the sharp peak lie outside the bound photon orbit.

In Figs.~\ref{fig12}--\ref{fig14}, the BH images of the first three order can be identified on the base of the shapes of the isoradial curves in Figs.~\ref{fig11} because in a given order BH image, isoradial curves essentially represent isobrightness curves. It is clearly exhibited that the shapes of the bright regions in the first and second BH images depend on the order of the image and the inclination $\theta_{0}$ of the observer, which is the same as that in the disk accretion model. In fact, this point can be confirmed by the previous statement that the shapes of isoradial curves in a given order BH image are dependant on the order of the image and the inclination angle of the observer. Thus, given that the isoradial curves in a BH image are depicted based on the transfer function~(\ref{equ3.19}), and the expressions of the transfer functions are only defined by the geometric configuration, the above point can be generalized to more general cases, and namely, for a BH image of arbitrary order in the CA models, the shape of the bright region  is determined by the geometric configuration, and it has noting to do with the emission profile. The images in  Figs.~\ref{fig12}--\ref{fig14} show that when the observer views the CA at a nonzero inclination angle $\theta_{0}$,  due to gravitational lensing, the shapes of the bright regions in different order BH images are completely different. The shape of the bright region in the first order BH image appears as a flattened oval, and as the angle $\theta_{0}$ increases, it becomes increasingly flattened. In the second order BH image, the bright region of the upper half-plane is confined to a thin semicircular ring, and that of the lower half-plane is expanded into a wide semicircular band. It is shown that with the increase of the angle $\theta_{0}$, the former becomes thinner, whereas the latter becomes wider. As to the third order BH image, the bright region forms an extremely narrow annular ring with radius around $b\approx5.2m$, and when the angle $\theta_{0}$ varies, this annular ring remains almost unchanged.

Partial geometric features of these bright regions can be qualitatively explained. When the observer views the CA at the face-on orientation, as mentioned earlier, there is $\iota_{k}(\alpha,0)=(k-1/2)\pi$, so that according to Eq.~(\ref{equ3.20}), once the boundaries of the CA are given, the inner and outer edge curves of the bright regions in the $k$th order BH image are provided by
\begin{eqnarray}
\label{equ3.70}h\left(\iota_{k}(\alpha,0),b\right)=r^{\text{CA}}_{\text{inn}},\qquad h\left(\iota_{k}(\alpha,0),b\right)=r^{\text{CA}}_{\text{out}}.
\end{eqnarray}
The solutions to the above two equations are $b^{\text{CA}}_{\text{min}}(\alpha,0,k)$ and $b^{\text{CA}}_{\text{max}}(\alpha,0,k)$, and they are just two numbers associated with $r^{\text{CA}}_{\text{inn}}$ and $r^{\text{CA}}_{\text{out}}$, so in this case,
the bright region in a given order BH image is an annular zone bounded by the circles with radius $b^{\text{CA}}_{\text{min}}(\alpha,0,k)$ and $b^{\text{CA}}_{\text{max}}(\alpha,0,k)$. For the case of $\theta_{0}\neq0$, the edge curves of the bright region in the $k$th order BH image are described by $b^{\text{CA}}_{\text{min}}(\alpha,\theta_{0},k)$ and $b^{\text{CA}}_{\text{max}}(\alpha,\theta_{0},k)$ in Eq.~(\ref{equ3.21}), and they depend on the angle $\alpha$. Aa a result, the bright region in a given order BH image is no longer an annular zone. For the third and higher order BH images, equation~(\ref{equ3.18}) implies that the total change of the azimuthal angle $\phi$ in the plane $P_{\text{e}}OO'$ along a light geodesic intersecting the screen of the observer is more than $2\pi$, and thus, according to Fig~2 in Ref.~\cite{Gralla:2019xty}, the bright regions in these high order images should form extremely narrow annular rings with radius around $b\approx5.2m$. The same reasoning can also be employed to explain why the width of the bright region in the upper half-plane is far more narrow than that in the lower half-plane for the second order BH image. As shown in Fig.~\ref{fig9}, along a second-order lightlike geodesic, the total change of the azimuthal angle $\phi$ is more than $3\pi/2$ in the half-plane $\pi/2<\alpha<3\pi/2$ and more than $\pi$ in the half-plane $-\pi/2<\alpha<\pi/2$. Thus, from Fig.~2 in Ref.~\cite{Gralla:2019xty}, it is not difficult to infer that in the second BH image, the bright region of the upper half-plane is confined to a thin semicircular ring, and that of the lower half-plane is expanded into a semicircular band. The shape of the bright region in the first order BH image can not be qualitatively interpreted as described above because in this case, a thorough understanding of the quantitative behaviors of the lightlike geodesics is required.

In view that the images in Figs.~\ref{fig12}--\ref{fig14} show the bright region in the third order BH image makes only a weak contribution to the total flux, the shadow of a BH image in the CA accretion models should refer to the union of the central dark areas in the first and second order BH images. According to Eqs.~(\ref{equ3.20}) and (\ref{equ3.21}), the inner edge curves of the bright regions in the first two order BH images are described by $b^{\text{CA}}_{\text{min}}(\alpha,\theta_{0},1)$ and $b^{\text{CA}}_{\text{min}}(\alpha,\theta_{0},2)$, and both of them are determined by the inner boundary of the CA, which means that the size of the shadow will vary as the inner boundary of the CA changes. In addition, the images in Figs.~\ref{fig12}--\ref{fig14} also imply that the geometric shape of the shadow is also dependent on the inner boundary of the CA. In Figs.~\ref{fig12} and~\ref{fig13}, since the inner boundary of the CA is sufficiently close to the BH, the central dark area of the first order BH images lies within that of the second order BH images, which results in that the shadow consists of two distinct dark parts. As a contrast, in Figs.~\ref{fig14}, the inner boundary of the CA is sufficiently far from the BH, so the central dark areas of the first two order BH images partially overlap, and thus, the shadow consists of three distinct dark parts. In this case, since the bright regions in the second and third order BH images do not touch each other, the shadow actually consists of four distinct dark parts. Based on the above discussions, we can conclude that the shadow of a BH image in the CA accretion models usually consists of some distinct dark parts. It should be noted that all these dark parts within the shadow of a BH image have a luminosity of zero because for each order BH image, the luminosity of the central dark area is zero, which is different from the case in the SS accretion models. Finally, another point that needs to be noted is that the inner boundary of the CA in the static or infalling model could extend to $r=2m$, whereas in the rotating model, it could only extend to $r=3m$ (cf.~Eq.~(\ref{equ3.31})), so for the first emission pattern in Fig.~\ref{fig12.0}, the inner boundary of the CA can only be set to be
$r_{\text{inn}}^{\text{CA}}=3m$ in the rotating CA model, which explains why in this case the shadow area in the static and infalling models is smaller than that in the rotating model.

\begin{figure}[tbp]
	\centering
	\begin{subfigure}{0.32\linewidth}
		\centering
		\includegraphics[width=1\linewidth]{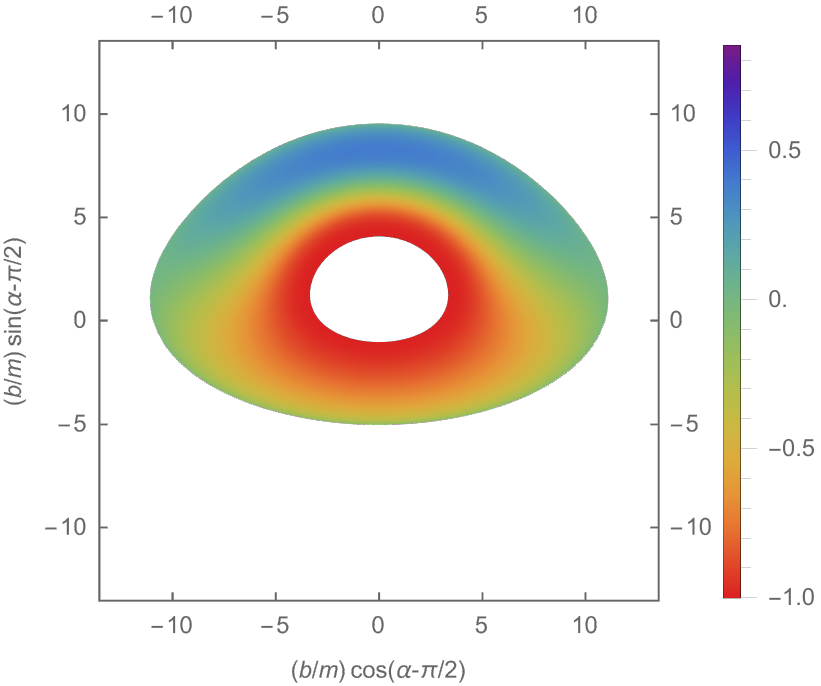}
		\caption{$Rc_{\text{fs}}(b,\alpha,\theta_{0},1)$}
		\label{fig15.1}
	\end{subfigure}
	\centering
	\begin{subfigure}{0.32\linewidth}
		\centering
		\includegraphics[width=1\linewidth]{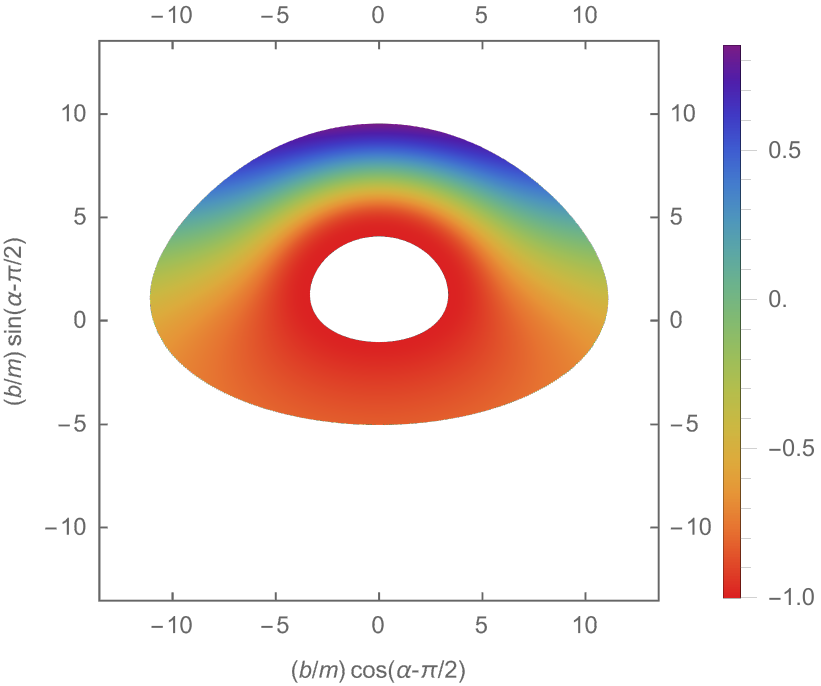}
		\caption{$Rc_{\text{fs}}(b,\alpha,\theta_{0},1)$}
		\label{fig15.2}
	\end{subfigure}
    \centering
	\begin{subfigure}{0.32\linewidth}
		\centering
		\includegraphics[width=1\linewidth]{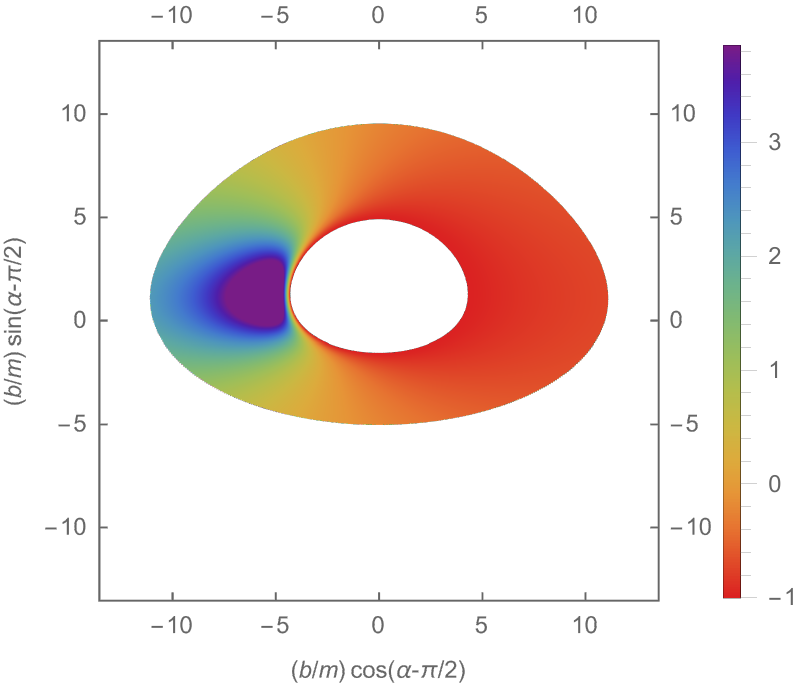}
		\caption{$Rc_{\text{rs}}(b,\alpha,\theta_{0},1)$}
		\label{fig15.3}
	\end{subfigure}
    \centering
	\begin{subfigure}{0.32\linewidth}
		\centering
		\includegraphics[width=1\linewidth]{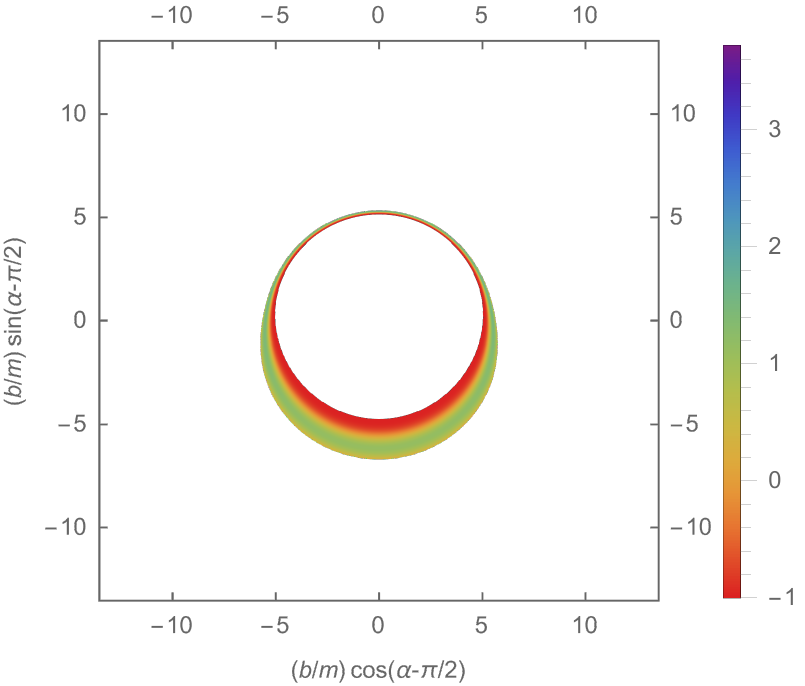}
		\caption{$Rc_{\text{fs}}(b,\alpha,\theta_{0},2)$}
		\label{fig15.4}
	\end{subfigure}
\centering
	\begin{subfigure}{0.32\linewidth}
		\centering
		\includegraphics[width=1\linewidth]{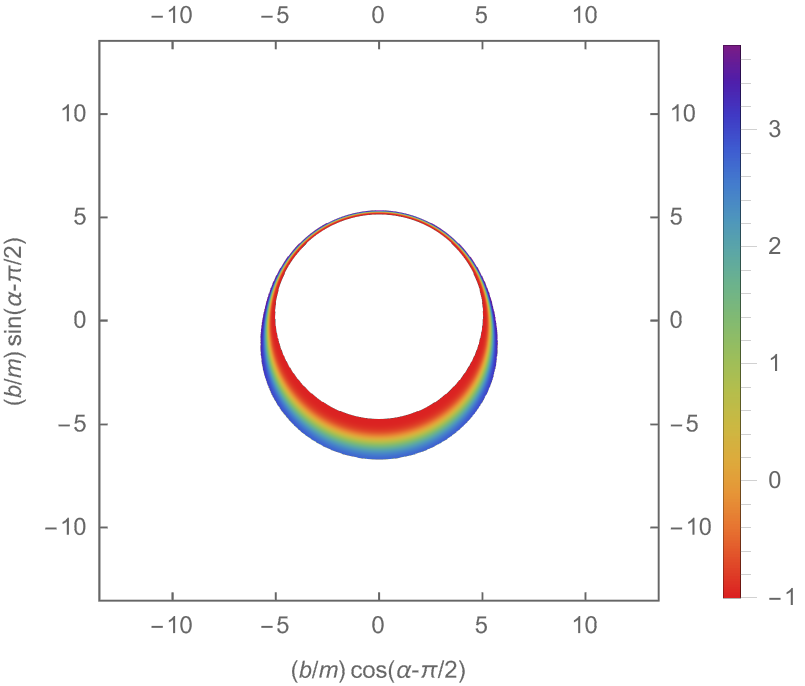}
		\caption{$Rc_{\text{fs}}(b,\alpha,\theta_{0},2)$}
		\label{fig15.5}
	\end{subfigure}
	\centering
	\begin{subfigure}{0.32\linewidth}
		\centering
		\includegraphics[width=1\linewidth]{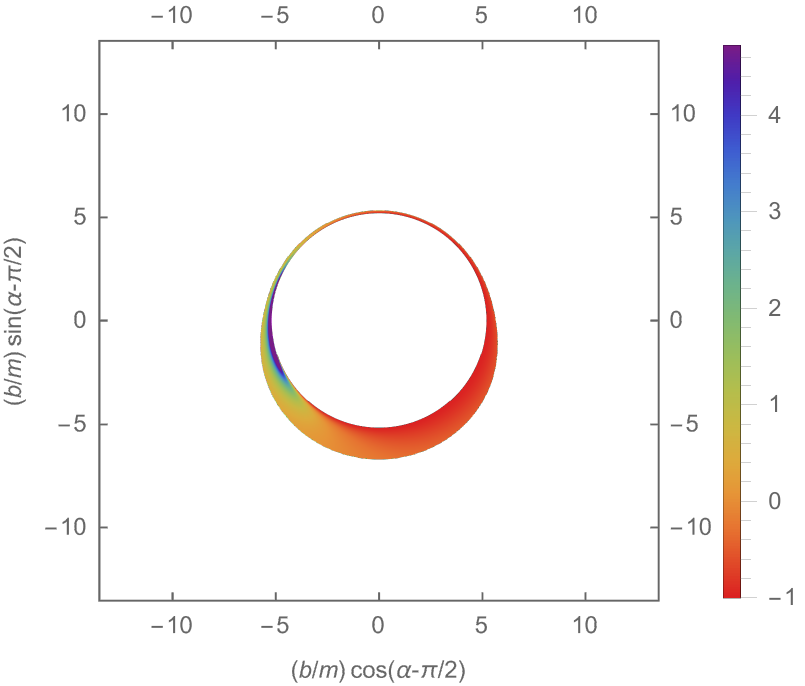}
		\caption{$Rc_{\text{rs}}(b,\alpha,\theta_{0},2)$}
		\label{fig15.6}
	\end{subfigure}
	\caption{Behaviors of the redshift comparison functions for $\theta_{0}=\pi/3$ in the first (top row) and second (bottom row) order BH images, where the range of $r_{\text{e}}$ appearing in the expressions of $Rc_{\text{fs}}(b,\alpha,\theta_{0},1)$ and $Rc_{\text{fs}}(b,\alpha,\theta_{0},2)$ is extended to $2m\leqslant r_{\text{e}}\leqslant10m$, and  the range of $r_{\text{e}}$ appearing in the expressions of $Rc_{\text{rs}}(b,\alpha,\theta_{0},1)$ and $Rc_{\text{rs}}(b,\alpha,\theta_{0},2)$ is extended to $3m\leqslant r_{\text{e}}\leqslant10m$.  The images in the first column are plotted for $r^{\text{CA}}_{\text{ini}}=10m$ and the images in the second column are plotted for $r^{\text{CA}}_{\text{ini}}=+\infty$. For these four images, the colorbars show the value ranges of the corresponding functions, whereas for the images in the third column, each colorbar only represents a portion of the data range. In each of these images, the horizontal axis corresponds to the $y'$-axis on the screen of the observer (with positive direction to the right), and the vertical axis corresponds to the $x'$-axis on the screen (with positive direction downward), where the polar angle $\alpha$ is measured from the $x'$-axis (cf.~Fig.~\ref{fig9}). $\phantom{11111111111111111111111111111111111111111111111111}$ }
	\label{fig15}
\end{figure}

Luminosity variations in the bright regions between different CA models are also an important research topic. Compared to the case in the SS accretion models, the comparison of luminosity between different CA models is straightforward. Based on the expressions of the observed integrated intensities in the static, infalling, and rotating CA models, namely $F_{\text{os}}$, $F_{\text{of}}$, and $F_{\text{or}}$, we just need to write down
\begin{eqnarray}
\label{equ3.71}&&F_{\text{of}}-F_{\text{os}}=\sum_{k=1}^{3}Rc_{\text{fs}}(b,\alpha,\theta_{0},k)g_{\text{s}}^4I_{\text{e}}(r_{\text{e}})\bigg|_{r_{\text{e}}=h\left(\iota_{k}(\alpha,\theta_{0}),b\right)},\displaystyle\quad\text{for}\quad b^{\text{CA}}_{\text{min}}(\alpha,\theta_{0},k)\leqslant b\leqslant b^{\text{CA}}_{\text{max}}(\alpha,\theta_{0},k),\qquad\quad\\
\label{equ3.72}&&F_{\text{or}}-F_{\text{os}}=\sum_{k=1}^{3}Rc_{\text{rs}}(b,\alpha,\theta_{0},k)g_{\text{s}}^4I_{\text{e}}(r_{\text{e}})\bigg|_{r_{\text{e}}=h\left(\iota_{k}(\alpha,\theta_{0}),b\right)},\displaystyle\quad \text{for}\quad  b^{\text{CA}}_{\text{min}}(\alpha,\theta_{0},k)\leqslant b \leqslant b^{\text{CA}}_{\text{max}}(\alpha,\theta_{0},k),\qquad\quad
\end{eqnarray}
where the two redshift comparison functions $Rc_{\text{fs}}(b,\alpha,\theta_{0},k)$ and $Rc_{\text{rs}}(b,\alpha,\theta_{0},k)$ are defined by
\begin{eqnarray}
\label{equ3.73}Rc_{\text{fs}}(b,\alpha,\theta_{0},k)&=&\frac{g_{\text{f}}^4}{g_{\text{s}}^4}-1\bigg|_{r_{\text{e}}=h\left(\iota_{k}(\alpha,\theta_{0}),b\right)},\qquad  Rc_{\text{rs}}(b,\alpha,\theta_{0},k)=\frac{g_{\text{r}}^4}{g_{\text{s}}^4}-1\bigg|_{r_{\text{e}}=h\left(\iota_{k}(\alpha,\theta_{0}),b\right)}.
\end{eqnarray}
From Eqs.~(\ref{equ3.45})--(\ref{equ3.48}) and (\ref{equ3.62})--(\ref{equ3.65}), the expressions of the above two functions are
\begin{eqnarray}
\label{equ3.74}Rc_{\text{fs}}(b,\alpha,\theta_{0},k)&=&\left\{
\begin{array}{ll}
\displaystyle\left(\frac{\displaystyle\sqrt{1-\frac{2m}{r_{\text{e}}}}}{\displaystyle\sqrt{1-\frac{2m}{r^{\text{CA}}_{\text{ini}}}}+\sqrt{1-\frac{b^2}{r_{\text{e}}^2}\left(1-\frac{2m}{r_{\text{e}}}\right)}\sqrt{\frac{2m}{r_{\text{e}}}-\frac{2m}{r^{\text{CA}}_{\text{ini}}}}}\right)^{4}-1,&\displaystyle\ \quad \text{for}\quad b\leqslant b_{\text{hal}}(\alpha,\theta_{0},k),\medskip\\
\displaystyle\left(\frac{\displaystyle\sqrt{1-\frac{2m}{r_{\text{e}}}}}{\displaystyle\sqrt{1-\frac{2m}{r^{\text{CA}}_{\text{ini}}}}-\sqrt{1-\frac{b^2}{r_{\text{e}}^2}\left(1-\frac{2m}{r_{\text{e}}}\right)}\sqrt{\frac{2m}{r_{\text{e}}}-\frac{2m}{r^{\text{CA}}_{\text{ini}}}}}\right)^{4}-1,&\displaystyle\ \quad \text{for}\quad b>b_{\text{hal}}(\alpha,\theta_{0},k),
\end{array}\right.
\end{eqnarray}
\begin{eqnarray}
\label{equ3.75}Rc_{\text{rs}}(b,\alpha,\theta_{0},k)&=&\left(\frac{\displaystyle\sqrt{1-\frac{3m}{r_{\text{e}}}}}{\displaystyle\sqrt{1-\frac{2m}{r_{\text{e}}}}\left(1+\displaystyle\sqrt{\frac{m}{r_{\text{e}}^{3}}}\,b\sin{\alpha}\sin{\theta_{0}}\right)}\right)^{4}-1,
\end{eqnarray}
where $r_{\text{e}}=h\left(\iota_{k}(\alpha,\theta_{0}),b\right)$. Obviously, Eqs.~(\ref{equ3.71})--(\ref{equ3.75}) imply that the redshift comparison functions can be directly used to compare the observed integrated intensities between different CA models, and their expressions can be directly calculated in the bright regions of the BH images without depending on the boundaries of the CA,
Thus, compared to the case in the SS accretion models, the comparison of luminosity between different CA models is indeed straightforward. The images of the two redshift comparison functions in the first and second order BH images for $\theta_{0}=\pi/3$ and $\theta_{0}=4\pi/9$ are presented in Figs.~\ref{fig15} and~\ref{fig16}, respectively, and in view that their expressions do not depend on the details of the emission profiles and the boundaries of the CA, the range of $r_{\text{e}}$ in Eq.~(\ref{equ3.74}) is extended to $2m\leqslant r_{\text{e}}\leqslant10m$, and that in Eq.~(\ref{equ3.75}) is extended to $3m\leqslant r_{\text{e}}\leqslant10m$. For the images of functions $Rc_{\text{fs}}(b,\alpha,\theta_{0},1)$ and $Rc_{\text{fs}}(b,\alpha,\theta_{0},2)$, the colorbars show the value ranges of the corresponding functions, whereas for the images of functions $Rc_{\text{rs}}(b,\alpha,\theta_{0},1)$ and $Rc_{\text{rs}}(b,\alpha,\theta_{0},2)$, in order to ensure a uniform distribution of the data on the colorbars, each colorbar only represents a portion of the data range.

The images of the redshift comparison functions in Figs.~\ref{fig15} and~\ref{fig16} illuminate the luminosity variations in the bright regions between different CA models, and in what follows, we will present a detailed analysis and summary.
\begin{enumerate}
\item Luminosity variations throughout the entire region $-\pi/2<\alpha<\pi/2$ for the first order BH images between the static and infalling CA models: From Eqs.~(\ref{equ3.71})--(\ref{equ3.73}), the images of $Rc_{\text{fs}}(b,\alpha,\theta_{0},1)$ in Figs.~\ref{fig15} and~\ref{fig16} indicate that within the entire region $-\pi/2<\alpha<\pi/2$ of the first order BH images, the luminosity in the infalling model is always lower than that in the static model, and it will be reduced as the initial radial position of accreting matters increases. These luminosity variations are clearly visible and confirmed by the relevant images in Figs.~\ref{fig12}--\ref{fig14}. For a lightlike geodesic intersecting the screen of the observer, when the impact parameter $b$ is larger than $b_{\text{cri}}$, as discussed previously below Eq.~(\ref{equ3.59}),  the azimuthal angle coordinate $\phi_{2}$ of the periastron of the geodesic in the plane $P_{\text{e}}OO'$ (cf.~Fig.~\ref{fig9}) is larger than $\pi/2$. For the first order BH image, within the entire region $-\pi/2<\alpha<\pi/2$, the luminosity is contributed by the lightlike geodesics that travel around the BH less than $\pi/2$, which means that newly emitted photons on the geodesic are located on the outward segments, so they are always moving away from the BH. In addition, even when $b\leqslant b_{\text{cri}}$, since the lightlike geodesic only has the outward segment, newly emitted photons on the geodesic are still moving away from the BH. In the infalling model, the fact that the emitters are moving closer to the BH means that the gravitational and Doppler redshift effects of the lightlike geodesics lead to the observed luminosity throughout the entire region $-\pi/2<\alpha<\pi/2$ in the first order BH image being reduced, which can also be validated from Eqs~(\ref{equ3.46}) and~(\ref{equ3.74}). In addition, with the increase of the initial radial position of accreting matters, the radial velocities of the emitters within the CA increase, which could strengthen the Doppler redshift effects, so that the luminosity in this case will be further reduced.
\begin{figure}[tbp]
	\centering
	\begin{subfigure}{0.32\linewidth}
		\centering
		\includegraphics[width=1\linewidth]{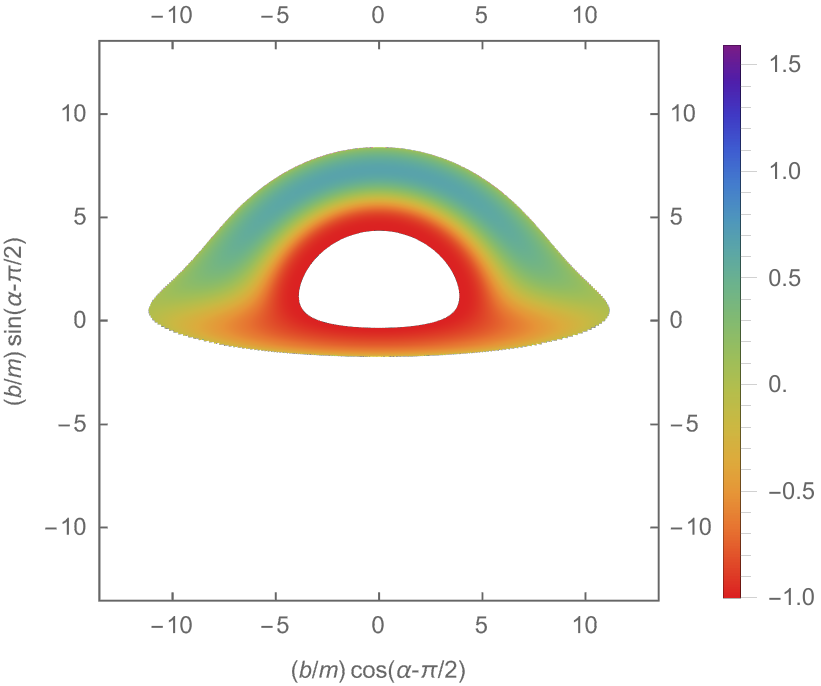}
		\caption{$Rc_{\text{fs}}(b,\alpha,\theta_{0},1)$}
		\label{fig16.1}
	\end{subfigure}
	\centering
	\begin{subfigure}{0.32\linewidth}
		\centering
		\includegraphics[width=1\linewidth]{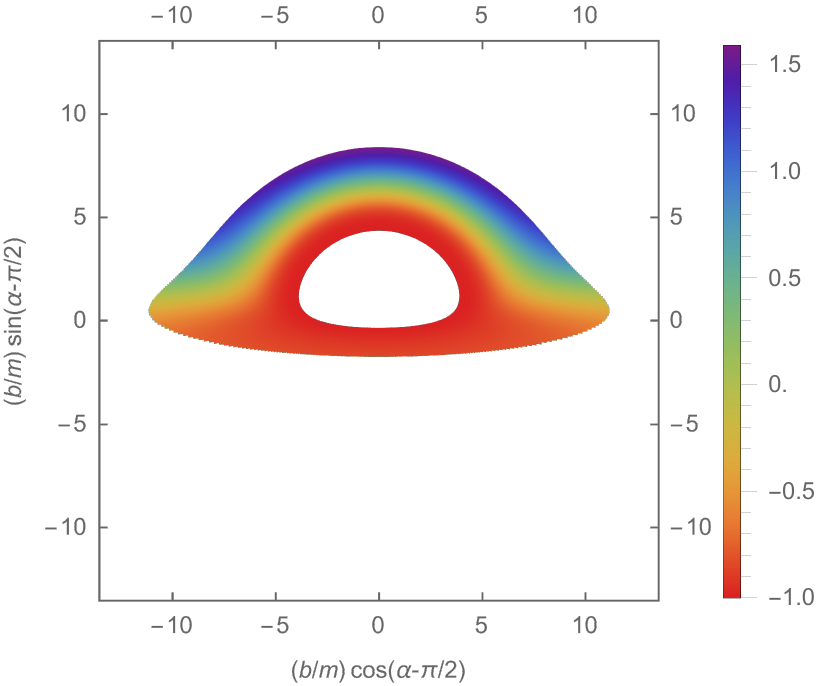}
		\caption{$Rc_{\text{fs}}(b,\alpha,\theta_{0},1)$}
		\label{fig16.2}
	\end{subfigure}
    \centering
	\begin{subfigure}{0.32\linewidth}
		\centering
		\includegraphics[width=1\linewidth]{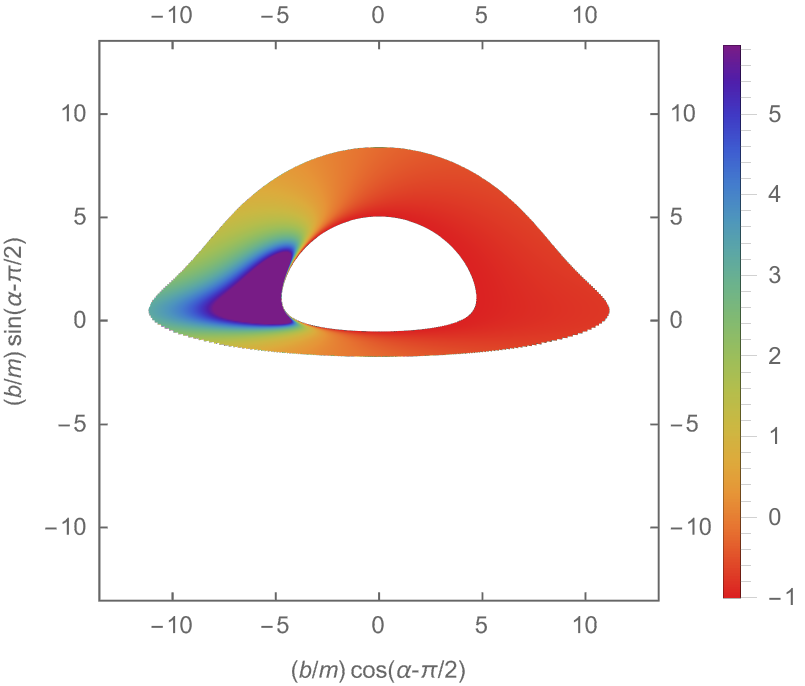}
		\caption{$Rc_{\text{rs}}(b,\alpha,\theta_{0},1)$}
		\label{fig16.3}
	\end{subfigure}
    \centering
	\begin{subfigure}{0.32\linewidth}
		\centering
		\includegraphics[width=1\linewidth]{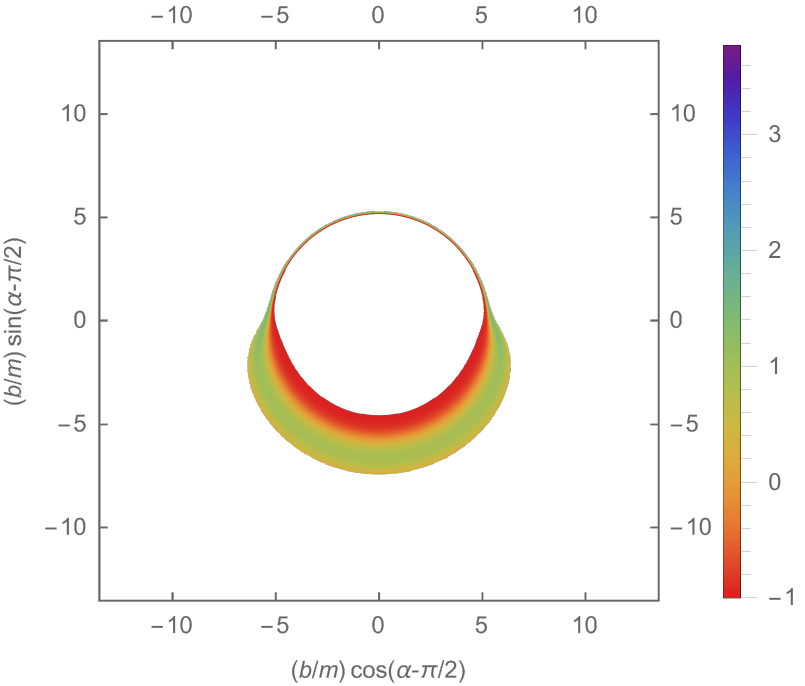}
		\caption{$Rc_{\text{fs}}(b,\alpha,\theta_{0},2)$}
		\label{fig16.4}
	\end{subfigure}
\centering
	\begin{subfigure}{0.32\linewidth}
		\centering
		\includegraphics[width=1\linewidth]{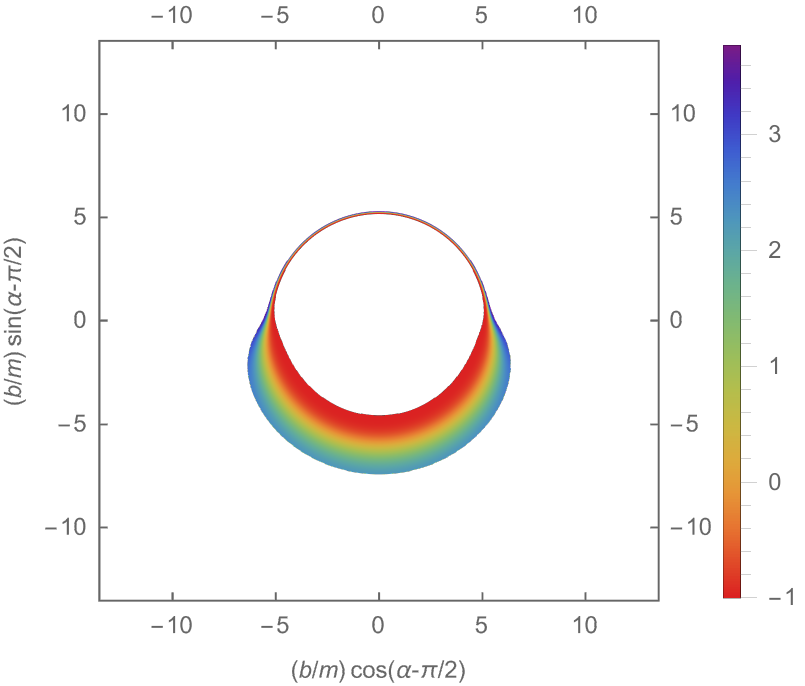}
		\caption{$Rc_{\text{fs}}(b,\alpha,\theta_{0},2)$}
		\label{fig16.5}
	\end{subfigure}
	\centering
	\begin{subfigure}{0.32\linewidth}
		\centering
		\includegraphics[width=1\linewidth]{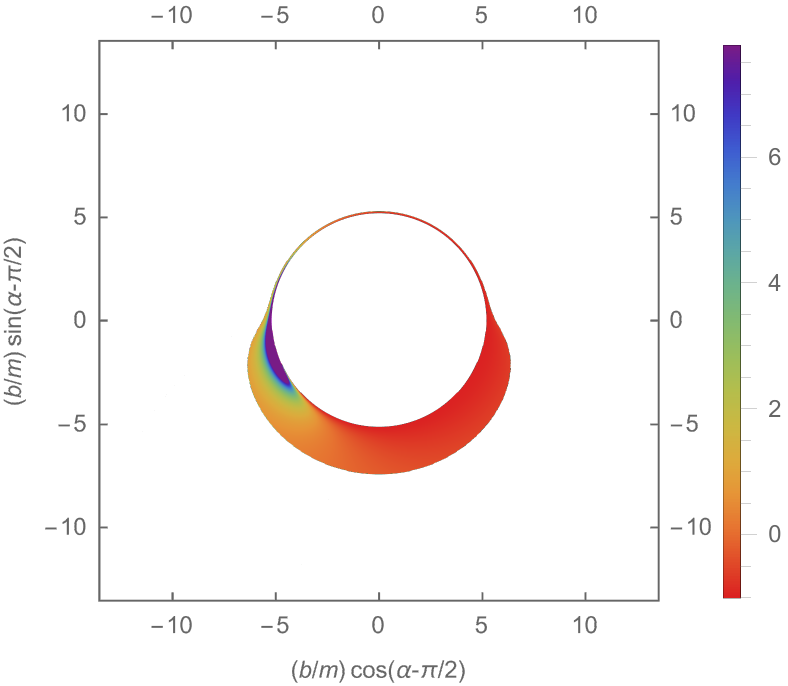}
		\caption{$Rc_{\text{rs}}(b,\alpha,\theta_{0},2)$}
		\label{fig16.6}
	\end{subfigure}
	\caption{Behaviors of the redshift comparison functions for $\theta_{0}=4\pi/9$ in the first (top row) and second (bottom row) order BH images, where the range of $r_{\text{e}}$ appearing in the expressions of $Rc_{\text{fs}}(b,\alpha,\theta_{0},1)$ and $Rc_{\text{fs}}(b,\alpha,\theta_{0},2)$ is extended to $2m\leqslant r_{\text{e}}\leqslant10m$, and  the range of $r_{\text{e}}$ appearing in the expressions of $Rc_{\text{rs}}(b,\alpha,\theta_{0},1)$ and $Rc_{\text{rs}}(b,\alpha,\theta_{0},2)$ is extended to $3m\leqslant r_{\text{e}}\leqslant10m$.  The images in the first column are plotted for $r^{\text{CA}}_{\text{ini}}=10m$ and the images in the second column are plotted for $r^{\text{CA}}_{\text{ini}}=+\infty$. In each of these images, the horizontal axis corresponds to the $y'$-axis on the screen of the observer (with positive direction to the right), and the vertical axis corresponds to the $x'$-axis on the screen (with positive direction downward), where the polar angle $\alpha$ is measured from the $x'$-axis (cf.~Fig.~\ref{fig9}). $\phantom{111111111111111111}$ }
	\label{fig16}
\end{figure}

\item Luminosity variations near the outer edge of the bright region for BH images of arbitrary order between the static and infalling CA models (Excluding the entire region $-\pi/2<\alpha<\pi/2$ for the first order BH image): From the images of $Rc_{\text{fs}}(b,\alpha,\theta_{0},1)$ and $Rc_{\text{fs}}(b,\alpha,\theta_{0},2)$ in Figs.~\ref{fig15} and~\ref{fig16}, if the outer boundary of the CA is sufficiently far from the bound photon orbit, the luminosities near the outer edge of the bright region for the first and second order BH images in the infalling model are always higher than those in the static model, and they will be enhanced as the initial radial position of accreting matters increases. These luminosity variations can be observed from the relevant images in Figs.~\ref{fig12}--\ref{fig14} and apply to BH images of arbitrary order.  For a lightlike geodesic intersecting the observational screen with $b>b_{\text{cri}}$, by following the previous discussion below Eq.~(\ref{equ3.59}),  the azimuthal angle coordinate $\phi_{2}$ of the periastron of the geodesic in the plane $P_{\text{e}}OO'$ will decrease as the impact factor $b$ increases. If the outer boundary of the CA is sufficiently large, when the impact parameter $b$ of a lightlike geodesic approaches the outer edge of the bright region, the azimuthal angle coordinate $\phi_{2}$ of the periastron will be much smaller than the azimuthal angle coordinate $\phi_{\text{e}}$ of the emitting point. As a result, newly emitted photons from the CA are on the inward segment of the lightlike geodesic and possess sufficiently large inward radial velocities. In this case, for the considered lightlike geodesics, given that the emitters in the infalling model are also moving inward, the Doppler blueshift effects could be stronger than the gravitational redshift effects, which leads to the observed luminosity near the outer edge of the bright region being enhanced. Furthermore, when the initial radial position of accreting matters increases, an increase in the radial velocities of the emitters means that the Doppler blushift effects are strengthened, and thus the luminosity will be further enhanced.

\item Luminosity variations near the inner edge of the bright region for BH images of arbitrary order between the static and infalling CA models (Excluding the entire region $-\pi/2<\alpha<\pi/2$ for the first order BH image): The images of $Rc_{\text{fs}}(b,\alpha,\theta_{0},1)$ and $Rc_{\text{fs}}(b,\alpha,\theta_{0},2)$ in Figs.~\ref{fig15} and~\ref{fig16} also show that if the inner boundary of the CA is extended to a region near the bound photon orbit, the luminosities near the inner edge of the bright region for the first and second order BH images in the infalling model are always lower than those in the static model, and they will be reduced as the initial radial position of accreting matters increases. These luminosity variations can be verified from the relevant images in Figs.~\ref{fig12} and \ref{fig13}, and they also apply to BH images of arbitrary order. If the inner boundary of the CA is near the bound photon orbit, for a lightlike geodesic with $b>b_{\text{cri}}$, the azimuthal angle coordinate $\phi_{2}$ of the periastron will be larger than the azimuthal angle coordinate $\phi_{\text{e}}$ of the emitting point while $b$ approaches the inner edge of the bright region, which implies that the emitting point of the geodesic is on the outward segment. In addition, as $b$ further moves toward the inner edge of the bright region so that it is less than $b_{\text{cri}}$, the emitting point of the geodesic is still on the outward segment. Consequently, in this case,  newly emitted photons from the CA will move away from the BH. The fact that the emitters in the infalling model are moving inward means that the gravitational and Doppler redshift effects of the considered lightlike geodesics lead to the observed luminosity near the inner edge of the bright region being reduced. Furthermore, when the initial radial position of accreting matters increases, the Doppler redshift effects will be strengthened so that the luminosity will be reduced accordingly.

 \item Luminosity variations for BH images of arbitrary order between the static and rotating CA models: From the images of $Rc_{\text{rs}}(b,\alpha,\theta_{0},1)$ and $Rc_{\text{rs}}(b,\alpha,\theta_{0},2)$ in Figs.~\ref{fig15} and~\ref{fig16}, it is displayed that for the first and second order BH images, in a region near $\alpha=\pi/2$, the observed luminosities in the rotating model are lower than those in the static model, and as the angle $\alpha$ tends to $\pi/2$, they will gradually decrease. Aa a contrast, in a region near $\alpha=3\pi/2$, the observed luminosities in the rotating model are higher than those in the static model, and as the angle $\alpha$ tends to $3\pi/2$,  they will gradually increase. One could easily recognize these luminosity variations from the relevant images in Figs.~\ref{fig12}--\ref{fig14}, and from the expression of $Rc_{\text{rs}}(b,\alpha,\theta_{0},k)$ in Eq.~(\ref{equ3.75}), it can be concluded that these luminosity variations are also applicable to BH images of arbitrary order. Since the CA rotates counterclockwise (viewed from the positive $z$-axis), the emitters on the positive side of $y$-axis possess velocity components pointing in the negative $x$-axis direction, whereas the emitters on the negative side of $y$-axis possess velocity components pointing in the positive $x$-axis direction (cf.~Fig.~\ref{fig9}). For a $k$th-order lightlike geodesic intersecting the observational screen in a region near $\alpha=\pi/2$,  one could find that the direction of the $x$-axis velocity components of the newly emitted photons is always opposite to that of the emitter. In this region, the gravitational and Doppler redshift effects of the considered lightlike geodesics lead to the observed luminosity in the rotating model being reduced, and as the angle $\alpha$ approaches $\pi/2$, the $x$-axis velocity components of the emitters will increase, which strengthens the Doppler redshift effects, so that the luminosity gradually decrease. On the contrary, for a $k$th-order lightlike geodesic intersecting the observational screen in a region near $\alpha=3\pi/2$, the direction of the $x$-axis velocity components of the newly emitted photons is the same as that of the emitter. Under this case, for the considered lightlike geodesics, since the $x$-axis velocity components of the emitters are large enough, the Doppler blueshift effects could be stronger than the gravitational redshift effects. Thus in this small region, the observed luminosity in the rotating model is enhanced, and as the angle $\alpha$ approaches $3\pi/2$, with the increase of the $x$-axis velocity components of the emitters, the Doppler blushift effects will also be strengthened, so the luminosity will gradually increase.
\end{enumerate}

The above summarized features of the BH images in the CA accretion models are generalizations of those in the
disk accretion models, and they strongly depend on the boundary positions of the CA. Firstly, the boundary
positions of the CA play crucial roles in defining the geometric features of the BH images.  For example, for a given order BH image, they determine the size and shape of the bright region by Eqs.~(\ref{equ3.20}), and in particular, for a complete BH image, the inner boundary position of the CA determines the internal structure of the shadow. Secondly, the luminosity within a specific region of the BH image in a CA model heavily relies on the nature of the lightlike geodesics and the motions of the emitters. To be specific, the Doppler shift effect of a lightlike geodesic depends on the relative motion between the newly emitted photons and the emitter, and  the luminosity in region where some lightlike geodesics intersect the observational screen is influenced by the comparison of the gravitational and Doppler shift effects. In the infalling or the rotating CA models, for the lightlike geodesics received by the observer, if the velocities of the emitters and new emitted photons are large enough, and they move in roughly the same direction, the Doppler blueshift effects  could be stronger than the gravitational redshift effects, which will enhance the luminosity of the region contributed by these lightlike geodesics. On the contrary, if the emitters and the newly emitted photons tend to move in roughly opposite direction, the gravitational and Doppler redshift effects will reduce the luminosity of the region contributed by these lightlike geodesics. The preceding summaries in this section confirm the above conclusions.

As discussed above, in the infalling and rotating CA models, the integrated intensity of a lightlike geodesic received by the observer is significantly influenced jointly by gravitational and Doppler shift effects. In the BH images for the static model, the contribution of the gravitational redshift effect has been considered. Thus, for the infalling and rotating models, we could mainly focus on the Doppler shift effect. In these two models, the luminosity in region where some lightlike geodesics intersect the observational screen can be enhanced only when the Doppler blueshift effects of these geodesics are strong enough, and otherwise, the luminosity of this region will be reduced. In the above summary, such typical regions have been pointed out in the BH images, and the luminosity variations within these regions between different CA models are analyzed. As to other regions in the BH images, the luminosity variations between different CA models depend on the combined effects of the gravitational and Doppler shifts along the corresponding lightlike geodesics.  Finally, it should be emphasized that Eqs. (\ref{equ3.57}), (\ref{equ3.58}), and  (\ref{equ3.61}) constitute the core ingredients for generating and analyzing the Schwarzschild BH images in the static, infalling, and rotating CA models of optically and geometrically thin accretion when the observer views the CA at an inclination angle, and the BH images in Figs.~\ref{fig12}--\ref{fig14} constitute an application example of these equations. For general emission pattern, one only needs to insert the expression of the emitted specific intensity into these equations so as to evaluate the observed integrated intensities in the static, infalling, and rotating CA models of optically and geometrically thin accretion on a Schwarzschild BH, and then the BH images can be plotted.
\section{Summary and discussions~\label{Sec:fourth}}
As the first purpose of this paper, the Schwarzschild BH images in the static and infalling SS models of optically thin accretion are studied by adopting the backward ray tracing method~\cite{Jaroszynski:1997bw}, where in these two models, accreting matters are located within a SS around the BH. In the SS accretion models, the calculation of the integrated intensity observed by a distant observer
always involves the integral along a lightlike geodesic within the boundaries of the SS, so the integrated intensity will vary
with the change of the boundaries of the SS.  In order to facilitate the applications, for the SS with different boundaries in the static and infalling SS models, the formulas for calculating the observed integrated intensities are presented, and by use of them, the Schwarzschild BH images for any emission pattern in these two models can be generated.

To study the properties of the BH images in the static and infalling SS models of optically thin accretion, the BH images for the monochromatic emission pattern with a $1/r^{2}$ radial profile are generated, and based on a comprehensive analysis,
the geometric and luminosity features of these images are summarized in detail. It is shown that these features strongly depend on the boundary positions of the SS. Firstly, the boundary positions of the SS play crucial roles in defining the geometric features of the BH images. Specifically, the outer boundary of the SS could determine the outer edge of the bright region in a BH image if it lies outside the bound photon orbit, and the inner boundary of the SS could determine the radial position of the luminosity peak if it lies outside the bound photon orbit. Secondly, the boundary positions of the SS significantly influence the luminosity variations between the static and infalling models. In general, the relative luminosity difference between these two models depends on the combined effects of the gravitational and Doppler shifts along the lightlike geodesics intersecting the screen of the observer.

The focus is first placed on the luminosity variation of the shadow between the static and infalling SS models. The shadow of a BH image in the SS models is contributed by the lightlike geodesics with $b<b_{\text{cri}}$, and they only have the outward segments. So in the infalling model, the fact that the emitters are always moving toward the BH means that the gravitational and Doppler redshift effects of the lightlike geodesics lead to the luminosity of the shadow being reduced, and the luminosity will be further reduced with the increase of the initial radial position of accreting matters. In view that the peak luminosity of a BH image is very important in observations, we also summarize the luminosity variations near the peak between the static and infalling SS models. The entire bright region outside the shadow of a BH image in the SS models is contributed by the lightlike geodesics with $b>b_{\text{cri}}$, and they have both inward and outward segments. When the SS is thin enough, for a lightlike geodesic with $b\approx b_{\text{pea}}$, the emitting point is very close to the periastron because the periastron is near the boundary of the SS (cf.~the discussion in page 14), which means that newly emitted photons from the SS have low radial velocities. This fact indicates that in the small region near the peak, the Doppler blueshift effects contributed by the inward segments of the geodesics with $b\approx b_{\text{pea}}$ are very weak, so that like the case of shadow, the gravitational and Doppler redshift effects of the lightlike geodesics lead to the observed luminosity near the peak in the infalling model being reduced, and the luminosity will be further reduced as the initial radial position of accreting matters increases. As a contrast, when the SS is thick enough, for a lightlike geodesic with $b\approx b_{\text{pea}}$, the Doppler shift effects contributed by the outward and inward segments approximately cancel each other out because the numerical calculations indicate that in this case, the observed luminosity near the peak in the infalling model shows negligible variation compared to that in the static model, and the luminosity is not sensitive to the initial radial position of accreting matters.

The above discussions suggest that under most situations, the observed luminosities of the BH images in the infalling SS model seem to be reduced. However, from our images, it is displayed that in the infalling SS model, the luminosity enhancement phenomenon does also exist as well. When the inner boundary of the SS is far from the bound photon orbit, for a lightlike geodesic with $b\gtrsim b_{\text{cri}}$, the emitting point is very far from the periastron because the periastron is situated at a great distance from the SS, which means that newly emitted photons from the SS have sufficiently large radial velocities. Thus, in the infalling SS model, the Doppler blueshift effects contributed by the inward segments of the lightlike geodesics with $b\gtrsim b_{\text{cri}}$ could become strong enough. As a result, the observed luminosity near the exterior of the shadow in the infalling SS model is enhanced, and it will be further enhanced with the increase of the initial radial position of accreting matters. Given that the unusual luminosity enhancement phenomenon is rare, it could be viewed as a notable luminosity feature of the BH images in the infalling SS model. Finally, it is important to stress that the images obtained on the basis of the monochromatic emission pattern in the SS accretion models only serve as an application example
of the formulas for calculating the observed integrated intensities. For more realistic emission patterns, by applying these formulas, one could generate more realistic BH images.

As the second purpose of this paper, the Schwarzschild BH images in the static, infalling, and rotating CA models of optically and geometrically thin accretion are studied by means of the equations of lightlike geodesics, where in these three models, accreting matters are located within a CA centered at the center of the BH. When a lightlike geodesic emitted from the CA is received by a distant observer viewing the CA at an inclination angle, how to determine the transfer function establishing the connection between the emitting and receiving points is important. With the equations of lightlike geodesics presented in the Appendix, the analytical forms of all order
transfer functions working for all impact parameter values are provided. It should be noted that before our results, the transfer functions for $b<b_{\text{cri}}$ are purely numerical, and our results present their complete analytical forms. In order to derive the integrated intensities observed by the distant observer in the static, infalling, and rotating CA models, the redshift factors in the three situations need to be evaluated. From the relevant results in Ref.~\cite{Gralla:2019xty,Luminet:1979nyg,Liu:2021lvk,Tian:2019yhn}, we only need to handle the redshift factor in the infalling model. In this paper, with the aid of the equations of lightlike geodesics, the redshift factor in the infalling model is given in the semianalytical form. With the transfer functions and redshift factors, after considering the contributions from the emissions at all orders, the formulas for deriving the observed integrated intensities in the static, infalling, and rotating CA models are achieved, and on the basis of these formulas, once emitted specific intensity is given, the observed integrated intensities in the three CA models can be evaluated.

To study the properties of the BH images in the static, infalling, and rotating CA models, for each emission pattern in Ref.~\cite{Gralla:2019xty},  the corresponding BH images are generated, where for these images, only contributions from the first three order emissions are considered because as stated in Ref.~\cite{Gralla:2019xty}, the contributions from the fourth and higher order emissions could be totally negligible. Based on a comprehensive analysis on these images,  the geometric and luminosity features for the BH image of arbitrary order in the three CA models are summarized.  These images display that the boundary positions of the CA also play crucial roles in defining the geometric features of the BH images. For example, for a given order BH image, they determine the size and shape of the bright region, and in particular, for a complete BH image, the inner boundary position of the CA determines the internal structure of the shadow. In general, the shadow of a BH image in the CA models consists of several distinct dark parts with zero luminosities, which is essentially different from the case in the SS accretion models because the luminosity of the shadow in a BH image for the SS models is usually nonzero.

The positions of the boundaries of the CA can also influence the luminosity variations of the BH images between different CA accretion models. Because the redshift comparison functions can be directly used to compare the observed integrated intensities between different CA models, and their expressions can be directly calculated in the bright regions of the BH images without depending on the boundaries of the CA, the comparison of the luminosity between different CA models is more straightforward compared to the case in the SS accretion models. If the outer boundary of the CA is sufficiently far from the bound photon orbit, for a lightlike geodesic with $b$ approaching the outer edge of the bright region, the azimuthal angle coordinate of the periastron in the plane $P_{\text{e}}OO'$ (cf.~Fig~\ref{fig9}) will be much smaller than that of the emitting point (cf.~the discussion in page 31), which means that newly emitted photons from the CA are on the inward segment of the lightlike geodesic and possess sufficiently large inward radial velocities.  In this case, given that the emitters in the infalling model are also moving inward, the Doppler blueshift effects of the considered lightlike geodesics could be stronger than the gravitational redshift effects, so the observed luminosity near the outer edge of the bright region is enhanced, and it will be further enhanced with the increase of the initial radial position of accreting matters. On the contrary, if the inner boundary of the CA is extended to a region near the bound photon orbit, for a lightlike geodesic with $b\gtrsim b_{\text{cri}}$, the azimuthal angle coordinate of the periastron in the plane $P_{\text{e}}OO'$ will be larger than that of the emitting point, which implies that the emitting point of the geodesic is on the outward segment. In addition, even when $b\lesssim b_{\text{cri}}$,  the emitting point of the geodesic is still on the outward segment. Thus, for a lightlike geodesic with $b\approx b_{\text{cri}}$, newly emitted photons from the CA will move away from the BH.  The fact that the emitters in the infalling model are moving inward means that the gravitational and Doppler redshift effects of the lightlike geodesics could lead to the observed luminosity near the inner edge of the bright region being reduced, and the luminosity will be further reduced as  the initial radial position of accreting matters increases.

It should be emphasized that the above two points do not apply to the entire region $-\pi/2<\alpha<\pi/2$ for the first order BH image. For a lightlike geodesic intersecting the screen of the observer with $b> b_{\text{cri}}$,  the azimuthal angle coordinate of the periastron in the plane $P_{\text{e}}OO'$ is always larger than $\pi/2$ (cf.~the discussion in page 30). Within the above region in the first order BH image, in view of the fact that the luminosity is contributed by the lightlike geodesics that travel around the BH less than $\pi/2$, newly emitted photons on the geodesics are located on the outward segments, and thus, they are always moving away from the BH. In addition, for a lightlike geodesic with $b<b_{\text{cri}}$, since it only has the outward segment, newly emitted photons are still moving away from the BH. In the infalling model, the emitters are moving closer to the BH which means that the gravitational and Doppler redshift effects of the considered lightlike geodesics lead to the observed luminosity throughout the above region in the first order BH image being always reduced, and the luminosity will be further reduced with the increase of the initial radial position of accreting matters. Between the static and rotating CA models, the luminosity variations of the BH images are obvious. For any order lightlike geodesics intersecting the observational screen, in a region near $\alpha=\pi/2$, one will find that the emitters and new emitted photons move in roughly opposite directions, so in this region, the gravitational and Doppler redshift effects lead to the observed luminosity in the rotating model being reduced, and as the angle $\alpha$ approaches $\pi/2$, the luminosity will gradually decrease because the motion directions of the emitters and new emitted photons gradually become increasingly opposite. Similar reasoning can also be used to explain the luminosity variation in a region near $\alpha=3\pi/2$. In this case, for any order lightlike geodesics intersecting the observational screen, the emitters and new emitted photons move in roughly the same directions, and thus, since the emitters have sufficiently large velocity components along the direction of photon propagation,  the Doppler blueshift effects could be stronger than the gravitational redshift effects. So, in this small region, the observed luminosity in the rotating model is enhanced, and as the angle $\alpha$ approaches $3\pi/2$, the luminosity will gradually increase because the motion directions of the emitters and new emitted photons gradually become increasingly aligned.

The BH images for the SS and CA accretion models in this paper are actually the generalizations of those for the spherical and disk accretion models in Refs.~\cite{Narayan:2019imo,Gralla:2019xty}, and they constitute application examples of the formulas for calculating the integrated intensities observed by a distant observer. Our images generalize those in Refs.~\cite{Narayan:2019imo,Gralla:2019xty}, and it is illuminated that the boundary positions of the SS and CA can influence the geometric and luminosity features of the BH images in the two types of models. In fact, as discussed above, the boundary positions of the SS and CA only have a limited effect on the geometry of the BH images, so in this paper, we emphasize the analysis of the luminosity variations between different SS or CA accretion models. In both the SS and CA models, if the emitters are moving, for the lightlike geodesics emitted from them, the Doppler shift effects can significantly influence the integrated intensities received by the observer. When
the velocities of the emitters and new emitted photons are large enough, and they move in roughly the same direction, the Doppler
blueshift effects could be stronger than the gravitational redshift effects, so that the luminosity of the region contributed by these lightlike geodesics can be enhanced compared to that in the static model. Such typical regions have been pointed out in our paper, and based on  these uncommon luminosity enhancement phenomena in the specific SS or CA model, people could effectively assess the motion states of the accretion flows. In other case, when the gravitational and Doppler redshift effects of the considered lightlike geodesics play a dominant role,  the luminosity of the region contributed by these geodesics will be reduced. Beyond the studies on luminosity variations between different SS or CA accretion models, this paper also provides semianalytical frameworks for calculating the observed integrated intensities in the SS and CA accretion models. In spite of being derived from the spherical and disk model frameworks in Refs.~\cite{Narayan:2019imo,Gralla:2019xty}, the semianalytical frameworks are constructed with distinct features tailored to the SS and CA accretion models. Based on the semianalytical frameworks, when the accreting matters undergo more complex motions, one only needs to derive the corresponding expression for the redshift factor so that the BH images can be generated.
\begin{acknowledgments}
This work was supported by the National Natural Science Foundation of China (Grants Nos.~12105039, 12133003, 12494574, and 12326602). This work was also supported by the Guangxi Talent Program (``Highland of Innovation Talent'').
\end{acknowledgments}
\appendix\label{appendix}
\section{The equations of the lightlike geodesics in Schwarzschild spacetime~\label{Sec:appfirst}}
Although the numerical approaches are able to provide the universal access to the BH images in various gravitational models, it is indicated that when analytical approaches are applied to problems that arise in the visualization of BH, the involved calculations are more accurate and considerably faster than commonly used numerical integrations~\cite{Cadez:2004cg}. In view that the derivations of the analytical formulas for calculating the observed integrated intensities under SS and CA accretion models require specialized knowledge on the equations of the lightlike geodesics in Schwarzschild spacetime, in this appendix, we plan to briefly review these equations, and rewrite them in a form suitable for practical application.

\subsection{The original lightlike geodesic equations}

The metric for Schwarzschild spacetime in spherical polar coordinates  $(ct,r,\theta,\varphi)$ is
\begin{equation}\label{equA1}
g_{\mu\nu}=\left(\begin{array}{cccc}
\displaystyle-\left(1-\frac{2m}{r}\right)&\displaystyle0&\displaystyle0&\displaystyle 0\\
\displaystyle0&\displaystyle\frac{\displaystyle 1}{\displaystyle 1-\frac{2m}{r}}\quad &\displaystyle0&\displaystyle 0\\
\displaystyle0&\displaystyle0&\displaystyle r^2&\displaystyle 0\\
\displaystyle0&\displaystyle0&\displaystyle0&\displaystyle\quad r^2\sin^{2}{\theta}
\end{array}\right)
\end{equation}
and four Killing vector fields in this spacetime are~\cite{Huang2023}
\begin{eqnarray}
\label{equA2}\varepsilon_{0}&=&\frac{\partial}{\partial t},\\
\label{equA3}\varepsilon_{1}&=&-\sin{\varphi}\frac{\partial}{\partial\theta}-\cot{\theta}\cos{\varphi}\frac{\partial}{\partial\varphi},\\
\label{equA4}\varepsilon_{2}&=&\cos{\varphi}\frac{\partial}{\partial\theta}-\cot{\theta}\sin{\varphi}\frac{\partial}{\partial\varphi},\\
\label{equA5}\varepsilon_{3}&=&\frac{\partial}{\partial\varphi}.
\end{eqnarray}
Consider a lightlike geodesic $x^{\mu}(\lambda)$ with $\lambda$ as an affine parameter, and since the geodesic equation does not fix the parameter $\lambda$, we normalize it so that the four-momentum of the photons is given by $p^{\mu}=\text{d}x^{\mu}/\text{d}\lambda$.
As in references, the entire geodesic path is in a plane,  which implies that we can rotate coordinates to set $\theta(\lambda)=\pi/2$. With this condition in mind, the equations satisfied by the four-momentum of the photons are
\begin{eqnarray}
\label{equA6} 0&=&\displaystyle g_{\alpha\beta}\frac{\text{d}x^{\alpha}}{\text{d}\lambda}\frac{\text{d}x^{\beta}}{\text{d}\lambda}=-\left(1-\frac{2m}{r}\right)c^2\left(\frac{\text{d}t}{\text{d}\lambda}\right)^2+\frac{\displaystyle\left(\frac{\text{d}r}{\text{d}\lambda}\right)^2}{\displaystyle 1-\frac{2m}{r}}+r^2\left(\frac{\displaystyle\text{d}\varphi}{\text{d}\lambda}\right)^2,\\
\label{equA7}-E&=&\displaystyle g_{\alpha\beta}\varepsilon_{0}^{\alpha}p^{\beta}=-c^2\left(1-\frac{2m}{r}\right)\frac{\text{d}t}{\text{d}\lambda},\\
\label{equA8}L&=&\displaystyle  g_{\alpha\beta}\varepsilon_{3}^{\alpha}p^{\beta}=r^2\frac{\text{d}\varphi}{\text{d}\lambda},
\end{eqnarray}
where the first equation is resulted from the fact that $p^{\mu}$ is a lightlike vector, and the last two equations arise from the Killing vector fields $\varepsilon_{0}$ and $\varepsilon_{3}$, respectively, with $E$ and $L$ as the conserved quantities related to the energy and the angular momentum of photons. From these three equations and $\theta(\lambda)=\pi/2$, the components of the vector $p^{\mu}$ are offered,
\begin{equation}\label{equA9}
\left\{\begin{array}{ll}
\displaystyle p^{0}&=\displaystyle \frac{E}{\displaystyle c\left(1-\frac{2m}{r}\right)},\smallskip\\
\displaystyle p^{r}&=\displaystyle \pm\frac{L}{b}\sqrt{1-\frac{b^2}{r^2}\left(1-\frac{2m}{r}\right)},\smallskip\\
\displaystyle p^{\theta}&=0,\smallskip\\
\displaystyle p^{\varphi}&=\displaystyle \frac{L}{r^2}
\end{array}\right.
\end{equation}
with $b:=cL/E$ as the impact parameter of the geodesic, and here, the sign $+(-)$ indicates that photons go away from (approach) the BH.

The differential equation of the lightlike geodesic can directly be read off from Eqs.~(\ref{equA9}),
\begin{equation}\label{equA10}
\frac{\text{d}r}{\text{d}\varphi}=V(r)=\pm\frac{r^2}{b}\sqrt{1-\frac{b^2}{r^2}\left(1-\frac{2m}{r}\right)}
\end{equation}
with $V(r)$ as the effective potential. By setting
\begin{equation}\label{equA11}
V(r)=0,\quad \frac{\text{d}V(r)}{\text{d}r}=0,
\end{equation}
the equation of the bound photon orbit and the corresponding critical impact parameter are given,
\begin{eqnarray}
\label{equA12}&&r_{\text{pho}}=3m,\\
\label{equA13}&&b_{\text{cri}}=3\sqrt{3}m.
\end{eqnarray}
The further discussions on the equation of the lightlike geodesic in the general case suggest that Eq.~(\ref{equA10}) ought to be recast as
\begin{equation}\label{equA14}
\frac{\text{d}\varphi}{\text{d}X}=\pm\frac{1}{\sqrt{2H(X)}}\quad \text{with}\quad H(X):=X^{3}-\frac{X^{2}}{2}+\frac{m^{2}}{2b^{2}},
\end{equation}
where $X:=m/r$. Assume that the lightlike geodesic passes through a spatial point $(r_{0},0,\varphi_{0})$, and then, the equation of the geodesic can be given by
\begin{eqnarray}
\label{equA15}\pm(\varphi-\varphi_{0})&=&\int_{m/r_{0}}^{m/r}\frac{1}{\sqrt{2H(\mathcal{X})}}\text{d}\mathcal{X},
\end{eqnarray}
where the positive (negative) sign on the left side indicates that while photons move towards the BH from the point $(r_{0},0,\varphi_{0})$, the lightlike geodesic is counterclockwise (clockwise). The integral on the right side of the above equation is an elliptical integral, and the integral result depends on roots of equation $H(X)=0$. When the impact parameter $b$ ranges from 0 to $+\infty$, the number of real roots of $H(X)=0$ varies.
By performing a detailed analysis, with the aid of the knowledge about elliptic integral~\cite{Abramowitz1965}, the expressions for real roots of $H(X)=0$ and the equation of the lightlike geodesic through the spatial point $(r_{0},0,\varphi_{0})$ are derived. For the sake of simplicity in writing, we immediately present the results below, and their validity can be easily proved.
\begin{enumerate}
\item When $0\leqslant b<b_{\text{cri}}$, the unique real root of $H(X)=0$ is
      \begin{eqnarray}
      \label{equA16}X_{1}&=&\frac{1}{6}-\frac{1}{3}\cosh{\left[\frac{2}{3}\arcsinh{\Bigg(\sqrt{\frac{b_{\text{cri}}^2}{b^2}-1}\Bigg)}\right]},
      \end{eqnarray}
      and the corresponding equation of the geodesic is
      \begin{eqnarray}
      \label{equA17}\pm(\varphi-\varphi_{0})&=&\sqrt{\frac{1}{2Y(X_{1})}}\left[-\F{\bigg(\vartheta_{\text{les}}\Big(\frac{m}{r}\Big),k_{\text{les}}\bigg)}+\F{\bigg(\vartheta_{\text{les}}\Big(\frac{m}{r_{0}}\Big),k_{\text{les}}\bigg)}\right]\qquad
      \end{eqnarray}
      with
      \begin{eqnarray}
      \label{equA18}\vartheta_{\text{les}}(X)&:=&\arccos{\Bigg(\frac{X-X_{1}-Y(X_{1})}{X-X_{1}+Y(X_{1})}\Bigg)},\\
      \label{equA19}k_{\text{les}}&:=&\sqrt{\frac{1}{2}+\frac{1-6X_{1}}{8Y(X_{1})}},
      \end{eqnarray}
      where
      \begin{eqnarray}
      \label{equA20}Y(X_{1})&:=&\sqrt{X_{1}(3X_{1}-1)},
      \end{eqnarray}
      and $\F$ is the incomplete elliptic integral of the first kind.
      Because of
      \begin{eqnarray}
      \label{equA21}\frac{\text{d}\varphi}{\text{d}r}&=&-\frac{1}{m}X^{2}\frac{\text{d}\varphi}{\text{d}X}\neq0\quad \text{for}\quad \frac{r}{m}\in[2,+\infty),
      \end{eqnarray}
      one end of the geodesic is at infinity, and the other is behind the horizon of the BH.
\item When $b>b_{\text{cri}}$, the three real roots of $H(X)=0$ are
     \begin{eqnarray}
     \label{equA22}X_{1}&=&\frac{1}{6}-\frac{1}{3}\cos{\left[\frac{2}{3}\arcsin{\Bigg(\sqrt{1-\frac{b_{\text{cri}}^2}{b^2}}\Bigg)}\right]},\\
     \label{equA23}X_{2}&=&\frac{1}{6}-\frac{1}{3}\cos{\left[\frac{2}{3}\arcsin{\Bigg(\sqrt{1-\frac{b_{\text{cri}}^2}{b^2}}\Bigg)}-\frac{2\pi}{3}\right]},\\
     \label{equA24}X_{3}&=&\frac{1}{6}-\frac{1}{3}\cos{\left[\frac{2}{3}\arcsin{\Bigg(\sqrt{1-\frac{b_{\text{cri}}^2}{b^2}}\Bigg)}-\frac{4\pi}{3}\right]},
     \end{eqnarray}
      and for them, the following inequalities
     \begin{eqnarray}
     \label{equA25}-\frac{1}{6}<X_{1}<0<X_{2}<\frac{1}{3}<X_{3}<\frac{1}{2}
     \end{eqnarray}
     hold. In this case, the form of the lightlike geodesic depends on the range of  $r_{0}$.
     If $r_{0}>r_{\text{pho}}$, both the ends of the lightlike geodesic are at infinity, and the periastron is at the spatial point
     $(r_{2},0,\varphi_{2})$ with
      \begin{eqnarray}
     \label{equA26}r_{2}&:=&m/X_{2}>r_{\text{pho}},\\
     \label{equA27}\varphi_{2}&:=&\varphi_{0}\pm\sqrt{\frac{2}{X_{3}-X_{1}}}\left[\K{(k_{\text{lar}})}-\F{\bigg(\vartheta_{\text{larI}}\Big(\frac{m}{r_{0}}\Big),k_{\text{lar}}\bigg)}\right],
     \end{eqnarray}
     where
     \begin{eqnarray}
     \label{equA28}\vartheta_{\text{larI}}(X)&:=&\arcsin{\left(\sqrt{\frac{X-X_{1}}{X_{2}-X_{1}}}\right)},\\
     \label{equA29}k_{\text{lar}}&:=&\sqrt{\frac{X_{2}-X_{1}}{X_{3}-X_{1}}},
     \end{eqnarray}
     $\K$ is the complete elliptic integral of the first kind. Because $X_{2}=m/r_{2}$ is a root of equation $H(X)=0$, from Eq.~(\ref{equA14}), it can be inferred that the impact parameter of the lightlike geodesic is expressed as
     \begin{eqnarray}
     \label{equA30}b=\frac{r_{2}}{\displaystyle \sqrt{1-\frac{2m}{r_{2}}}}.
     \end{eqnarray}
     The equation of the geodesic is
    \begin{eqnarray}
     \label{equA31}\pm(\varphi-\varphi_{0})&=&\left\{
     \begin{array}{l}
     \displaystyle\sqrt{\frac{2}{X_{3}-X_{1}}}\left[\F{\bigg(\vartheta_{\text{larI}}\Big(\frac{m}{r}\Big),k_{\text{lar}}\bigg)}-\F{\bigg(\vartheta_{\text{larI}}\Big(\frac{m}{r_{0}}\Big),k_{\text{lar}}\bigg)}\right],\\
     \displaystyle \sqrt{\frac{2}{X_{3}-X_{1}}}\left[2\K{(k_{\text{lar}})}-\F{\bigg(\vartheta_{\text{larI}}\Big(\frac{m}{r}\Big),k_{\text{lar}}\bigg)}-\F{\bigg(\vartheta_{\text{larI}}\Big(\frac{m}{r_{0}}\Big),k_{\text{lar}}\bigg)}\right].
     \end{array}\right.
     \end{eqnarray}
     Here, the upper and lower terms on the right side stand for the pre- and post-periastron branches, respectively.
     As a contrast, if $2m\leqslant r_{0}<r_{\text{pho}}$, both the ends of the lightlike geodesic are behind the horizon of the BH, and the apastron is at the spatial point
     $(r_{3},0,\varphi_{3})$ with
      \begin{eqnarray}
     \label{equA32}r_{3}&:=&m/X_{3}<r_{\text{pho}},\\
     \label{equA33}\varphi_{3}&:=&\varphi_{0}\mp\sqrt{\frac{2}{X_{3}-X_{1}}}\F{\bigg(\vartheta_{\text{larII}}\Big(\frac{m}{r_{0}}\Big),k_{\text{lar}}\bigg)},
     \end{eqnarray}
     where
     \begin{eqnarray}
     \label{equA34}\vartheta_{\text{larII}}(X)&:=&\arctan{\left(\sqrt{\frac{X-X_{3}}{X_{3}-X_{2}}}\right)}.
     \end{eqnarray}
     Because $X_{3}=m/r_{3}$ is also a root of equation $H(X)=0$, from Eq.~(\ref{equA14}), the impact parameter of the lightlike geodesic can be expressed as
     \begin{eqnarray}
     \label{equA35}b=\frac{r_{3}}{\displaystyle \sqrt{1-\frac{2m}{r_{3}}}}.
     \end{eqnarray}
     Now, the equation of the geodesic is
     \begin{eqnarray}
     \label{equA36}\pm(\varphi-\varphi_{0})&=&\left\{
     \begin{array}{l}
     \displaystyle\sqrt{\frac{2}{X_{3}-X_{1}}}\left[\F{\bigg(\vartheta_{\text{larII}}\Big(\frac{m}{r}\Big),k_{\text{lar}}\bigg)}-\F{\bigg(\vartheta_{\text{larII}}\Big(\frac{m}{r_{0}}\Big),k_{\text{lar}}\bigg)}\right],\\
     \displaystyle\sqrt{\frac{2}{X_{3}-X_{1}}}\left[-\F{\bigg(\vartheta_{\text{larII}}\Big(\frac{m}{r}\Big),k_{\text{lar}}\bigg)}-\F{\bigg(\vartheta_{\text{larII}}\Big(\frac{m}{r_{0}}\Big),k_{\text{lar}}\bigg)}\right],\qquad
     \end{array}\right.
     \end{eqnarray}
     where the upper and lower terms on the right side stand for the post- and pre-apastron branches, respectively.
\item When $b=b_{\text{cri}}$, the three real roots of $H(X)=0$ under the above case reduce to
     \begin{eqnarray}
     \label{equA37}X_{1}=-\frac{1}{6},\quad X_{2}=X_{3}=\frac{1}{3}.
     \end{eqnarray}
     The two equations~(\ref{equA31}) and (\ref{equA36}) are degenerated into
     \begin{eqnarray}
     \label{equA38}\pm(\varphi-\varphi_{0})&=&2\left(\arctanh{\sqrt{\frac{2m}{r}+\frac{1}{3}}}-\arctanh{\sqrt{\frac{2m}{r_{0}}+\frac{1}{3}}}\right),\\
     \label{equA39}\pm(\varphi-\varphi_{0})&=&2\left(-\arccoth{\sqrt{\frac{2m}{r}+\frac{1}{3}}}+\arccoth{\sqrt{\frac{2m}{r_{0}}+\frac{1}{3}}}\right).\
     \end{eqnarray}
\end{enumerate}

With the above results in hand, for a lightlike geodesic through a spatial point $(r_{0},0,\varphi_{0})$, we can calculate the total change of the azimuthal angle $\varphi$ and express it in terms of the impact parameter $b$.
According to the previous equations~(\ref{equA17}), (\ref{equA31}), and (\ref{equA36}), when $r_{0}>r_{\text{pho}}$ and $2m\leqslant r_{0}<r_{\text{pho}}$, the total changes of the azimuthal angle $\varphi$ are, respectively,
\begin{eqnarray}
\label{equA40}\pm\Delta\varphi&=&\left\{
\begin{array}{ll}
\displaystyle\sqrt{\frac{1}{2Y(X_{1})}}\left[-\F{\left(\vartheta_{\text{les}}(1/2),k_{\text{les}}\right)}+\F{\left(\vartheta_{\text{les}}(0),k_{\text{les}}\right)}\right],&\quad\text{for}\ \ 0\leqslant b<b_{\text{cri}},\qquad\medskip \\
\displaystyle+\infty,&\quad\text{for}\ \ b=b_{\text{cri}},\medskip\\
\displaystyle2\sqrt{\frac{2}{X_{3}-X_{1}}}\left[\K{(k_{\text{lar}})}-\F{\left(\vartheta_{\text{larI}}(0),k_{\text{lar}}\right)}\right],&\quad\text{for}\ \ b>b_{\text{cri}}
\end{array}\right.
\end{eqnarray}
and
\begin{eqnarray}
\label{equA41}\pm\Delta\varphi&=&\left\{
\begin{array}{ll}
\displaystyle\sqrt{\frac{1}{2Y(X_{1})}}\left[-\F{\left(\vartheta_{\text{les}}(1/2),k_{\text{les}}\right)}+\F{\left(\vartheta_{\text{les}}(0),k_{\text{les}}\right)}\right],&\quad\text{for}\ \ 0\leqslant b<b_{\text{cri}},\qquad\medskip \\
\displaystyle+\infty,&\quad\text{for}\ \ b=b_{\text{cri}},\medskip\\
\displaystyle2\sqrt{\frac{2}{X_{3}-X_{1}}}\left[\F{\left(\vartheta_{\text{larII}}(1/2),k_{\text{lar}}\right)}\right],&\quad\text{for}\ \ b>b_{\text{cri}}.
\end{array}\right.
\end{eqnarray}
One could find that whether $r_{0}$ is larger or less than $r_{\text{pho}}$, the total change of the azimuthal angle do not depend on the value of $r_{0}$. In addition, another basic fact that one could recognize is that the total changes of the azimuthal angle go to infinity when $b=b_{\text{cri}}$, and the reason for this is that the lightlike geodesic with $b=b_{\text{cri}}$ could travel around the BH an infinite number of times.  In many application scenarios, it is useful to have simple approximations to the total change of the azimuthal angle near some particular values of the impact parameter $b$. By applying the techniques of series expansion to the right sides of  Eqs.~(\ref{equA40}) and (\ref{equA41}), the corresponding approximations are
\begin{eqnarray}
\label{equA42}\pm\Delta\varphi&\approx&\left\{
\begin{array}{ll}
\displaystyle\frac{b}{2m},&\quad\text{for}\ \ b\rightarrow0,\qquad\medskip \\
\displaystyle \ln{\left(\frac{C^{\mp}_{\text{I}}m}{|b-b_{\text{cri}}|}\right)},&\quad\text{for}\ \ b\rightarrow b_{\text{cri}}^{\mp},\medskip\\
\displaystyle\pi+\frac{4m}{b},&\quad\text{for}\ \ b\rightarrow +\infty
\end{array}\right.
\end{eqnarray}
and
\begin{eqnarray}
\label{equA43}\pm\Delta\varphi&\approx&\left\{
\begin{array}{ll}
\displaystyle\frac{b}{2m},&\quad\text{for}\ \ b\rightarrow0,\qquad\medskip \\
\displaystyle \ln{\left(\frac{C^{\mp}_{\text{II}}m}{|b-b_{\text{cri}}|}\right)},&\quad\text{for}\ \ b\rightarrow b_{\text{cri}}^{\mp},\medskip\\
\displaystyle\frac{8m}{b},&\quad\text{for}\ \ b\rightarrow +\infty
\end{array}\right.
\end{eqnarray}
with
\begin{eqnarray}
\label{equA44}C^{-}_{\text{I}}=C^{-}_{\text{II}}:=648 (26 \sqrt{3}-45),\quad C^{+}_{\text{I}}:=\frac{648 \sqrt{3}}{\left(2+\sqrt{3}\right)^2},\quad C^{+}_{\text{II}}:=0.74229\times\frac{9 \sqrt{3}}{2}.
\end{eqnarray}
One is able to confirm that the results in Eq.~(\ref{equA42}) for $b\rightarrow b_{\text{cri}}^{\mp}$ under the situation of $r_{0}>r_{\text{pho}}$ are identical to those in Ref.~\cite{Gralla:2019xty}. Furthermore, in this situation,  the result for $b\rightarrow +\infty$ shows that under the weak field approximation, the total deflection angle of the geodesic is $\pm\Delta\varphi-\pi=4m/b$, which is just the famous deflection angle formula for a lightlike geodesic passing near a spherical body in GR.

\subsection{Rewriting the original lightlike geodesic equations in a form suitable for practical application}

At present, all the ingredients to generate and analyze the BH images are available, but inspection of Eqs.~(\ref{equA31}) and (\ref{equA36}) reveals that when $b>b_{\text{cri}}$, two branches of the lightlike geodesic through the spatial point $(r_{0},0,\varphi_{0})$ are described by different equations, which could cause ambiguities in applications~\cite{Cadez:2004cg}. In fact, even though one adopts numerical integration method to describe the geodesic, this problem will also emerge, and to overcome it, special numerical skills are needed. Now, in this section, we will show that the analytical approach can endow us with the ability to circumvent the problem. Equations~(\ref{equA31}) and (\ref{equA36}) could be rewritten as
\begin{eqnarray}
\label{equA45}\pm(\varphi-\varphi_{0})\sqrt{\frac{X_{3}-X_{1}}{2}}+\F{\bigg(\vartheta_{\text{larI}}\Big(\frac{m}{r_{0}}\Big),k_{\text{lar}}\bigg)}&=&\left\{
\begin{array}{l}
\displaystyle\F{\bigg(\vartheta_{\text{larI}}\Big(\frac{m}{r}\Big),k_{\text{lar}}\bigg)},\\
\displaystyle2\K{(k_{\text{lar}})}-\F{\bigg(\vartheta_{\text{larI}}\Big(\frac{m}{r}\Big),k_{\text{lar}}\bigg)}\quad\ \,
\end{array}\right.
\end{eqnarray}
and
\begin{eqnarray}
\label{equA46}\pm(\varphi-\varphi_{0})\sqrt{\frac{X_{3}-X_{1}}{2}}+\F{\bigg(\vartheta_{\text{larII}}\Big(\frac{m}{r_{0}}\Big),k_{\text{lar}}\bigg)}&=&\left\{
\begin{array}{l}
\displaystyle\F{\bigg(\vartheta_{\text{larII}}\Big(\frac{m}{r}\Big),k_{\text{lar}}\bigg)},\\
\displaystyle-\F{\bigg(\vartheta_{\text{larII}}\Big(\frac{m}{r}\Big),k_{\text{lar}}\bigg)},\qquad
\end{array}\right.
\end{eqnarray}
respectively. It is known that the elliptic sine, elliptic cosine, and elliptic tangent functions of modulus $k$ are defined by~\cite{Abramowitz1965,Gradshteyn2007,boyd2016tracing}
\begin{eqnarray}
\label{equA47}\ssn\hspace{-0.05cm}(u,k)&=&\sin\hspace{-0.05cm}{[\am\hspace{-0.05cm}(u,k)]},\quad\scn\hspace{-0.05cm}(u,k)=\cos\hspace{-0.05cm}{[\am\hspace{-0.05cm}(u,k)]},\quad\ssc\hspace{-0.05cm}(u,k)=\tan\hspace{-0.05cm}{[\am\hspace{-0.05cm}(u,k)]}\qquad
\end{eqnarray}
with $\am(u,k)$ as the Jacobi amplitude function satisfying
\begin{eqnarray}
\label{equA48}\varphi=\am\hspace{-0.05cm}{[\F{(\varphi,k)},k]},
\end{eqnarray}
and these Jacobi elliptic functions bear the following properties
\begin{eqnarray}
\label{equA49}\ssn\hspace{-0.05cm}(-u,k)&=&-\ssn(u,k),\quad\ssn\hspace{-0.1cm}\left(u+2\K{(k)},k\right)=-\ssn\hspace{-0.05cm}(u,k),\\
\label{equA50}\scn\hspace{-0.05cm}(-u,k)&=&\scn(u,k),\quad\phantom{-}\scn\hspace{-0.1cm}\left(u+2\K{(k)},k\right)=-\scn\hspace{-0.05cm}(u,k),\\
\label{equA51}\ssc\hspace{-0.05cm}(-u,k)&=&-\ssc(u,k),\quad\ \ssc\hspace{-0.1cm}\left(u+2\K{(k)},k\right)=\ssc\hspace{-0.05cm}(u,k).
\end{eqnarray}
With these definitions and properties of Jacobi elliptic functions, by use of Eqs.~(\ref{equA28}) and (\ref{equA34}), one could prove that
 \begin{eqnarray}
\label{equA52}\ssn\hspace{-0.1cm}^{2}\hspace{-0.05cm}{\left[\F{\bigg(\vartheta_{\text{larI}}\Big(\frac{m}{r}\Big),k_{\text{lar}}\bigg)},k_{\text{lar}}\right]}\hspace{-0.05cm}&=&\ssn\hspace{-0.1cm}^{2}\hspace{-0.05cm}{\left[2\K{(k_{\text{lar}})}-\F{\bigg(\vartheta_{\text{larI}}\Big(\frac{m}{r}\Big),k_{\text{lar}}\bigg)},k_{\text{lar}}\right]}=\frac{m/r-X_{1}}{X_{2}-X_{1}},\qquad\ \\
\label{equA53}\ssc\hspace{-0.1cm}^{2}\hspace{-0.05cm}{\left[\F{\bigg(\vartheta_{\text{larII}}\Big(\frac{m}{r}\Big),k_{\text{lar}}\bigg)},k_{\text{lar}}\right]}\hspace{-0.05cm}&=&\ssc\hspace{-0.1cm}^{2}\hspace{-0.05cm}{\left[-\F{\bigg(\vartheta_{\text{larII}}\Big(\frac{m}{r}\Big),k_{\text{lar}}\bigg)},k_{\text{lar}}\right]}=\frac{m/r-X_{3}}{X_{3}-X_{2}},
\end{eqnarray}
and then substituting these two identities in Eqs.~(\ref{equA45}) and (\ref{equA46}) gives rise to
\begin{eqnarray}
\label{equA54}r&=&m\left[X_{1}+(X_{2}-X_{1})\ssn\hspace{-0.1cm}^{2}{\left(\pm(\varphi-\varphi_{0})\sqrt{\frac{X_{3}-X_{1}}{2}}+\F{\bigg(\vartheta_{\text{larI}}\Big(\frac{m}{r_{0}}\Big),k_{\text{lar}}\bigg)}\right)}\right]^{-1},\qquad\\
\label{equA55}r&=&m\left[X_{3}+(X_{3}-X_{2})\ssc\hspace{-0.1cm}^{2}{\left(\pm(\varphi-\varphi_{0})\sqrt{\frac{X_{3}-X_{1}}{2}}+\F{\bigg(\vartheta_{\text{larII}}\Big(\frac{m}{r_{0}}\Big),k_{\text{lar}}\bigg)}\right)}\right]^{-1}.\qquad
\end{eqnarray}
Obviously, Eqs.~(\ref{equA54}) and (\ref{equA55}), originating from Eqs.~(\ref{equA31}) and (\ref{equA36}), respectively, are the equations of the lightlike geodesic through the spatial point $(r_{0},0,\varphi_{0})$ when $b>b_{\text{cri}}$ under the situations of $r_{0}>r_{\text{pho}}$ and $2m\leqslant r_{0}<r_{\text{pho}}$. The derivations explicitly indicate that for either situation, compared with the original form of the equation, namely Eq.~(\ref{equA31}) or (\ref{equA36}), two branches of the geodesic now can be described by the identical equation, which results in the vanishing of
the branch ambiguities and thus will facilitate generating and analyzing the BH images. In the same manner, the original equations of the lightlike geodesic when $b<b_{\text{cri}}$ and $b=b_{\text{cri}}$ could also be recast as the similar form,
\begin{eqnarray}
\label{equA56}r&=&m\left[X_{1}+Y(X_{1})\left(\frac{1+\scn\hspace{-0.05cm}{[\pm(\varphi-\varphi_{0})\sqrt{2Y(X_{1})}}-\F{(\vartheta_{\text{les}}(m/r_{0}),k_{\text{les}})]}}{1-\scn\hspace{-0.05cm}{[\pm(\varphi-\varphi_{0})\sqrt{2Y(X_{1})}}-\F{(\vartheta_{\text{les}}(m/r_{0}),k_{\text{les}})]}}\right)\right]^{-1},\\
\label{equA57}r&=&m\left[\frac{1}{2}\tanh^{2}{\left(\pm\frac{1}{2}(\varphi-\varphi_{0})+\arctanh{\left(\sqrt{\frac{2m}{r_{0}}+\frac{1}{3}}\right)}\right)}-\frac{1}{6}\right]^{-1},\\
\label{equA58}r&=&m\left[\frac{1}{2}\coth^{2}{\left(\pm\frac{1}{2}(\varphi-\varphi_{0})-\arccoth{\left(\sqrt{\frac{2m}{r_{0}}+\frac{1}{3}}\right)}\right)}-\frac{1}{6}\right]^{-1},
\end{eqnarray}
and it is not difficult to verify that they are equivalent to their original forms. As mentioned earlier, the equation with the sign ``$+\ (-)$'' before $(\varphi-\varphi_{0})$ represents the counterclockwise (clockwise) lightlike geodesic, while photons move towards the BH from the point $(r_{0},0,\varphi_{0})$.

Up to now, in Schwarzschild spacetime, all the equations to describe the lightlike geodesics have been achieved, and they are composed by Eqs.~(\ref{equA54})--(\ref{equA58}) or equivalently by Eqs.~(\ref{equA17}),  (\ref{equA31}),  (\ref{equA36}),  (\ref{equA38}), and  (\ref{equA39}). Any lightlike geodesic in Schwarzschild spacetime could be described by one of these equations, where it should be reminded that the bound photon orbit is only a segment of the lightlike geodesic with $b=b_{\text{cri}}$.

\end{document}